\documentclass[english,aps,prl,reprint,superscriptaddress,numerical,showpacs,floatfix,nofootinbib,twocolumn]{revtex4-2}
\usepackage[T1]{fontenc}
\usepackage[latin9]{inputenc}
\usepackage[pdftex]{color}
\usepackage{babel}
\usepackage{amssymb}
\usepackage[pdftex]{graphicx}
\usepackage{tabularx}
\usepackage{footnote}
\usepackage{tabu}
\usepackage{amsmath}
\usepackage{mathrsfs}
\usepackage{dcolumn}

\begin{document}

\newcommand{\dm}[1]{\textcolor{green}{#1}}

\title{Nuclear \boldmath$\beta$ decay as a probe for physics beyond the Standard Model}

\author{M.~Brodeur}
\affiliation{Department of Physics and Astronomy, University of Notre Dame, Notre Dame, IN 46556 USA}
\author{N.~Buzinsky}
\affiliation{Department of Physics, University of Washington, Seattle, Washington 98195, USA}
\author{M.A.~Caprio}
\affiliation{Department of Physics and Astronomy, University of Notre Dame, Notre Dame, IN 46556 USA}
\author{V.~Cirigliano}
\affiliation{Institute for Nuclear Theory, University of Washington, Seattle, WA 98195, USA}
\author{J.A.~Clark}
\affiliation{Physics Division, Argonne National Laboratory, Lemont, Illinois 60439, USA}
\author{P.J.~Fasano}
\affiliation{Department of Physics and Astronomy, University of Notre Dame, Notre Dame, IN 46556 USA}
\author{J.A.~Formaggio}
\affiliation{Laboratory for Nuclear Science, Massachusetts Institute of Technology, 77 Mass. Ave., Cambridge, MA 02139}
\author{A.T.~Gallant}
\affiliation{Nuclear and Chemical Sciences Division, Lawrence Livermore National Laboratory, Livermore, California 94550, USA}
\author{A.~Garcia}
\affiliation{Department of Physics, University of Washington, Seattle, Washington 98195, USA}
\author{S.~Gandolfi}
\affiliation{Theoretical Division, Los Alamos National Laboratory}
\author{S.~Gardner}
\affiliation{Department of Physics and Astronomy, University of Kentucky, Lexington, KY 40506-0055}
\author{A.~Glick-Magid}
\affiliation{Department of Physics, University of Washington, Seattle, Washington 98195, USA}
\author{L.~Hayen}
\affiliation{Department of Physics, North Carolina State University, Raleigh, North Carolina 27695, USA}
\affiliation{Triangle Universities Nuclear Laboratory, Durham, North Carolina 27708, USA}
\author{H.~Hergert}
\affiliation{Facility for Rare Isotope Beams, Michigan State University, East Lansing, Michigan 48824, USA}
\affiliation{Department of Physics \& Astronomy, Michigan State University, East Lansing, Michigan 48824, USA}
\author{J.~D.~Holt}
 \affiliation{TRIUMF, Vancouver, BC V6T 2A3, Canada}%
\affiliation{Department of Physics, McGill University, Montr\'eal, QC H3A 2T8, Canada}
\author{M.~Horoi}
\affiliation{Department of Physics, Central Michigan University, Mount Pleasant, MI 48859, USA}
\author{M.Y.~Huang}
\affiliation{Department of Physics and Astronomy, Iowa State University, Ames, IA 50011, USA}
\author{K.D.~Launey}
\affiliation{Department of Physics and Astronomy, Louisiana State University, Baton Rouge, LA 70803, USA}
\author{K.G.~Leach}
\affiliation{Department of Physics, Colorado School of Mines, Golden, CO 80401, USA}
\affiliation{Facility for Rare Isotope Beams, Michigan State University, East Lansing, MI 48824, USA}
\author{B.~Longfellow}
\affiliation{Nuclear and Chemical Sciences Division, Lawrence Livermore National Laboratory, Livermore, California 94550, USA}
\author{A.~Lovato}
\affiliation{Physics Division, Argonne National Laboratory, Lemont IL 60439, USA}

\author{A.E.~McCoy}
\affiliation{Facility for Rare Isotope Beams, Michigan State University, East Lansing, MI 48824, USA}
\affiliation{Department of Physics, Washington University in Saint Louis, Saint Louis, MO 63130, USA}
\author{D.~Melconian}
\affiliation{Cyclotron Institute, Texas A\&M University, 3366 TAMU, College Station, Texas 77843-3366, USA}
\affiliation{Department of Physics and Astronomy, Texas A\&M University, 4242 TAMU, College Station, Texas 77843-4242, USA}
\author{P.~Mohanmurthy}
\affiliation{Laboratory for Nuclear Science, Massachusetts Institute of Technology, 77 Mass. Ave., Cambridge, MA 02139}
\author{D.C.~Moore}
\affiliation{Wright Laboratory, Department of Physics, Yale University, New Haven, CT 06520, USA}
\author{P.~Mueller}
\affiliation{Physics Division, Argonne National Laboratory, Lemont, Illinois 60439, USA}

\author{E.~Mereghetti}
\affiliation{Los Alamos National Laboratory,
Los Alamos, NM 87545, USA}

\author{W.~Mittig}
\affiliation{Department of Physics and Astronomy, Michigan State University, East Lansing 48824, USA}
\affiliation{Facility for Rare Isotope Beams, Michigan State University, East Lansing, MI 48824, USA}

\author{P.~Navratil}
\affiliation{TRIUMF, Vancouver, BC V6T 2A3, Canada}

\author{S.~Pastore}
\affiliation{Department of Physics, Washington University in Saint Louis, Saint Louis, MO 63130, USA}
\affiliation{McDonnell Center for the Space Sciences at Washington University in St. Louis, MO 63130, USA}

\author{M.~Piarulli}
\affiliation{Department of Physics, Washington University in Saint Louis, Saint Louis, MO 63130, USA}
\affiliation{McDonnell Center for the Space Sciences at Washington University in St. Louis, MO 63130, USA}

\author{D.~Puentes}
\affiliation{Department of Physics and Astronomy, Michigan State University, East Lansing 48824, USA}
\affiliation{Facility for Rare Isotope Beams, Michigan State University, East Lansing, MI 48824, USA}

\author{B.C.~Rasco}
\affiliation{Physics Division, Oak Ridge National Laboratory, Oak Ridge, TN 37830, USA}

\author{M.~Redshaw}
\affiliation{Department of Physics, Central Michigan University, Mount Pleasant, MI 48859, USA}

\author{G.H.~Sargsyan}
\affiliation{Nuclear and Chemical Sciences Division, Lawrence Livermore National Laboratory, Livermore, California 94550, USA}

\author{G.~Savard}
\affiliation{Physics Division, Argonne National Laboratory, Lemont, Illinois 60439, USA}
\affiliation{Department of Physics, University of Chicago, Chicago, Illinois 60637, USA}
\author{N.D.~Scielzo}
\affiliation{Nuclear and Chemical Sciences Division, Lawrence Livermore National Laboratory, Livermore, California 94550, USA}
\author{C.-Y.~Seng}
\affiliation{Department of Physics, University of Washington, Seattle, Washington 98195, USA}
\affiliation{Facility for Rare Isotope Beams, Michigan State University, East Lansing, MI 48824, USA}
\author{A.~Shindler}
\affiliation{Facility for Rare Isotope Beams, Michigan State University, East Lansing, Michigan 48824, USA}
\affiliation{Department of Physics \& Astronomy, Michigan State University, East Lansing, Michigan 48824, USA}

\author{S.R.~Stroberg}
\affiliation{Department of Physics and Astronomy, University of Notre Dame, Notre Dame, IN 46556 USA}

\author{J.~Surbrook}
\affiliation{Department of Physics and Astronomy, Michigan State University, East Lansing 48824, USA}
\affiliation{Facility for Rare Isotope Beams, Michigan State University, East Lansing, MI 48824, USA}

\author{A.~Walker-Loud}
\affiliation{Nuclear Science Division, Lawrence Berkeley National Laboratory, Berkeley, CA, 94720, USA}

\author{R.~B.~Wiringa}
\affiliation{Physics Division, Argonne National Laboratory, Lemont, IL 60439, USA}

\author{C.~Wrede}
\affiliation{Department of Physics and Astronomy, Michigan State University, East Lansing 48824, USA}
\affiliation{Facility for Rare Isotope Beams, Michigan State University, East Lansing, MI 48824, USA}

\author{A.~R.~Young}
\affiliation{Department of Physics, North Carolina State University, Raleigh 27695, USA}
\affiliation{Triangle Universities Nuclear Laboratory, Duke University, Durham 27708, USA}

\author{V.~Zelevinsky}
\affiliation{Department of Physics and Astronomy, Michigan State University, East Lansing 48824, USA}
\affiliation{Facility for Rare Isotope Beams, Michigan State University, East Lansing, MI 48824, USA}

\date{\today}

\begin{abstract}
    This white paper was submitted to the 2022 Fundamental Symmetries, Neutrons, and Neutrinos (FSNN) Town Hall Meeting
    in preparation for the next NSAC Long Range Plan.
    We advocate to support current and future theoretical and experimental searches for physics beyond the Standard Model using nuclear $\beta$ decay.
\end{abstract}

\maketitle

\section{Recommendations}
The nuclear $\beta$-decay research community has made impressive progress, both in theory and experiment, since the previous long range plan was implemented.  A wide variety of nuclear systems provide enhanced sensitivity to phenomena that can arise in beyond the Standard Model (BSM) contributions to the electroweak interaction. At the level of precision currently being achieved for this type of fundamental symmetries research, numerous higher-order corrections need to be accounted for through nuclear theory. In addition to progress on improving the determination of electroweak radiative corrections, one of the most dramatic developments has been the emergence of new many-body methods with controllable uncertainties that can be applied in a wide range of relevant nuclei. This progress has set the stage for productive interactions between theory and experiment to reassess and sharpen theoretical uncertainties and to interpret the high-quality data being collected with advanced experimental systems. 

One of the central goals for the coming years is to place  the on-going work with allowed decays on a more rigorous theoretical footing and push the precision frontier for the experimental campaigns targeting a set of key nuclei. These activities should produce improved results both for the "top-row" unitarity test and for the search for exotic couplings, allowing nuclear $\beta$ decay to continue to 
be at the forefront in probing new physics in charged current weak interactions. 

The remarkable progress on theory has been accompanied by a number of successful experimental programs that highlight the impact of relatively small-scale research efforts. These experiments have achieved the highest precision measurements to date of the $\beta$-asymmetry with $^{37}$K and the $\beta$-$\nu$ angular correlation with $^{8}$Li. At the same time, we have developed new spectroscopic techniques using cyclotron resonance spectroscopy (CRES), superconducting tunnel junctions, and novel ion-trapping systems. These advances are poised to yield a wealth of results with unprecedented precision that provide multiple stringent constraints on BSM scenarios.

The community's recommendations for advancing fundamental symmetries research using nuclear $\beta$ decays are:
\begin{itemize}
\item 
Accelerating progress on improving constraints on BSM scenarios and clarifying the status of the Cabibbo angle anomaly through the formation of a center to study fundamental symmetries with $\beta$ decay that strengthens collaboration between experiment and theory.
\item 
Increasing investment in small-scale and mid-scale projects and initiatives at universities, the ARUNA laboratories, and national laboratories to exploit the strengths of existing facilities and create pipelines for highly-qualified personnel into the field.
\item
Establishing robust and sustained support for the nuclear theory effort, which entails workforce development at all levels (students, post-docs, faculty) and viable career paths.

\item 
Pursuing and developing the cutting-edge experimental techniques used to perform \textit{model-independent} searches for BSM neutrino physics via $\beta$ decay, with the goal of a direct absolute neutrino mass measurement.

\item
The nuclear beta decay community strives to promote a diverse and inclusive environment while advocating for an increased financial support that would improve our graduate student standard of living.

\item
We recommend several items that would facilitate fundamental symmetry research at FRIB. These include the development of a solid stopper, a helium-jet ion source for commensal beam operation, means and staff support for producing beams and sources at FRIB from harvested isotopes, for example by using the existing batch mode ion source (BMIS), and more usable space for such future large precision experiments.

\end{itemize}

\section{Introduction}

Over the past decade, $\beta$-decay experiments have significantly extended the frontiers of knowledge on physics beyond the Standard Model. This has been done using a variety of probes including angular-correlation measurements (e.g., \cite{Behr2014, Asner2015, Mardor2018, burkey2022}) and precise $ft$-value measurements \cite{Towner2010, Hardy2014}. Further planned  experimental efforts using different techniques will  reliably discover or set stronger constraints on BSM physics effects \cite{Gonzalez2019}. Such measurements are done both at National and University laboratories. New experiments will achieve precision of 0.1\% or better, enabling them to reach TeV-scale physics \cite{Gonzalez2019}, and in most case to compete with high-energy searches at the Large Hadron Collider~\cite{Cirigliano:2012ab}. Important recent  developments in nuclear theory now allow fundamental physics to be extracted from experiments at this level of precision. 

Nuclear $\beta$ decay transitions can be used to extract $V_{ud}$ to test the CKM matrix unitarity, probe the presence of scalar and tensor currents, tensor coupling to right-handed neutrinos, search for sterile neutrino, and contribute to the neutrino mass determination.

The type of transition used to probe BSM physics include superallowed pure Fermi between 0$^+$ states, superallowed mixed between mirror nuclei, allowed pure Gamow-Teller, first-forbidden, as well as some specific electron-capture and ultra-low $Q$-value transitions. 

In this white paper we will highlight recent achievements, concisely articulate the scientific opportunities, and give the path towards discoveries using each of these nuclear $\beta$ decay transitions in the next decade and beyond. 

\section{CKM matrix unitarity tests}

\subsection{Motivation}

Precise tests of the CKM  matrix top-row unitarity are a unique, model-independent probe having discovery potential for new physics at the 10 TeV scale with current precision levels \cite{Gonzalez2019}.  $\beta$ decay supplies over 95\% of the top-row sum data. Following recent broad theory progress and signs of top-row unitarity violation at the $>3\sigma$ level, we are in a most opportune time to deliver a comprehensive advancement in the determination of the $V_{ud}$ matrix element using nuclear $\beta$ decays. In particular, mature ab initio nuclear theory efforts and a high-precision nuclear data set underscore the potential/need for a concerted effort between theory and experiment. Formation of a
 topical group, the VUDU (Vud Unitarity) alliance, would foster further collaboration, strengthen theory benchmarking efforts, amplify the impact of focused experimental efforts and sustain the leadership role of the nuclear $\beta$ decay community in precision CKM unitarity tests.

\subsection{Progress}

\begin{figure}[ht]
    \centering
    \includegraphics[width=0.48\textwidth]{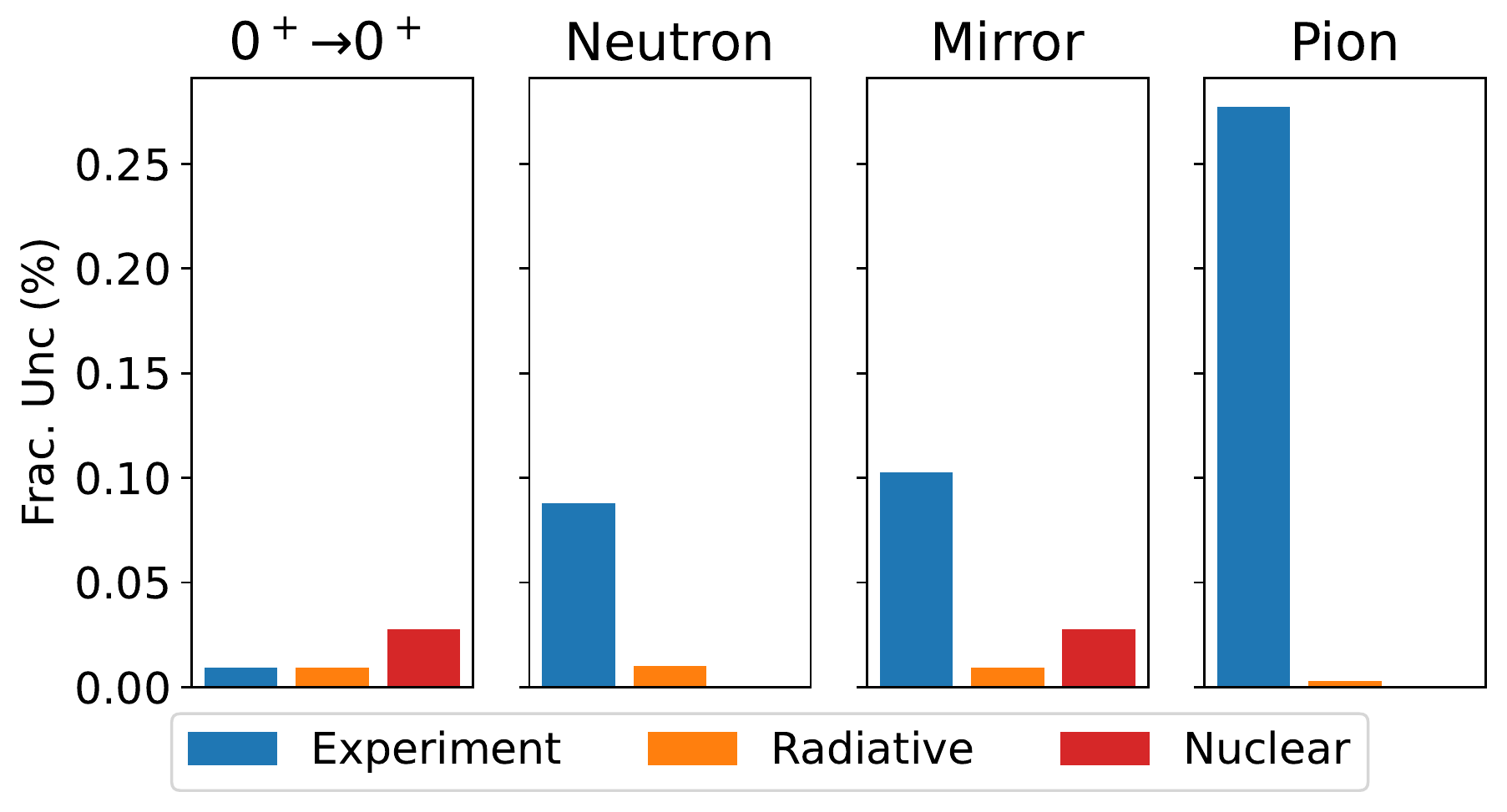}
    \caption{Fractional uncertainties to the $V_{ud}$ extraction from the most precise channels.}
    \label{fig:frac_unc_Vud}
\end{figure}

The $V_{ud}$ element of the CKM matrix can be determined from the $\beta$ decays of the pion, neutron, pure Fermi $0^+$ to $0^+$ and mixed, mirror nuclear decays \cite{Towner2010-Vud}. The current status is summarized in Fig. \ref{fig:frac_unc_Vud} showing the fractional uncertainty due to experimental input, electroweak radiative corrections, and nuclear structure corrections where relevant. Currently, the most precise determination of $V_{ud}$ is obtained from $0^+\to 0^+$ nuclear decays, where a collection of more than 200 measurements in 21 different nuclei allow for a significant statistical advantage. Its precision is currently limited by uncertainties in sub-percent level nuclear structure corrections, the improvement of which is a strong driver for theoretical advances and comprises one of the major goals in the field.

Significant theoretical progress in the calculation of radiative corrections using dispersion relations was made possible by a critical paradigm shift and served to reduce uncertainties due to radiative corrections in all systems by at least a factor of two \cite{Seng2018, Seng2019b}. The resultant $3\sigma$ shift compared to the previous state of the art for the neutron and nuclear decays has been confirmed by several independent calculations \cite{Czarnecki2019, Hayen2021, Shiells2021}, and have been shown to be systematically improvable using lattice QCD calculations \cite{Feng2020, Seng2020,Seng:2019plg}.

Leveraging improvements in the theory framework, nuclear structure effects in electroweak radiative corrections (denoted $\delta_{NS}$) were reevaluated and resulted in substantial changes when including quasi-elastic and nuclear polarization effects. These were initially treated in simplified models and increased the uncertainty on the resultant $V_{ud}$ determination from $0^+\to 0^+$ decays by 50\% due to fully correlated theoretical uncertainties~\cite{Seng2019b,Gorchtein:2018fxl,Hardy2020}. A fully-relativistic framework~\cite{Seng:2022cnq} permits rigorous studies of $\delta_{NS}$ using \textit{ab initio} methods;  the past two decades have witnessed a tremendous progress of the latter in the description of nuclei. This was due to the advent of effective field theories that link  nuclear many-body interactions and electroweak currents to the fundamental symmetries of  the underlying theory of Quantum Chromodynamics~\cite{Weinberg:1990rz,Weinberg:1991rw,Weinberg:1992yk,Machleidt:2016yo,King_2020,Epelbaum:2020rr,Kolck:2020dn,Krebs:2020pii,Baroni:2021vrc, Gysbers2019}; the development of new algorithms suited to solve the many-body nuclear problem for nuclei in the medium mass region and beyond~\cite{10.3389/fphy.2020.00379};
and the increased availability of computational resources~\cite{osti_1369223}. Capitalizing on these recent developments, the theory community 
is poised to provide improved theoretical estimates for the nuclear structure effects in radiative corrections ($\delta_{NS}$) and isospin breaking effects ($\delta_{C}$).

There has been a surge of experimental activity in the past years around mirror transitions at various institutions world-wide including: half-life ($^{37}$K~\cite{Shidling2014}, $^{21}$Na~\cite{Shidling2018} and $^{29}$P~\cite{Iacob-InPrep}) and branching ratio ($^{37}$K~\cite{Ozmetin-InPrep}) measurements at Texas A\&M University; half-life measurements of $^{11}$C~\cite{Valverde2018},
$^{13}$N~\cite{Long2022}, $^{15}$O~\cite{Burdette2020}, $^{25}$Al~\cite{Long2017} and $^{29}$P~\cite{Long2020} at the University of Notre Dame; $Q_{EC}$-value measurements of $^{11}$C \cite{Gulyuz2016},  $^{21}$Na and $^{29}$P using LEBIT at NSCL~\cite{Eibach2015}; and with significant development of $99.13(9)\%$ nuclear polarization via optical pumping~\cite{fenkerNJP}, a precise $\beta$-asymmetry measurement of $^{37}$K using TRINAT at TRIUMF improved the value of $V_{ud}$ for this isotope by a factor of 4~\cite{Fenker2018}.

\subsection{Prospects}

Significant progress in the determination of nuclear structure correction using nuclear \textit{ab initio} methods are paramount in order to maximize the potential of the superallowed global data set for $V_{ud}$ extraction. In particular, a benchmarking effort centered around low mass nuclei with high precision experimental data ($^6$He, $^{10, 11}$C, $^{14}$O, $^{19}$Ne) that are accessible to nuclear many-body methods with a minimum number of approximations (No Core Shell Model, Quantum Monte Carlo, Lattice Effective Field Theory, $\ldots$) and methods with a wider mass reach (Coupled Cluster, In Medium Similarity Renormalization Group, and hybrid models) will allow one to reliably compute corrections for the full data set. Supplemented by focused experimental measurements of $0^+ \to 0^+$ and mirror decays, the community foresees a synergistic approach with maximal impact.  Given that the uncertainties for the value of $V_{ud}$ determined from the superallowed decays are dominated by the uncertainty in theoretical corrections due to nuclear structure effects in the electroweak radiative corrections, we can anticipate significant progress in the precision with which these are calculated and a reduction in the uncertainty budget for the superallowed data set during the next long range planning period.  This should shift uncertainties in $V_{ud}$ back to the electroweak radiative corrections for the nucleon (and presumably kaon decay). The more challenging goal of quantifying and reducing uncertainties in the analysis of isospin-mixing effects will proceed in parallel, with an immediate increment in precision for some BSM tests when this goal is achieved.

In particular, there is a strong need for more precise measurements of the branching ratio of the $0^+\rightarrow0^+$ transitions of $^{10}$C and $^{14}$O. Both of these isotopes weigh in most on searches for exotic scalar currents through a non-zero Fierz interference term. 
Through the development of quantum sensors at radioactive ion beam facilities (e.g.\ using superconducting tunnel junctions), measurements of the branching ratios of both could be performed through recoil spectroscopy as a way of avoiding common systematic effects. Additional information on recoil-order and isospin breaking corrections may be obtained using precise electroweak nuclear radii measurements in several isotriplet systems \cite{Seng:2022epj,Seng:2022inj}. On the theory side, a reliable determination by \emph{ab initio} methods of both the $\delta_C$ and $\delta_{NS}$ nuclear structure corrections for $^{10}$C and $^{14}$O transitions is now within the reach.

The $\beta$-delayed proton decays of $^{20}$Mg, $^{24}$Si, $^{28}$S, $^{32}$Ar and $^{36}$Ca, to be studied at TAMUTRAP~\cite{shidling2021}, will provide alternate $0^+\rightarrow0^+$ cases once the $^{3}$He-LIG system at the Cyclotron Institute is fully commissioned.  These near-proton-dripline cases will have vastly different experimental systematics and provide a demanding test of isospin-symmetry-breaking corrections ($\delta_C$).

Superallowed mixed $\beta$ decay transitions between mirror nuclei have been proposed as an independent means to extract the $V_{ud}$ element of the CKM matrix \cite{Naviliat2009} by measuring the mixing ratio through, e.g., angular correlations. While this requires an additional experimental input, substantial enhancements are available through near-cancellation of the observable, exceeding that of the neutron (e.g.$^{17}$F) to up to a factor 13 ($^{19}$Ne) \cite{Hayen2020}. Besides multiple on-going analysis and future half-life, branching ratio, and $Q_{EC}$-value measurements, several efforts to measure correlation parameters in mirror transitions are underway. These includes more precise angular correlation measurements ($a_{\beta\nu}$, $A_\beta$, recoil-asymmetry, \ldots) of K and Rb isotopes with TRINAT, and the St.\ Benedict ion trapping system~\cite{Brodeur2016, OMalley2020} at the University of Notre Dame that will be devoted to measuring $\beta$-$\nu$ angular correlations in multiple mirror transitions including the very sensitive $^{17}$F\@. 

Recently, the use of superconducting tunnel junctions has shown tremendous promise for precision spectroscopy of recoiling ions following nuclear beta decay with vastly different systematic corrections to traditional approaches \cite{Fretwell2020, Friedrich2021}. Unlike other quantum sensors, the microsecond(s) response time enables both high precision and high count rate spectroscopy, making them an ideal emerging technology for use at radioactive ion beam facilities. The Superconducting Array for Low Energy Radiation (SALER) targets precision recoil spectroscopy of short-lived mirror isotopes such as $^{11}$C, which can open additional channels for precision $V_{ud}$ determinations. This includes isotopes typically inaccessible using ion or atom trap technology due to their long lifetimes, thereby providing complementary input.

Besides a strong potential for a competitive $V_{ud}$ extraction, additional precision measurements in the low mass range ($A < 20$) provide critical input for nuclear ab initio theory efforts to benchmark and improve the nuclear structure corrections limiting the $0^+ \to 0^+$ $V_{ud}$ determination.

\section{Searches for tensor and scalar currents}

\subsection{Motivation}

Detailed studies of angular correlations in nuclear $\beta$ decay played a key role in elucidating the ``vector-minus-axial vector'' (V$-$A) structure of the 
charged current electroweak interaction, which is mediated  by the W boson. 
Today, nuclear $\beta$ decay efforts remain at the forefront in searches for evidence of the additional scalar (S) and tensor (T) Lorentz-invariant interactions that naturally arise in SM extensions. Measurements of the $\beta$-$\nu$ angular correlation, the $\beta$ asymmetry, and the Fierz interference term provide important constraints on BSM physics.  Ongoing and planned experiments are poised to further refine these measurements, continuing to reach sensitivities that surpass that of the LHC.

Nuclei that undergo allowed nuclear $\beta$ decay provide excellent laboratories for these types of measurements. The underlying nature of the electroweak interaction can be isolated because the uncertainties associated with the nuclear medium are minimized. The nuclear-physics corrections that arise are typically of order 1\%; as these measurements now aim for 0.1\% precision or beyond, these corrections need to be carefully understood through nuclear-theory calculations and experimental constraints when possible. In addition, in certain select cases, precise $\beta$-shape functions for forbidden $\beta$ decays can access exotic couplings that are not accessible with allowed decays \cite{Glick-Magid2017}. Measurements of spin asymmetry in EC decay of polarized nuclei are also proposed, taking advantage of its linear dependence on exotic couplings.

\subsection{Progress}

Over the past decade, experimental developments paired with modern nuclear-theory calculations have yielded a new generation of $\beta$-decay studies that continue to reach unprecedented sensitivity. 

Atom-trap and ion-trap techniques have been used to collect and suspend samples of $\beta$-emitting isotopes in vacuum, allowing both the measurement of the low-energy nuclear recoils, from which the neutrino momentum can be inferred, and the opportunity to polarize the confined nuclei. Experiments with the BPT at the ATLAS facility at ANL, have achieved increasingly precise results for the $\beta$-$\nu$ angular correlations in $^{8}$Li~\cite{sternberg2015,burkey2022} and $^{8}$B~\cite{gallant2022}. Atom traps have been used to determine this correlation in $^{6}$He~\cite{mueller2022} and to polarize $^{37}$K atoms to measure the $\beta$ asymmetry~\cite{fenkerNJP,Fenker2018}. These experiments have achieved precision as good  as $0.3\%$, placing stringent limits on the possible existence of tensor interactions and right-handed currents, and there are well-defined paths to further improve the precision.

In addition, new ultra-sensitive detection techniques have been demonstrated that will undoubtedly increase the precision of angular-correlation measurements. The CRES measurement approach, first demonstrated with low-energy electrons, has recently been applied to the study of higher-energy $\beta$ particles from the decays of $^{6}$He and $^{19}$Ne~\cite{CRES-arXiv}. The implantation of $\beta$-emitters in ultrahigh-resolution cryogenic detectors and scintillator detectors also show great promise by capturing the full energy of the recoiling nucleus or the emitted $\beta$ particle. 

At the same time, significant theoretical progress has been made in nuclear \textit{ab initio} methods for  precision calculations of $\beta$-decay observables, including energy spectra and angular correlations. Combined with a recent review and extension of allowed $\beta$ decay spectroscopy corrections \cite{Hayen2018}, as well as shape and recoil corrections for allowed and forbidden $\beta$ decays \cite{Glick-Magid2022formalism}, several independent calculations have provided precision input for interpretation of $\beta$ decays in $^{6}$He~\cite{King2022, Glick-Magid2022} and $^{8}$Li~\cite{Sargsyan2022}, which vastly improves upon the previous state of the art. As shown in Ref. \cite{Sargsyan2022}, highly reduced theoretical uncertainties have been achieved by identifying a strong correlation between the recoil-order terms and quadrupole moments, which emphasizes the significance of the proper treatment of collective features in nuclei, including \textit{ab initio} predictions of quadrupole moments and E2 transitions without effective charges. Similar efforts are underway for additional decays to ensure theoretical precision at the 0.01\% level, thereby enabling discovery potential at the 10-TeV level.
Comparison of $\beta$ decay in consecutive isotopes is very important as there is no good understanding of interplay between the weak and strong interaction that is responsible for collective vibrations and rotations that can influence Gamow-Teller decays \cite{Zelevinsky2017}.

\subsection{Prospects}

This is a particularly exciting time for the search for BSM physics with nuclear $\beta$ decay. Existing efforts have matured to the point where they are providing probes of new physics that are competitive (and complementary) with the reach of the LHC, and additional reach is imminent. In addition to increased sensitivity to be obtained by atom and ion trap based experiments, a set of new approaches are being developed to determine the $\beta$ spectral shape to unprecedented precision, therefore greatly increasing the ability to determine the Fierz interference term.

Further increase in sensitivity using the mass-8 system is being pursued and will require access to high-intensity beams of $^{8}$Li and $^{8}$B. New trap-structure designs to minimize $\beta$-particle scattering and efforts to better characterize the detector-array performance will further reduce uncertainties. In addition, a better understanding of the low-lying continuum level structure of $^8$Be, including resolving the question of the existence of low-lying intruder states, and the associated recoil-order contributions will be needed. 

Additional devices like TAMUTRAP and St. Benedict are poised to further extend the reach of precision angular-correlation measurements. A recent global fit of nuclear and neutron beta decay data show a hint of BSM tensor coupling to right-handed neutrinos at the 3$\sigma$-level \cite{Falkowski2021}. This effect, that could be generated by various BSM effects such as a TeV-range leptoquark coupling to light quarks, positrons, and right-handed neutrinos, can only be seen with the inclusion of correlation measurement data of mirror transitions in the data set. Hence, there is a critical need to expend the mirror transition correlation measurement data set using instruments such as St. Benedict to better constrain this effect.
Finally, the HUNTER collaboration is proposing a precision EC spin asymmetry measurement for which linear dependence on tensor couplings offers strong potential for advances. 

The CRES approach can be used to determine the spectral shapes of the $^6$He and $^{19}$Ne decays. By studying both $\beta^{-}$ and $\beta^{+}$ decays, the sign of the Fierz interference term changes sign, and therefore measuring both significantly reduces most systematic effects. In addition, by confining $\beta$-emitters in a specially-designed Penning trap for CRES measurements, the approach can be extended to study any isotope, including nuclei that decay by pure Fermi transitions and enable increased sensitivity to scalar interactions.  As high precision methods for spectrum measurement are refined, sub 0.1\% precision in spectral observables can enable a direct analysis of endpoint-related effects in the electroweak radiative corrections as well.

Quantum sensors will enable high-precision nuclear-recoil spectroscopy following $\beta$ decay for a wide range of accessible isotopes in experiments such as SALER at FRIB. In addition to competitive determinations of $V_{ud}$, precision measurements recoil spectra following beta decay and the relative decay fractions into electron capture and $\beta^+$ branches provide a sensitivity enhancement to a non-zero Fierz interference term with substantially reduced nuclear corrections.

Similar to the $^8$Li beta decay \cite{Sargsyan2022}, \textit{ab initio} calculations of $^8$B beta decay recoil-order corrections are ongoing to help reduce the uncertainty on the tensor current limits from the $\beta$-$\nu$ angular correlation measurements in this nucleus. Given the collective and cluster nature of the decay product $^{8}$Be, as well as the near-threshold ground state of $^{8}$B, \textit{ab initio} approaches that use hybrid basis such as symmetry-adapted and continuum bases allow one to reliably compute such contributions. 
Furthermore, with the help of the symmetry-adapted basis one can extend the calculations to heavier systems such as $^{19}$Ne. Future experiments would benefit from having the energy dependence of the recoil-order terms, which can be obtained by computing their response functions.

\section{Beta decays for neutrino physics}

\subsection{Motivation}
The lepton sector of the SM provides a unique window into BSM physics given the confirmed observation of non-zero neutrino masses~\cite{Fuk98,Ahm01}, 
the persisting hint of the muon $g-2$ anomaly~\cite{PhysRevLett.126.141801},
and several other outstanding claims of leptonic BSM physics.  As a result, extensions to the SM description of leptons are unavoidable and must account for the fact that neutrinos have at least two non-zero mass eigenstates.  The search for how to extend the SM in this area may indeed lead us to a wide range of BSM physics, including a connection to the dark sector.

\subsection{Progress and future - BSM neutrino masses}
Energy and momentum conservation in nuclear $\beta$ decay allows  model-independent searches for  \textit{any} new neutrino mass physics coupled to the electron flavor, and is a uniquely powerful method for BSM physics searches in this area.  This includes the absolute neutrino mass measurements of the light mass states via $\beta$ decay endpoint measurements as well as the search for new, heavy (mostly sterile) mass states as an expansion to the $3\times3$ PMNS matrix.

\subsubsection{Absolute neutrino mass measurements}
Tritium $\beta$ decay remains one of the most sensitive measurements of the absolute neutrino mass scale, independent of the nature of the neutrino mass (Majorana or Dirac).  Currently, the KATRIN neutrino experiment has set the most stringent limit on the neutrino mass scale of $m_\beta \leq 0.8$ eV/c$^2$ at 90\% C.L. using molecular tritium~\cite{Aker2022}, with a final target mass sensitivity of  $m_\beta \leq 0.2$ eV/c$^2$ at 90\% C.L.  The Project 8 experiment, which uses the CRES technique to measure electrons from $\beta$ decay, is developing an R\&D program for an {\em atomic} tritium source, with a target sensitivity of $m_\beta \leq 0.04$ eV/c$^2$ at 90\% C.L.~\cite{Project8:2022wqh}.  This is among the highest-impact physics cases in the $\beta$ decay community, however since a separate, dedicated whitepaper will be forthcoming from these collaborations we do not emphasize it here.

To go beyond $m_\beta \leq 0.04$ eV/c$^2$, however, new experimental paradigms must be considered.  Of growing interest are ultra-low Q value $\beta$-decays that would occur from the ground state of the parent isotope to an excited nuclear state in the daughter with $Q_{ES} = Q_{GS} - E^{*} \lesssim 1$ keV. Such decays could provide new candidates for direct neutrino mass determination experiments~\cite{Cattadori2007_115In} and further insight into atomic interference effects in $\beta$-decay at low energies~\cite{Mustonen2010_ULQCalcs}.  A number of isotopes have been found that could have an ultra-low Q value transition~\cite{Mustonen2010_ULQs,Mustonen2011_135Cs,Haaranen2013_115Cd,Suhonen2014_ULQs,Gamage2019_ULQs,Keblbeck2022_ULQs}, but more precise Q value (from Penning traps), and in some cases energy level data is needed~\cite{Horana2022_75As,Ramalho2022_75As,Ge2022_111In,Eronen2022_131I,Ge2021_159Dy,deRoubin2020_135Cs}.  Experimentally, these ultra-low $Q$ values are challenging to implement, however recent work with trapped nanoscale objects may permit a variety of isotopes to be characterized while reaching sensitivities that are sufficient to resolve the requisite momenta in a single nuclear decay~\cite{Carney:2022pku}. Solid materials allow a high density of nuclei to be confined in a trap, and enable control and readout of the motional state of the particle using tools from quantum optomechanics.  Although challenging, it is plausible that smaller particles may eventually reach the momentum sensitivity needed to detect the mass of the light SM neutrinos. If an ultra-low $Q$ value EC or $\beta$ transition ($\leq0.1$~keV) were also identified with sufficiently high decay rate, detection of the light neutrino masses with this technique may be possible~\cite{Carney:2022pku}.

\subsubsection{Direct search for sub-MeV sterile neutrinos}
\begin{figure}[t!]
    \centering
    \includegraphics[width=\linewidth]{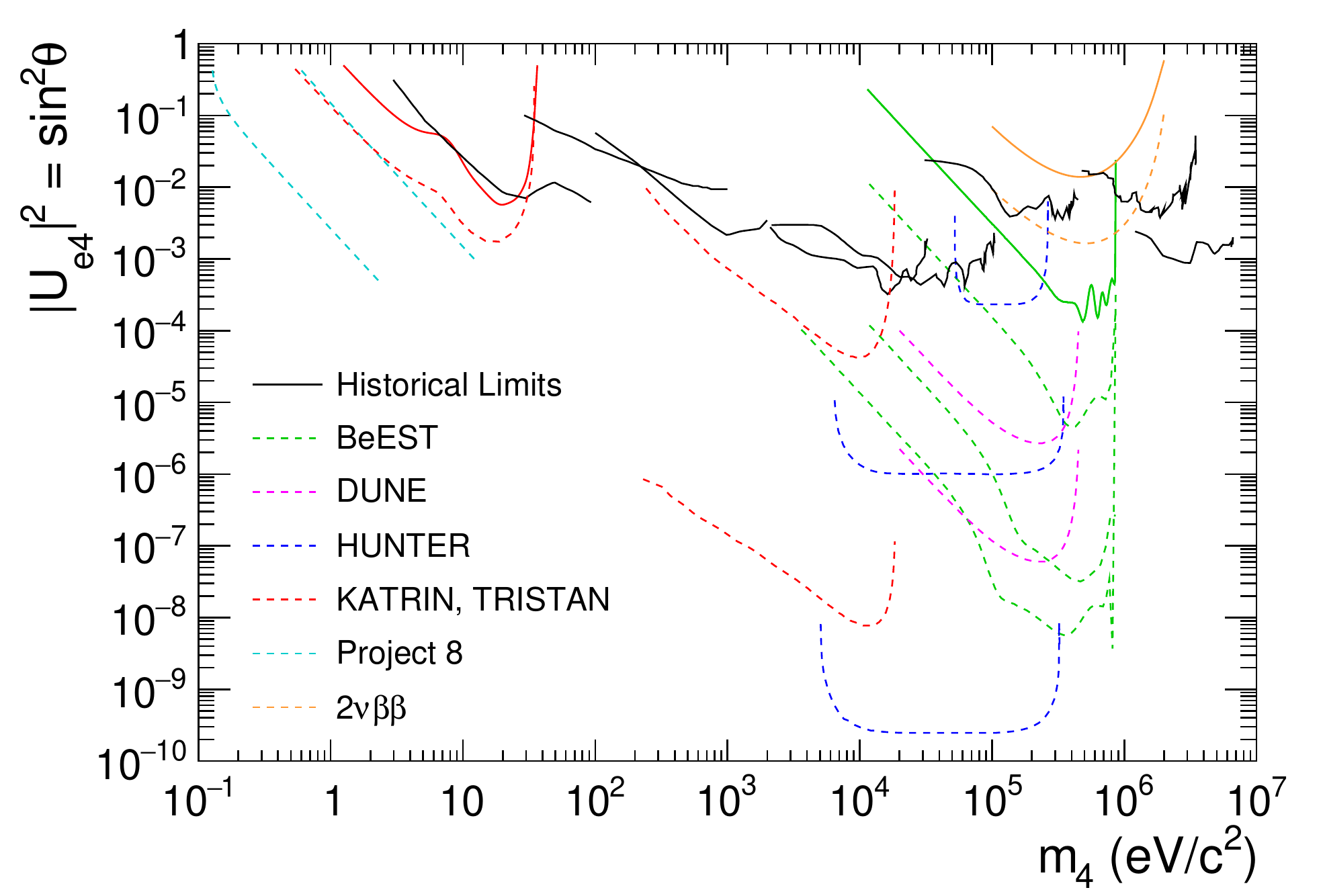}
    \caption{Projected sensitivities for heavy sterile neutrino searches in the eV - MeV mass range for current and planned experiments including the KATRIN/TRISTAN, Project-8, HUNTER, and the BeEST.  Figure from Snowmass 2022~\cite{Acero:2022wqg}.}
    \label{fig:limits}
\end{figure}
The search for sub-MeV sterile neutrinos via precision nuclear decay measurements is among the most powerful methods for BSM massive-neutrino searches since it relies only on the existence of a heavy neutrino admixture to the active neutrinos, and not on the model-dependent details of their interactions.  Sub-MeV sterile neutrinos are well motivated, natural extensions to the Standard Model (SM) that have been extensively studied over the past 25 years~\cite{Dod94,Adh17,Boy19}.  To date, the vast majority of laboratory-based experimental searches for neutrinos in this mass range have been performed using momentum and energy conservation in nuclear $\beta$ decay.  The neutrino ``missing mass" is reconstructed using precise momentum measurement of all other products (including the recoil nucleus) from decay of a nucleus at rest.  The experimental situation is simplified dramatically in neutron-deficient nuclei where the 3-body $\beta$ decay mode is energetically forbidden, and thus the parent nucleus \textit{only} undergoes nuclear electron capture (EC) decay.  Precision measurement of the low-energy nuclear
recoil and all the other (low energy) decay products allows the neutrino four-momentum and mass to be directly probed.  Measurements of this type are currently being performed using $^7$Be decay by the BeEST experiment~\cite{Leach:2021bvh,Fretwell2020,Friedrich2021} and planned for $^{131}$Cs by the HUNTER experiment~\cite{Martoff:2021vxp}.  The projected limits from these experiments are impressive (Fig.~\ref{fig:limits}) and will provide the most stringent constraints on the existence of these particles.  In fact, the BeEST experiment currently sets the best laboratory limits in the 100 - 850~keV mass range of any experimental method~\cite{Friedrich2021}.

\subsection{Other prospects}
$\beta$ decay measurements not only provide direct probes of BSM physics but also support other exotic neutrino physics searches, including:
\begin{itemize}
    \item Reactor-Antineutrino Anomaly (RAA) -- Experimental $\beta$-shape functions are needed in order to get to \% level $\Bar{\nu}$ flux predictions~\cite{fallot2012,hayes2014,sonzogni2015,rasco2016,fijalkowska2017,estienne2019,hayen2019}.
    \item Precisely measure dominant backgrounds for dark matter searches and for neutrino-less-double $\beta$ decays ($0\nu\beta\beta$)~\cite{stuckel2022}. 
\end{itemize}

Nuclear theory also plays a critical role in this area.  In particular, neutrinoless double beta ($0\nu\beta\beta$) decay nuclear matrix elements~\cite{HoroiNeacsu-prc18,Horoi-prd16,SenkovHoroi-prc16,BrownFangHoroi2015}, precision $\beta$- and EC-decay spectral calculations~\cite{Glick-Magid2017,Hayen2018,hayen2019,ramalho2022}, and neutrino-nucleus interactions. Both experimental and theoretical spectroscopy efforts to resolve shape factors of forbidden beta decays in the fission fragment region and their inclusion in nuclear databases has been recognized as a significant goal for the reactor neutrino field \cite{Fallot2019}.

\section{Rare decays}

\subsection{Motivation}

Rare decays in the nuclear domain may represent a window for the study of unknown phenomena. As an example, it was suggested that the neutron lifetime anomaly could be caused by a dark decay branch \cite{Fornal2018}, on which neutron star observables can provide strong constraints \cite{Berryman2022}. This could have as a consequence that very loosely bound neutrons in exotic nuclei could decay in a similar way, and the residual nucleus would be the signature of such a decay \cite{Pfutzner2018}. 

\subsection{Progress and future prospects}

A promising candidate is the decay of $^{11}$Be where the dark decay would produce $^{10}$Be as residue. $^{11}$Be also has a $(\beta, p)$ decay branching leading to $^{10}$Be (an unusual decay for a neutron rich nucleus) that was measured recently \cite{Ayyad2019}.  It was confirmed that this decay proceeds via a near threshold resonance in $^{11}$B \cite{Ayyad2022,Lopez2022}. This resonance near threshold can be considered as an example of an open quantum system resonance. The question if there is, or if there is not, a signature for a dark decay is not yet solved because of scattered results on the $^{10}$Be production ratio \cite{Riisager2020}. Another case that was measured in 2021 is the $\beta$ decay of  $^{6}$He and search of a dark neutron decay to the unbound $^{5}$He that can be probed by the following neutron decay to $^{4}$He. A very low upper limit was found in this case \cite{Savajols2022}. More generally, rare decays will favor visibility of higher order effects because they may become competitive with respect to standard interactions.

\bibliographystyle{apsrev4-2}
\bibliography{main}

\begin{thebibliography}{114}%
\makeatletter
\providecommand \@ifxundefined [1]{%
 \@ifx{#1\undefined}
}%
\providecommand \@ifnum [1]{%
 \ifnum #1\expandafter \@firstoftwo
 \else \expandafter \@secondoftwo
 \fi
}%
\providecommand \@ifx [1]{%
 \ifx #1\expandafter \@firstoftwo
 \else \expandafter \@secondoftwo
 \fi
}%
\providecommand \natexlab [1]{#1}%
\providecommand \enquote  [1]{``#1''}%
\providecommand \bibnamefont  [1]{#1}%
\providecommand \bibfnamefont [1]{#1}%
\providecommand \citenamefont [1]{#1}%
\providecommand \href@noop [0]{\@secondoftwo}%
\providecommand \href [0]{\begingroup \@sanitize@url \@href}%
\providecommand \@href[1]{\@@startlink{#1}\@@href}%
\providecommand \@@href[1]{\endgroup#1\@@endlink}%
\providecommand \@sanitize@url [0]{\catcode `\\12\catcode `\$12\catcode
  `\&12\catcode `\#12\catcode `\^12\catcode `\_12\catcode `\%12\relax}%
\providecommand \@@startlink[1]{}%
\providecommand \@@endlink[0]{}%
\providecommand \url  [0]{\begingroup\@sanitize@url \@url }%
\providecommand \@url [1]{\endgroup\@href {#1}{\urlprefix }}%
\providecommand \urlprefix  [0]{URL }%
\providecommand \Eprint [0]{\href }%
\providecommand \doibase [0]{https://doi.org/}%
\providecommand \selectlanguage [0]{\@gobble}%
\providecommand \bibinfo  [0]{\@secondoftwo}%
\providecommand \bibfield  [0]{\@secondoftwo}%
\providecommand \translation [1]{[#1]}%
\providecommand \BibitemOpen [0]{}%
\providecommand \bibitemStop [0]{}%
\providecommand \bibitemNoStop [0]{.\EOS\space}%
\providecommand \EOS [0]{\spacefactor3000\relax}%
\providecommand \BibitemShut  [1]{\csname bibitem#1\endcsname}%
\let\auto@bib@innerbib\@empty
\bibitem [{\citenamefont {Behr}\ \emph {et~al.}(2014)\citenamefont {Behr},
  \citenamefont {Gorelov}, \citenamefont {Jackson}, \citenamefont {Pearson},
  \citenamefont {Anholm}, \citenamefont {Kong}, \citenamefont {Behling},
  \citenamefont {Fenker}, \citenamefont {Melconian}, \citenamefont {Ashery}
  \emph {et~al.}}]{Behr2014}%
  \BibitemOpen
  \bibfield  {author} {\bibinfo {author} {\bibfnamefont {J.}~\bibnamefont
  {Behr}}, \bibinfo {author} {\bibfnamefont {A.}~\bibnamefont {Gorelov}},
  \bibinfo {author} {\bibfnamefont {K.}~\bibnamefont {Jackson}}, \bibinfo
  {author} {\bibfnamefont {M.}~\bibnamefont {Pearson}}, \bibinfo {author}
  {\bibfnamefont {M.}~\bibnamefont {Anholm}}, \bibinfo {author} {\bibfnamefont
  {T.}~\bibnamefont {Kong}}, \bibinfo {author} {\bibfnamefont {R.}~\bibnamefont
  {Behling}}, \bibinfo {author} {\bibfnamefont {B.}~\bibnamefont {Fenker}},
  \bibinfo {author} {\bibfnamefont {D.}~\bibnamefont {Melconian}}, \bibinfo
  {author} {\bibfnamefont {D.}~\bibnamefont {Ashery}}, \emph {et~al.},\
  }\href@noop {} {\bibfield  {journal} {\bibinfo  {journal} {Hyperfine
  Interactions}\ }\textbf {\bibinfo {volume} {225}},\ \bibinfo {pages} {115}
  (\bibinfo {year} {2014})}\BibitemShut {NoStop}%
\bibitem [{\citenamefont {Asner}\ \emph {et~al.}(2015)\citenamefont {Asner},
  \citenamefont {Bradley}, \citenamefont {de~Viveiros}, \citenamefont {Doe},
  \citenamefont {Fernandes}, \citenamefont {Fertl}, \citenamefont {Finn},
  \citenamefont {Formaggio}, \citenamefont {Furse}, \citenamefont {Jones},
  \citenamefont {Kofron}, \citenamefont {LaRoque}, \citenamefont {Leber},
  \citenamefont {McBride}, \citenamefont {Miller}, \citenamefont {Mohanmurthy},
  \citenamefont {Monreal}, \citenamefont {Oblath}, \citenamefont {Robertson},
  \citenamefont {Rosenberg}, \citenamefont {Rybka}, \citenamefont {Rysewyk},
  \citenamefont {Sternberg}, \citenamefont {Tedeschi}, \citenamefont
  {Th\"ummler}, \citenamefont {VanDevender},\ and\ \citenamefont
  {Woods}}]{Asner2015}%
  \BibitemOpen
  \bibfield  {author} {\bibinfo {author} {\bibfnamefont {D.~M.}\ \bibnamefont
  {Asner}}, \bibinfo {author} {\bibfnamefont {R.~F.}\ \bibnamefont {Bradley}},
  \bibinfo {author} {\bibfnamefont {L.}~\bibnamefont {de~Viveiros}}, \bibinfo
  {author} {\bibfnamefont {P.~J.}\ \bibnamefont {Doe}}, \bibinfo {author}
  {\bibfnamefont {J.~L.}\ \bibnamefont {Fernandes}}, \bibinfo {author}
  {\bibfnamefont {M.}~\bibnamefont {Fertl}}, \bibinfo {author} {\bibfnamefont
  {E.~C.}\ \bibnamefont {Finn}}, \bibinfo {author} {\bibfnamefont {J.~A.}\
  \bibnamefont {Formaggio}}, \bibinfo {author} {\bibfnamefont {D.}~\bibnamefont
  {Furse}}, \bibinfo {author} {\bibfnamefont {A.~M.}\ \bibnamefont {Jones}},
  \bibinfo {author} {\bibfnamefont {J.~N.}\ \bibnamefont {Kofron}}, \bibinfo
  {author} {\bibfnamefont {B.~H.}\ \bibnamefont {LaRoque}}, \bibinfo {author}
  {\bibfnamefont {M.}~\bibnamefont {Leber}}, \bibinfo {author} {\bibfnamefont
  {E.~L.}\ \bibnamefont {McBride}}, \bibinfo {author} {\bibfnamefont {M.~L.}\
  \bibnamefont {Miller}}, \bibinfo {author} {\bibfnamefont {P.}~\bibnamefont
  {Mohanmurthy}}, \bibinfo {author} {\bibfnamefont {B.}~\bibnamefont
  {Monreal}}, \bibinfo {author} {\bibfnamefont {N.~S.}\ \bibnamefont {Oblath}},
  \bibinfo {author} {\bibfnamefont {R.~G.~H.}\ \bibnamefont {Robertson}},
  \bibinfo {author} {\bibfnamefont {L.~J.}\ \bibnamefont {Rosenberg}}, \bibinfo
  {author} {\bibfnamefont {G.}~\bibnamefont {Rybka}}, \bibinfo {author}
  {\bibfnamefont {D.}~\bibnamefont {Rysewyk}}, \bibinfo {author} {\bibfnamefont
  {M.~G.}\ \bibnamefont {Sternberg}}, \bibinfo {author} {\bibfnamefont {J.~R.}\
  \bibnamefont {Tedeschi}}, \bibinfo {author} {\bibfnamefont {T.}~\bibnamefont
  {Th\"ummler}}, \bibinfo {author} {\bibfnamefont {B.~A.}\ \bibnamefont
  {VanDevender}},\ and\ \bibinfo {author} {\bibfnamefont {N.~L.}\ \bibnamefont
  {Woods}} (\bibinfo {collaboration} {Project 8 Collaboration}),\ }\href
  {https://doi.org/10.1103/PhysRevLett.114.162501} {\bibfield  {journal}
  {\bibinfo  {journal} {Phys. Rev. Lett.}\ }\textbf {\bibinfo {volume} {114}},\
  \bibinfo {pages} {162501} (\bibinfo {year} {2015})}\BibitemShut {NoStop}%
\bibitem [{\citenamefont {Mardor}\ \emph {et~al.}(2018)\citenamefont {Mardor},
  \citenamefont {Aviv}, \citenamefont {Avrigeanu}, \citenamefont {Berkovits},
  \citenamefont {Dahan}, \citenamefont {Dickel}, \citenamefont {Eliyahu},
  \citenamefont {Gai}, \citenamefont {Gavish-Segev}, \citenamefont {Halfon}
  \emph {et~al.}}]{Mardor2018}%
  \BibitemOpen
  \bibfield  {author} {\bibinfo {author} {\bibfnamefont {I.}~\bibnamefont
  {Mardor}}, \bibinfo {author} {\bibfnamefont {O.}~\bibnamefont {Aviv}},
  \bibinfo {author} {\bibfnamefont {M.}~\bibnamefont {Avrigeanu}}, \bibinfo
  {author} {\bibfnamefont {D.}~\bibnamefont {Berkovits}}, \bibinfo {author}
  {\bibfnamefont {A.}~\bibnamefont {Dahan}}, \bibinfo {author} {\bibfnamefont
  {T.}~\bibnamefont {Dickel}}, \bibinfo {author} {\bibfnamefont
  {I.}~\bibnamefont {Eliyahu}}, \bibinfo {author} {\bibfnamefont
  {M.}~\bibnamefont {Gai}}, \bibinfo {author} {\bibfnamefont {I.}~\bibnamefont
  {Gavish-Segev}}, \bibinfo {author} {\bibfnamefont {S.}~\bibnamefont
  {Halfon}}, \emph {et~al.},\ }\href@noop {} {\bibfield  {journal} {\bibinfo
  {journal} {The European Physical Journal A}\ }\textbf {\bibinfo {volume}
  {54}},\ \bibinfo {pages} {1} (\bibinfo {year} {2018})}\BibitemShut {NoStop}%
\bibitem [{\citenamefont {Burkey}\ \emph {et~al.}(2022)\citenamefont {Burkey},
  \citenamefont {Savard}, \citenamefont {Gallant}, \citenamefont {Scielzo},
  \citenamefont {Clark}, \citenamefont {Hirsh}, \citenamefont {Varriano},
  \citenamefont {Sargsyan}, \citenamefont {Launey}, \citenamefont {Brodeur},
  \citenamefont {Burdette}, \citenamefont {Heckmaier}, \citenamefont {Joerres},
  \citenamefont {Klimes}, \citenamefont {Kolos}, \citenamefont {Laminack},
  \citenamefont {Leach}, \citenamefont {Levand}, \citenamefont {Longfellow},
  \citenamefont {Maa\ss{}}, \citenamefont {Marley}, \citenamefont {Morgan},
  \citenamefont {Mueller}, \citenamefont {Orford}, \citenamefont {Padgett},
  \citenamefont {P\'erez~Galv\'an}, \citenamefont {Pierce}, \citenamefont
  {Ray}, \citenamefont {Segel}, \citenamefont {Siegl}, \citenamefont {Sharma},\
  and\ \citenamefont {Wang}}]{burkey2022}%
  \BibitemOpen
  \bibfield  {author} {\bibinfo {author} {\bibfnamefont {M.~T.}\ \bibnamefont
  {Burkey}}, \bibinfo {author} {\bibfnamefont {G.}~\bibnamefont {Savard}},
  \bibinfo {author} {\bibfnamefont {A.~T.}\ \bibnamefont {Gallant}}, \bibinfo
  {author} {\bibfnamefont {N.~D.}\ \bibnamefont {Scielzo}}, \bibinfo {author}
  {\bibfnamefont {J.~A.}\ \bibnamefont {Clark}}, \bibinfo {author}
  {\bibfnamefont {T.~Y.}\ \bibnamefont {Hirsh}}, \bibinfo {author}
  {\bibfnamefont {L.}~\bibnamefont {Varriano}}, \bibinfo {author}
  {\bibfnamefont {G.~H.}\ \bibnamefont {Sargsyan}}, \bibinfo {author}
  {\bibfnamefont {K.~D.}\ \bibnamefont {Launey}}, \bibinfo {author}
  {\bibfnamefont {M.}~\bibnamefont {Brodeur}}, \bibinfo {author} {\bibfnamefont
  {D.~P.}\ \bibnamefont {Burdette}}, \bibinfo {author} {\bibfnamefont
  {E.}~\bibnamefont {Heckmaier}}, \bibinfo {author} {\bibfnamefont
  {K.}~\bibnamefont {Joerres}}, \bibinfo {author} {\bibfnamefont {J.~W.}\
  \bibnamefont {Klimes}}, \bibinfo {author} {\bibfnamefont {K.}~\bibnamefont
  {Kolos}}, \bibinfo {author} {\bibfnamefont {A.}~\bibnamefont {Laminack}},
  \bibinfo {author} {\bibfnamefont {K.~G.}\ \bibnamefont {Leach}}, \bibinfo
  {author} {\bibfnamefont {A.~F.}\ \bibnamefont {Levand}}, \bibinfo {author}
  {\bibfnamefont {B.}~\bibnamefont {Longfellow}}, \bibinfo {author}
  {\bibfnamefont {B.}~\bibnamefont {Maa\ss{}}}, \bibinfo {author}
  {\bibfnamefont {S.~T.}\ \bibnamefont {Marley}}, \bibinfo {author}
  {\bibfnamefont {G.~E.}\ \bibnamefont {Morgan}}, \bibinfo {author}
  {\bibfnamefont {P.}~\bibnamefont {Mueller}}, \bibinfo {author} {\bibfnamefont
  {R.}~\bibnamefont {Orford}}, \bibinfo {author} {\bibfnamefont {S.~W.}\
  \bibnamefont {Padgett}}, \bibinfo {author} {\bibfnamefont {A.}~\bibnamefont
  {P\'erez~Galv\'an}}, \bibinfo {author} {\bibfnamefont {J.~R.}\ \bibnamefont
  {Pierce}}, \bibinfo {author} {\bibfnamefont {D.}~\bibnamefont {Ray}},
  \bibinfo {author} {\bibfnamefont {R.}~\bibnamefont {Segel}}, \bibinfo
  {author} {\bibfnamefont {K.}~\bibnamefont {Siegl}}, \bibinfo {author}
  {\bibfnamefont {K.~S.}\ \bibnamefont {Sharma}},\ and\ \bibinfo {author}
  {\bibfnamefont {B.~S.}\ \bibnamefont {Wang}},\ }\href
  {https://doi.org/10.1103/PhysRevLett.128.202502} {\bibfield  {journal}
  {\bibinfo  {journal} {Phys. Rev. Lett.}\ }\textbf {\bibinfo {volume} {128}},\
  \bibinfo {pages} {202502} (\bibinfo {year} {2022})}\BibitemShut {NoStop}%
\bibitem [{\citenamefont {Towner}\ and\ \citenamefont
  {Hardy}(2010)}]{Towner2010}%
  \BibitemOpen
  \bibfield  {author} {\bibinfo {author} {\bibfnamefont {I.~S.}\ \bibnamefont
  {Towner}}\ and\ \bibinfo {author} {\bibfnamefont {J.~C.}\ \bibnamefont
  {Hardy}},\ }\href {https://doi.org/10.1103/PhysRevC.82.065501} {\bibfield
  {journal} {\bibinfo  {journal} {Phys. Rev. C}\ }\textbf {\bibinfo {volume}
  {82}},\ \bibinfo {pages} {065501} (\bibinfo {year} {2010})}\BibitemShut
  {NoStop}%
\bibitem [{\citenamefont {Hardy}\ and\ \citenamefont
  {Towner}(2014)}]{Hardy2014}%
  \BibitemOpen
  \bibfield  {author} {\bibinfo {author} {\bibfnamefont {J.~C.}\ \bibnamefont
  {Hardy}}\ and\ \bibinfo {author} {\bibfnamefont {I.~S.}\ \bibnamefont
  {Towner}},\ }\href {https://doi.org/10.1088/0954-3899/41/11/114004}
  {\bibfield  {journal} {\bibinfo  {journal} {Journal of Physics G: Nuclear and
  Particle Physics}\ }\textbf {\bibinfo {volume} {41}},\ \bibinfo {pages}
  {114004} (\bibinfo {year} {2014})}\BibitemShut {NoStop}%
\bibitem [{\citenamefont {Gonzalez-Alonso}\ \emph {et~al.}(2019)\citenamefont
  {Gonzalez-Alonso}, \citenamefont {Naviliat-Cuncic},\ and\ \citenamefont
  {Severijns}}]{Gonzalez2019}%
  \BibitemOpen
  \bibfield  {author} {\bibinfo {author} {\bibfnamefont {M.}~\bibnamefont
  {Gonzalez-Alonso}}, \bibinfo {author} {\bibfnamefont {O.}~\bibnamefont
  {Naviliat-Cuncic}},\ and\ \bibinfo {author} {\bibfnamefont {N.}~\bibnamefont
  {Severijns}},\ }\href
  {https://doi.org/https://doi.org/10.1016/j.ppnp.2018.08.002} {\bibfield
  {journal} {\bibinfo  {journal} {Progress in Particle and Nuclear Physics}\
  }\textbf {\bibinfo {volume} {104}},\ \bibinfo {pages} {165 } (\bibinfo {year}
  {2019})}\BibitemShut {NoStop}%
\bibitem [{\citenamefont {Cirigliano}\ \emph {et~al.}(2013)\citenamefont
  {Cirigliano}, \citenamefont {Gonzalez-Alonso},\ and\ \citenamefont
  {Graesser}}]{Cirigliano:2012ab}%
  \BibitemOpen
  \bibfield  {author} {\bibinfo {author} {\bibfnamefont {V.}~\bibnamefont
  {Cirigliano}}, \bibinfo {author} {\bibfnamefont {M.}~\bibnamefont
  {Gonzalez-Alonso}},\ and\ \bibinfo {author} {\bibfnamefont {M.~L.}\
  \bibnamefont {Graesser}},\ }\href {https://doi.org/10.1007/JHEP02(2013)046}
  {\bibfield  {journal} {\bibinfo  {journal} {JHEP}\ }\textbf {\bibinfo
  {volume} {02}},\ \bibinfo {pages} {046}},\ \Eprint
  {https://arxiv.org/abs/1210.4553} {arXiv:1210.4553 [hep-ph]} \BibitemShut
  {NoStop}%
\bibitem [{\citenamefont {Hardy}\ and\ \citenamefont
  {Towner}(2010)}]{Towner2010-Vud}%
  \BibitemOpen
  \bibfield  {author} {\bibinfo {author} {\bibfnamefont {J.~C.}\ \bibnamefont
  {Hardy}}\ and\ \bibinfo {author} {\bibfnamefont {I.~S.}\ \bibnamefont
  {Towner}},\ }\href {http://stacks.iop.org/0034-4885/73/i=4/a=046301}
  {\bibfield  {journal} {\bibinfo  {journal} {Reports on Progress in Physics}\
  }\textbf {\bibinfo {volume} {73}},\ \bibinfo {pages} {046301} (\bibinfo
  {year} {2010})}\BibitemShut {NoStop}%
\bibitem [{\citenamefont {Seng}\ \emph {et~al.}(2018)\citenamefont {Seng},
  \citenamefont {Gorchtein}, \citenamefont {Patel},\ and\ \citenamefont
  {Ramsey-Musolf}}]{Seng2018}%
  \BibitemOpen
  \bibfield  {author} {\bibinfo {author} {\bibfnamefont {C.-Y.}\ \bibnamefont
  {Seng}}, \bibinfo {author} {\bibfnamefont {M.}~\bibnamefont {Gorchtein}},
  \bibinfo {author} {\bibfnamefont {H.~H.}\ \bibnamefont {Patel}},\ and\
  \bibinfo {author} {\bibfnamefont {M.~J.}\ \bibnamefont {Ramsey-Musolf}},\
  }\href {https://doi.org/10.1103/PhysRevLett.121.241804} {\bibfield  {journal}
  {\bibinfo  {journal} {Physical Review Letters}\ }\textbf {\bibinfo {volume}
  {121}},\ \bibinfo {pages} {241804} (\bibinfo {year} {2018})},\ \Eprint
  {https://arxiv.org/abs/1807.10197} {arXiv:1807.10197} \BibitemShut {NoStop}%
\bibitem [{\citenamefont {Seng}\ \emph {et~al.}(2019)\citenamefont {Seng},
  \citenamefont {Gorchtein},\ and\ \citenamefont {Ramsey-Musolf}}]{Seng2019b}%
  \BibitemOpen
  \bibfield  {author} {\bibinfo {author} {\bibfnamefont {C.~Y.}\ \bibnamefont
  {Seng}}, \bibinfo {author} {\bibfnamefont {M.}~\bibnamefont {Gorchtein}},\
  and\ \bibinfo {author} {\bibfnamefont {M.~J.}\ \bibnamefont
  {Ramsey-Musolf}},\ }\href {https://doi.org/10.1103/PhysRevD.100.013001}
  {\bibfield  {journal} {\bibinfo  {journal} {Physical Review D}\ }\textbf
  {\bibinfo {volume} {100}},\ \bibinfo {pages} {013001} (\bibinfo {year}
  {2019})},\ \Eprint {https://arxiv.org/abs/1812.03352} {arXiv:1812.03352}
  \BibitemShut {NoStop}%
\bibitem [{\citenamefont {Czarnecki}\ \emph {et~al.}(2019)\citenamefont
  {Czarnecki}, \citenamefont {Marciano},\ and\ \citenamefont
  {Sirlin}}]{Czarnecki2019}%
  \BibitemOpen
  \bibfield  {author} {\bibinfo {author} {\bibfnamefont {A.}~\bibnamefont
  {Czarnecki}}, \bibinfo {author} {\bibfnamefont {W.~J.}\ \bibnamefont
  {Marciano}},\ and\ \bibinfo {author} {\bibfnamefont {A.}~\bibnamefont
  {Sirlin}},\ }\href {https://doi.org/10.1103/PhysRevD.100.073008} {\bibfield
  {journal} {\bibinfo  {journal} {Physical Review D}\ }\textbf {\bibinfo
  {volume} {100}},\ \bibinfo {pages} {73008} (\bibinfo {year} {2019})},\
  \Eprint {https://arxiv.org/abs/1907.06737} {arXiv:1907.06737} \BibitemShut
  {NoStop}%
\bibitem [{\citenamefont {Hayen}(2021)}]{Hayen2021}%
  \BibitemOpen
  \bibfield  {author} {\bibinfo {author} {\bibfnamefont {L.}~\bibnamefont
  {Hayen}},\ }\href {https://doi.org/10.1103/PhysRevD.103.113001} {\bibfield
  {journal} {\bibinfo  {journal} {Physical Review D}\ }\textbf {\bibinfo
  {volume} {103}},\ \bibinfo {pages} {113001} (\bibinfo {year} {2021})},\
  \Eprint {https://arxiv.org/abs/2010.07262} {arXiv:2010.07262} \BibitemShut
  {NoStop}%
\bibitem [{\citenamefont {Shiells}\ \emph {et~al.}(2021)\citenamefont
  {Shiells}, \citenamefont {Blunden},\ and\ \citenamefont
  {Melnitchouk}}]{Shiells2021}%
  \BibitemOpen
  \bibfield  {author} {\bibinfo {author} {\bibfnamefont {K.}~\bibnamefont
  {Shiells}}, \bibinfo {author} {\bibfnamefont {P.~G.}\ \bibnamefont
  {Blunden}},\ and\ \bibinfo {author} {\bibfnamefont {W.}~\bibnamefont
  {Melnitchouk}},\ }\href {https://doi.org/10.1103/PhysRevD.104.033003}
  {\bibfield  {journal} {\bibinfo  {journal} {Physical Review D}\ }\textbf
  {\bibinfo {volume} {104}},\ \bibinfo {pages} {33003} (\bibinfo {year}
  {2021})}\BibitemShut {NoStop}%
\bibitem [{\citenamefont {Feng}\ \emph {et~al.}(2020)\citenamefont {Feng},
  \citenamefont {Gorchtein}, \citenamefont {Jin}, \citenamefont {Ma},\ and\
  \citenamefont {Seng}}]{Feng2020}%
  \BibitemOpen
  \bibfield  {author} {\bibinfo {author} {\bibfnamefont {X.}~\bibnamefont
  {Feng}}, \bibinfo {author} {\bibfnamefont {M.}~\bibnamefont {Gorchtein}},
  \bibinfo {author} {\bibfnamefont {L.~C.}\ \bibnamefont {Jin}}, \bibinfo
  {author} {\bibfnamefont {P.~X.}\ \bibnamefont {Ma}},\ and\ \bibinfo {author}
  {\bibfnamefont {C.~Y.}\ \bibnamefont {Seng}},\ }\href
  {https://doi.org/10.1103/PhysRevLett.124.192002} {\bibfield  {journal}
  {\bibinfo  {journal} {Physical Review Letters}\ }\textbf {\bibinfo {volume}
  {124}},\ \bibinfo {pages} {192002} (\bibinfo {year} {2020})},\ \Eprint
  {https://arxiv.org/abs/2003.09798} {arXiv:2003.09798} \BibitemShut {NoStop}%
\bibitem [{\citenamefont {Seng}\ \emph {et~al.}(2020)\citenamefont {Seng},
  \citenamefont {Feng}, \citenamefont {Gorchtein},\ and\ \citenamefont
  {Jin}}]{Seng2020}%
  \BibitemOpen
  \bibfield  {author} {\bibinfo {author} {\bibfnamefont {C.~Y.}\ \bibnamefont
  {Seng}}, \bibinfo {author} {\bibfnamefont {X.}~\bibnamefont {Feng}}, \bibinfo
  {author} {\bibfnamefont {M.}~\bibnamefont {Gorchtein}},\ and\ \bibinfo
  {author} {\bibfnamefont {L.~C.}\ \bibnamefont {Jin}},\ }\href
  {https://doi.org/10.1103/PhysRevD.101.111301} {\bibfield  {journal} {\bibinfo
   {journal} {Physical Review D}\ }\textbf {\bibinfo {volume} {101}},\ \bibinfo
  {pages} {111301} (\bibinfo {year} {2020})}\BibitemShut {NoStop}%
\bibitem [{\citenamefont {Seng}\ and\ \citenamefont
  {Mei\ss{}ner}(2019)}]{Seng:2019plg}%
  \BibitemOpen
  \bibfield  {author} {\bibinfo {author} {\bibfnamefont {C.-Y.}\ \bibnamefont
  {Seng}}\ and\ \bibinfo {author} {\bibfnamefont {U.-G.}\ \bibnamefont
  {Mei\ss{}ner}},\ }\href {https://doi.org/10.1103/PhysRevLett.122.211802}
  {\bibfield  {journal} {\bibinfo  {journal} {Phys. Rev. Lett.}\ }\textbf
  {\bibinfo {volume} {122}},\ \bibinfo {pages} {211802} (\bibinfo {year}
  {2019})},\ \Eprint {https://arxiv.org/abs/1903.07969} {arXiv:1903.07969
  [hep-ph]} \BibitemShut {NoStop}%
\bibitem [{\citenamefont {Gorchtein}(2019)}]{Gorchtein:2018fxl}%
  \BibitemOpen
  \bibfield  {author} {\bibinfo {author} {\bibfnamefont {M.}~\bibnamefont
  {Gorchtein}},\ }\href {https://doi.org/10.1103/PhysRevLett.123.042503}
  {\bibfield  {journal} {\bibinfo  {journal} {Phys. Rev. Lett.}\ }\textbf
  {\bibinfo {volume} {123}},\ \bibinfo {pages} {042503} (\bibinfo {year}
  {2019})},\ \Eprint {https://arxiv.org/abs/1812.04229} {arXiv:1812.04229
  [nucl-th]} \BibitemShut {NoStop}%
\bibitem [{\citenamefont {Hardy}\ and\ \citenamefont
  {Towner}(2020)}]{Hardy2020}%
  \BibitemOpen
  \bibfield  {author} {\bibinfo {author} {\bibfnamefont {J.~C.}\ \bibnamefont
  {Hardy}}\ and\ \bibinfo {author} {\bibfnamefont {I.~S.}\ \bibnamefont
  {Towner}},\ }\href {https://doi.org/10.1103/PhysRevC.102.045501} {\bibfield
  {journal} {\bibinfo  {journal} {Phys. Rev. C}\ }\textbf {\bibinfo {volume}
  {102}},\ \bibinfo {pages} {045501} (\bibinfo {year} {2020})}\BibitemShut
  {NoStop}%
\bibitem [{\citenamefont {Seng}\ and\ \citenamefont
  {Gorchtein}(2022{\natexlab{a}})}]{Seng:2022cnq}%
  \BibitemOpen
  \bibfield  {author} {\bibinfo {author} {\bibfnamefont {C.-Y.}\ \bibnamefont
  {Seng}}\ and\ \bibinfo {author} {\bibfnamefont {M.}~\bibnamefont
  {Gorchtein}},\ }\href@noop {} {\  (\bibinfo {year} {2022}{\natexlab{a}})},\
  \Eprint {https://arxiv.org/abs/2211.10214} {arXiv:2211.10214 [nucl-th]}
  \BibitemShut {NoStop}%
\bibitem [{\citenamefont {Weinberg}(1990)}]{Weinberg:1990rz}%
  \BibitemOpen
  \bibfield  {author} {\bibinfo {author} {\bibfnamefont {S.}~\bibnamefont
  {Weinberg}},\ }\href {https://doi.org/10.1016/0370-2693(90)90938-3}
  {\bibfield  {journal} {\bibinfo  {journal} {Phys. Lett.}\ }\textbf {\bibinfo
  {volume} {B251}},\ \bibinfo {pages} {288} (\bibinfo {year}
  {1990})}\BibitemShut {NoStop}%
\bibitem [{\citenamefont {Weinberg}(1991)}]{Weinberg:1991rw}%
  \BibitemOpen
  \bibfield  {author} {\bibinfo {author} {\bibfnamefont {S.}~\bibnamefont
  {Weinberg}},\ }\href
  {https://doi.org/https://doi.org/10.1016/0550-3213(91)90231-L} {\bibfield
  {journal} {\bibinfo  {journal} {Nuclear Physics B}\ }\textbf {\bibinfo
  {volume} {363}},\ \bibinfo {pages} {3} (\bibinfo {year} {1991})}\BibitemShut
  {NoStop}%
\bibitem [{\citenamefont {Weinberg}(1992)}]{Weinberg:1992yk}%
  \BibitemOpen
  \bibfield  {author} {\bibinfo {author} {\bibfnamefont {S.}~\bibnamefont
  {Weinberg}},\ }\href {https://doi.org/10.1016/0370-2693(92)90099-P}
  {\bibfield  {journal} {\bibinfo  {journal} {Phys. Lett.}\ }\textbf {\bibinfo
  {volume} {B295}},\ \bibinfo {pages} {114} (\bibinfo {year} {1992})},\ \Eprint
  {https://arxiv.org/abs/hep-ph/9209257} {arXiv:hep-ph/9209257 [hep-ph]}
  \BibitemShut {NoStop}%
\bibitem [{\citenamefont {Machleidt}\ and\ \citenamefont
  {Sammarruca}(2016)}]{Machleidt:2016yo}%
  \BibitemOpen
  \bibfield  {author} {\bibinfo {author} {\bibfnamefont {R.}~\bibnamefont
  {Machleidt}}\ and\ \bibinfo {author} {\bibfnamefont {F.}~\bibnamefont
  {Sammarruca}},\ }\href {http://stacks.iop.org/1402-4896/91/i=8/a=083007}
  {\bibfield  {journal} {\bibinfo  {journal} {Physica Scripta}\ }\textbf
  {\bibinfo {volume} {91}},\ \bibinfo {pages} {083007} (\bibinfo {year}
  {2016})}\BibitemShut {NoStop}%
\bibitem [{\citenamefont {King}\ \emph {et~al.}(2020)\citenamefont {King},
  \citenamefont {Andreoli}, \citenamefont {Pastore}, \citenamefont {Piarulli},
  \citenamefont {Schiavilla}, \citenamefont {Wiringa}, \citenamefont
  {Carlson},\ and\ \citenamefont {Gandolfi}}]{King_2020}%
  \BibitemOpen
  \bibfield  {author} {\bibinfo {author} {\bibfnamefont {G.~B.}\ \bibnamefont
  {King}}, \bibinfo {author} {\bibfnamefont {L.}~\bibnamefont {Andreoli}},
  \bibinfo {author} {\bibfnamefont {S.}~\bibnamefont {Pastore}}, \bibinfo
  {author} {\bibfnamefont {M.}~\bibnamefont {Piarulli}}, \bibinfo {author}
  {\bibfnamefont {R.}~\bibnamefont {Schiavilla}}, \bibinfo {author}
  {\bibfnamefont {R.~B.}\ \bibnamefont {Wiringa}}, \bibinfo {author}
  {\bibfnamefont {J.}~\bibnamefont {Carlson}},\ and\ \bibinfo {author}
  {\bibfnamefont {S.}~\bibnamefont {Gandolfi}},\ }\bibfield  {journal}
  {\bibinfo  {journal} {Physical Review C}\ }\textbf {\bibinfo {volume}
  {102}},\ \href {https://doi.org/10.1103/physrevc.102.025501}
  {10.1103/physrevc.102.025501} (\bibinfo {year} {2020})\BibitemShut {NoStop}%
\bibitem [{\citenamefont {Epelbaum}\ \emph {et~al.}(2020)\citenamefont
  {Epelbaum}, \citenamefont {Krebs},\ and\ \citenamefont
  {Reinert}}]{Epelbaum:2020rr}%
  \BibitemOpen
  \bibfield  {author} {\bibinfo {author} {\bibfnamefont {E.}~\bibnamefont
  {Epelbaum}}, \bibinfo {author} {\bibfnamefont {H.}~\bibnamefont {Krebs}},\
  and\ \bibinfo {author} {\bibfnamefont {P.}~\bibnamefont {Reinert}},\ }\href
  {https://doi.org/10.3389/fphy.2020.00098} {\bibfield  {journal} {\bibinfo
  {journal} {Frontiers in Physics}\ }\textbf {\bibinfo {volume} {8}},\ \bibinfo
  {pages} {98} (\bibinfo {year} {2020})}\BibitemShut {NoStop}%
\bibitem [{\citenamefont {van Kolck}(2020)}]{Kolck:2020dn}%
  \BibitemOpen
  \bibfield  {author} {\bibinfo {author} {\bibfnamefont {U.}~\bibnamefont {van
  Kolck}},\ }\href {https://doi.org/10.3389/fphy.2020.00079} {\bibfield
  {journal} {\bibinfo  {journal} {Frontiers in Physics}\ }\textbf {\bibinfo
  {volume} {8}},\ \bibinfo {pages} {79} (\bibinfo {year} {2020})}\BibitemShut
  {NoStop}%
\bibitem [{\citenamefont {Krebs}(2020)}]{Krebs:2020pii}%
  \BibitemOpen
  \bibfield  {author} {\bibinfo {author} {\bibfnamefont {H.}~\bibnamefont
  {Krebs}},\ }\href {https://doi.org/10.1140/epja/s10050-020-00230-9}
  {\bibfield  {journal} {\bibinfo  {journal} {Eur. Phys. J. A}\ }\textbf
  {\bibinfo {volume} {56}},\ \bibinfo {pages} {234} (\bibinfo {year} {2020})},\
  \Eprint {https://arxiv.org/abs/2008.00974} {arXiv:2008.00974 [nucl-th]}
  \BibitemShut {NoStop}%
\bibitem [{\citenamefont {Baroni}\ \emph {et~al.}(2021)\citenamefont {Baroni},
  \citenamefont {King},\ and\ \citenamefont {Pastore}}]{Baroni:2021vrc}%
  \BibitemOpen
  \bibfield  {author} {\bibinfo {author} {\bibfnamefont {A.}~\bibnamefont
  {Baroni}}, \bibinfo {author} {\bibfnamefont {G.~B.}\ \bibnamefont {King}},\
  and\ \bibinfo {author} {\bibfnamefont {S.}~\bibnamefont {Pastore}},\ }\href
  {https://doi.org/10.1007/s00601-021-01700-6} {\bibfield  {journal} {\bibinfo
  {journal} {Few Body Syst.}\ }\textbf {\bibinfo {volume} {62}},\ \bibinfo
  {pages} {114} (\bibinfo {year} {2021})},\ \Eprint
  {https://arxiv.org/abs/2107.10721} {arXiv:2107.10721 [nucl-th]} \BibitemShut
  {NoStop}%
\bibitem [{\citenamefont {Gysbers}\ \emph {et~al.}(2019)\citenamefont
  {Gysbers}, \citenamefont {Hagen}, \citenamefont {Holt}, \citenamefont
  {Jansen}, \citenamefont {Morris}, \citenamefont {Navr{\'{a}}til},
  \citenamefont {Papenbrock}, \citenamefont {Quaglioni}, \citenamefont
  {Schwenk}, \citenamefont {Stroberg},\ and\ \citenamefont
  {Wendt}}]{Gysbers2019}%
  \BibitemOpen
  \bibfield  {author} {\bibinfo {author} {\bibfnamefont {P.}~\bibnamefont
  {Gysbers}}, \bibinfo {author} {\bibfnamefont {G.}~\bibnamefont {Hagen}},
  \bibinfo {author} {\bibfnamefont {J.~D.}\ \bibnamefont {Holt}}, \bibinfo
  {author} {\bibfnamefont {G.~R.}\ \bibnamefont {Jansen}}, \bibinfo {author}
  {\bibfnamefont {T.~D.}\ \bibnamefont {Morris}}, \bibinfo {author}
  {\bibfnamefont {P.}~\bibnamefont {Navr{\'{a}}til}}, \bibinfo {author}
  {\bibfnamefont {T.}~\bibnamefont {Papenbrock}}, \bibinfo {author}
  {\bibfnamefont {S.}~\bibnamefont {Quaglioni}}, \bibinfo {author}
  {\bibfnamefont {A.}~\bibnamefont {Schwenk}}, \bibinfo {author} {\bibfnamefont
  {S.~R.}\ \bibnamefont {Stroberg}},\ and\ \bibinfo {author} {\bibfnamefont
  {K.~A.}\ \bibnamefont {Wendt}},\ }\href
  {https://doi.org/10.1038/s41567-019-0450-7} {\bibfield  {journal} {\bibinfo
  {journal} {Nature Physics}\ }\textbf {\bibinfo {volume} {15}},\ \bibinfo
  {pages} {428} (\bibinfo {year} {2019})},\ \Eprint
  {https://arxiv.org/abs/arXiv:1903.00047v1} {arXiv:arXiv:1903.00047v1}
  \BibitemShut {NoStop}%
\bibitem [{\citenamefont {Hergert}(2020)}]{10.3389/fphy.2020.00379}%
  \BibitemOpen
  \bibfield  {author} {\bibinfo {author} {\bibfnamefont {H.}~\bibnamefont
  {Hergert}},\ }\bibfield  {journal} {\bibinfo  {journal} {Frontiers in
  Physics}\ }\textbf {\bibinfo {volume} {8}},\ \href
  {https://doi.org/10.3389/fphy.2020.00379} {10.3389/fphy.2020.00379} (\bibinfo
  {year} {2020})\BibitemShut {NoStop}%
\bibitem [{\citenamefont {Carlson}\ \emph {et~al.}(2017)\citenamefont
  {Carlson}, \citenamefont {Savage}, \citenamefont {Gerber}, \citenamefont
  {Antypas}, \citenamefont {Bard}, \citenamefont {Coffey}, \citenamefont
  {Dart}, \citenamefont {Dosanjh}, \citenamefont {Hack}, \citenamefont {Monga},
  \citenamefont {Papka}, \citenamefont {Riley}, \citenamefont {Rotman},
  \citenamefont {Straatsma}, \citenamefont {Wells}, \citenamefont {Avakian},
  \citenamefont {Ayyad}, \citenamefont {Bass}, \citenamefont {Bazin},
  \citenamefont {Boehnlein}, \citenamefont {Bollen}, \citenamefont {Broussard},
  \citenamefont {Calder}, \citenamefont {Couch}, \citenamefont {Couture},
  \citenamefont {Cromaz}, \citenamefont {Detmold}, \citenamefont {Detwiler},
  \citenamefont {Duan}, \citenamefont {Edwards}, \citenamefont {Engel},
  \citenamefont {Fryer}, \citenamefont {Fuller}, \citenamefont {Gandolfi},
  \citenamefont {Gavalian}, \citenamefont {Georgobiani}, \citenamefont {Gupta},
  \citenamefont {Gyurjyan}, \citenamefont {Hausmann}, \citenamefont {Heyes},
  \citenamefont {Hix}, \citenamefont {ito}, \citenamefont {Jansen},
  \citenamefont {Jones}, \citenamefont {Joo}, \citenamefont {Kaczmarek},
  \citenamefont {Kasen}, \citenamefont {Kostin}, \citenamefont {Kurth},
  \citenamefont {Lauret}, \citenamefont {Lawrence}, \citenamefont {Lin},
  \citenamefont {Lin}, \citenamefont {Mantica}, \citenamefont {Maris},
  \citenamefont {Messer}, \citenamefont {Mittig}, \citenamefont {Mosby},
  \citenamefont {Mukherjee}, \citenamefont {Nam}, \citenamefont {Navratil},
  \citenamefont {Nazarewicz}, \citenamefont {Ng}, \citenamefont {O'Donnell},
  \citenamefont {Orginos}, \citenamefont {Pellemoine}, \citenamefont
  {Petreczky}, \citenamefont {Pieper}, \citenamefont {Pinkenburg},
  \citenamefont {Plaster}, \citenamefont {Porter}, \citenamefont {Portillo},
  \citenamefont {Pratt}, \citenamefont {Purschke}, \citenamefont {Qiang},
  \citenamefont {Quaglioni}, \citenamefont {Richards}, \citenamefont {Roblin},
  \citenamefont {Schenke}, \citenamefont {Schiavilla}, \citenamefont
  {Schlichting}, \citenamefont {Schunck}, \citenamefont {Steinbrecher},
  \citenamefont {Strickland}, \citenamefont {Syritsyn}, \citenamefont {Terzic},
  \citenamefont {Varner}, \citenamefont {Vary}, \citenamefont {Wild},
  \citenamefont {Winter}, \citenamefont {Zegers}, \citenamefont {Zhang},
  \citenamefont {Ziegler},\ and\ \citenamefont {Zingale}}]{osti_1369223}%
  \BibitemOpen
  \bibfield  {author} {\bibinfo {author} {\bibfnamefont {J.}~\bibnamefont
  {Carlson}}, \bibinfo {author} {\bibfnamefont {M.~J.}\ \bibnamefont {Savage}},
  \bibinfo {author} {\bibfnamefont {R.}~\bibnamefont {Gerber}}, \bibinfo
  {author} {\bibfnamefont {K.}~\bibnamefont {Antypas}}, \bibinfo {author}
  {\bibfnamefont {D.}~\bibnamefont {Bard}}, \bibinfo {author} {\bibfnamefont
  {R.}~\bibnamefont {Coffey}}, \bibinfo {author} {\bibfnamefont
  {E.}~\bibnamefont {Dart}}, \bibinfo {author} {\bibfnamefont {S.}~\bibnamefont
  {Dosanjh}}, \bibinfo {author} {\bibfnamefont {J.}~\bibnamefont {Hack}},
  \bibinfo {author} {\bibfnamefont {I.}~\bibnamefont {Monga}}, \bibinfo
  {author} {\bibfnamefont {M.~E.}\ \bibnamefont {Papka}}, \bibinfo {author}
  {\bibfnamefont {K.}~\bibnamefont {Riley}}, \bibinfo {author} {\bibfnamefont
  {L.}~\bibnamefont {Rotman}}, \bibinfo {author} {\bibfnamefont
  {T.}~\bibnamefont {Straatsma}}, \bibinfo {author} {\bibfnamefont
  {J.}~\bibnamefont {Wells}}, \bibinfo {author} {\bibfnamefont
  {H.}~\bibnamefont {Avakian}}, \bibinfo {author} {\bibfnamefont
  {Y.}~\bibnamefont {Ayyad}}, \bibinfo {author} {\bibfnamefont {S.~A.}\
  \bibnamefont {Bass}}, \bibinfo {author} {\bibfnamefont {D.}~\bibnamefont
  {Bazin}}, \bibinfo {author} {\bibfnamefont {A.}~\bibnamefont {Boehnlein}},
  \bibinfo {author} {\bibfnamefont {G.}~\bibnamefont {Bollen}}, \bibinfo
  {author} {\bibfnamefont {L.~J.}\ \bibnamefont {Broussard}}, \bibinfo {author}
  {\bibfnamefont {A.}~\bibnamefont {Calder}}, \bibinfo {author} {\bibfnamefont
  {S.}~\bibnamefont {Couch}}, \bibinfo {author} {\bibfnamefont
  {A.}~\bibnamefont {Couture}}, \bibinfo {author} {\bibfnamefont
  {M.}~\bibnamefont {Cromaz}}, \bibinfo {author} {\bibfnamefont
  {W.}~\bibnamefont {Detmold}}, \bibinfo {author} {\bibfnamefont
  {J.}~\bibnamefont {Detwiler}}, \bibinfo {author} {\bibfnamefont
  {H.}~\bibnamefont {Duan}}, \bibinfo {author} {\bibfnamefont {R.}~\bibnamefont
  {Edwards}}, \bibinfo {author} {\bibfnamefont {J.}~\bibnamefont {Engel}},
  \bibinfo {author} {\bibfnamefont {C.}~\bibnamefont {Fryer}}, \bibinfo
  {author} {\bibfnamefont {G.~M.}\ \bibnamefont {Fuller}}, \bibinfo {author}
  {\bibfnamefont {S.}~\bibnamefont {Gandolfi}}, \bibinfo {author}
  {\bibfnamefont {G.}~\bibnamefont {Gavalian}}, \bibinfo {author}
  {\bibfnamefont {D.}~\bibnamefont {Georgobiani}}, \bibinfo {author}
  {\bibfnamefont {R.}~\bibnamefont {Gupta}}, \bibinfo {author} {\bibfnamefont
  {V.}~\bibnamefont {Gyurjyan}}, \bibinfo {author} {\bibfnamefont
  {M.}~\bibnamefont {Hausmann}}, \bibinfo {author} {\bibfnamefont
  {G.}~\bibnamefont {Heyes}}, \bibinfo {author} {\bibfnamefont {W.~R.}\
  \bibnamefont {Hix}}, \bibinfo {author} {\bibfnamefont {M.}~\bibnamefont
  {ito}}, \bibinfo {author} {\bibfnamefont {G.}~\bibnamefont {Jansen}},
  \bibinfo {author} {\bibfnamefont {R.}~\bibnamefont {Jones}}, \bibinfo
  {author} {\bibfnamefont {B.}~\bibnamefont {Joo}}, \bibinfo {author}
  {\bibfnamefont {O.}~\bibnamefont {Kaczmarek}}, \bibinfo {author}
  {\bibfnamefont {D.}~\bibnamefont {Kasen}}, \bibinfo {author} {\bibfnamefont
  {M.}~\bibnamefont {Kostin}}, \bibinfo {author} {\bibfnamefont
  {T.}~\bibnamefont {Kurth}}, \bibinfo {author} {\bibfnamefont
  {J.}~\bibnamefont {Lauret}}, \bibinfo {author} {\bibfnamefont
  {D.}~\bibnamefont {Lawrence}}, \bibinfo {author} {\bibfnamefont {H.-W.}\
  \bibnamefont {Lin}}, \bibinfo {author} {\bibfnamefont {M.}~\bibnamefont
  {Lin}}, \bibinfo {author} {\bibfnamefont {P.}~\bibnamefont {Mantica}},
  \bibinfo {author} {\bibfnamefont {P.}~\bibnamefont {Maris}}, \bibinfo
  {author} {\bibfnamefont {B.}~\bibnamefont {Messer}}, \bibinfo {author}
  {\bibfnamefont {W.}~\bibnamefont {Mittig}}, \bibinfo {author} {\bibfnamefont
  {S.}~\bibnamefont {Mosby}}, \bibinfo {author} {\bibfnamefont
  {S.}~\bibnamefont {Mukherjee}}, \bibinfo {author} {\bibfnamefont {H.~A.}\
  \bibnamefont {Nam}}, \bibinfo {author} {\bibfnamefont {P.}~\bibnamefont
  {Navratil}}, \bibinfo {author} {\bibfnamefont {W.}~\bibnamefont
  {Nazarewicz}}, \bibinfo {author} {\bibfnamefont {E.}~\bibnamefont {Ng}},
  \bibinfo {author} {\bibfnamefont {T.}~\bibnamefont {O'Donnell}}, \bibinfo
  {author} {\bibfnamefont {K.}~\bibnamefont {Orginos}}, \bibinfo {author}
  {\bibfnamefont {F.}~\bibnamefont {Pellemoine}}, \bibinfo {author}
  {\bibfnamefont {P.}~\bibnamefont {Petreczky}}, \bibinfo {author}
  {\bibfnamefont {S.~C.}\ \bibnamefont {Pieper}}, \bibinfo {author}
  {\bibfnamefont {C.~H.}\ \bibnamefont {Pinkenburg}}, \bibinfo {author}
  {\bibfnamefont {B.}~\bibnamefont {Plaster}}, \bibinfo {author} {\bibfnamefont
  {R.~J.}\ \bibnamefont {Porter}}, \bibinfo {author} {\bibfnamefont
  {M.}~\bibnamefont {Portillo}}, \bibinfo {author} {\bibfnamefont
  {S.}~\bibnamefont {Pratt}}, \bibinfo {author} {\bibfnamefont {M.~L.}\
  \bibnamefont {Purschke}}, \bibinfo {author} {\bibfnamefont {J.}~\bibnamefont
  {Qiang}}, \bibinfo {author} {\bibfnamefont {S.}~\bibnamefont {Quaglioni}},
  \bibinfo {author} {\bibfnamefont {D.}~\bibnamefont {Richards}}, \bibinfo
  {author} {\bibfnamefont {Y.}~\bibnamefont {Roblin}}, \bibinfo {author}
  {\bibfnamefont {B.}~\bibnamefont {Schenke}}, \bibinfo {author} {\bibfnamefont
  {R.}~\bibnamefont {Schiavilla}}, \bibinfo {author} {\bibfnamefont
  {S.}~\bibnamefont {Schlichting}}, \bibinfo {author} {\bibfnamefont
  {N.}~\bibnamefont {Schunck}}, \bibinfo {author} {\bibfnamefont
  {P.}~\bibnamefont {Steinbrecher}}, \bibinfo {author} {\bibfnamefont
  {M.}~\bibnamefont {Strickland}}, \bibinfo {author} {\bibfnamefont
  {S.}~\bibnamefont {Syritsyn}}, \bibinfo {author} {\bibfnamefont
  {B.}~\bibnamefont {Terzic}}, \bibinfo {author} {\bibfnamefont
  {R.}~\bibnamefont {Varner}}, \bibinfo {author} {\bibfnamefont
  {J.}~\bibnamefont {Vary}}, \bibinfo {author} {\bibfnamefont {S.}~\bibnamefont
  {Wild}}, \bibinfo {author} {\bibfnamefont {F.}~\bibnamefont {Winter}},
  \bibinfo {author} {\bibfnamefont {R.}~\bibnamefont {Zegers}}, \bibinfo
  {author} {\bibfnamefont {H.}~\bibnamefont {Zhang}}, \bibinfo {author}
  {\bibfnamefont {V.}~\bibnamefont {Ziegler}},\ and\ \bibinfo {author}
  {\bibfnamefont {M.}~\bibnamefont {Zingale}}\ }\href
  {https://doi.org/10.2172/1369223} {10.2172/1369223} (\bibinfo {year}
  {2017})\BibitemShut {NoStop}%
\bibitem [{\citenamefont {Shidling}\ \emph {et~al.}(2014)\citenamefont
  {Shidling}, \citenamefont {Melconian}, \citenamefont {Behling}, \citenamefont
  {Fenker}, \citenamefont {Hardy}, \citenamefont {Iacob}, \citenamefont
  {McCleskey}, \citenamefont {McCleskey}, \citenamefont {Mehlman},
  \citenamefont {Park},\ and\ \citenamefont {Roeder}}]{Shidling2014}%
  \BibitemOpen
  \bibfield  {author} {\bibinfo {author} {\bibfnamefont {P.~D.}\ \bibnamefont
  {Shidling}}, \bibinfo {author} {\bibfnamefont {D.}~\bibnamefont {Melconian}},
  \bibinfo {author} {\bibfnamefont {S.}~\bibnamefont {Behling}}, \bibinfo
  {author} {\bibfnamefont {B.}~\bibnamefont {Fenker}}, \bibinfo {author}
  {\bibfnamefont {J.~C.}\ \bibnamefont {Hardy}}, \bibinfo {author}
  {\bibfnamefont {V.~E.}\ \bibnamefont {Iacob}}, \bibinfo {author}
  {\bibfnamefont {E.}~\bibnamefont {McCleskey}}, \bibinfo {author}
  {\bibfnamefont {M.}~\bibnamefont {McCleskey}}, \bibinfo {author}
  {\bibfnamefont {M.}~\bibnamefont {Mehlman}}, \bibinfo {author} {\bibfnamefont
  {H.~I.}\ \bibnamefont {Park}},\ and\ \bibinfo {author} {\bibfnamefont
  {B.~T.}\ \bibnamefont {Roeder}},\ }\href
  {https://doi.org/10.1103/PhysRevC.90.032501} {\bibfield  {journal} {\bibinfo
  {journal} {Phys. Rev. C}\ }\textbf {\bibinfo {volume} {90}},\ \bibinfo
  {pages} {032501} (\bibinfo {year} {2014})}\BibitemShut {NoStop}%
\bibitem [{\citenamefont {Shidling}\ \emph {et~al.}(2018)\citenamefont
  {Shidling}, \citenamefont {Behling}, \citenamefont {Fenker}, \citenamefont
  {Hardy}, \citenamefont {Iacob}, \citenamefont {Mehlman}, \citenamefont
  {Park}, \citenamefont {Roeder},\ and\ \citenamefont
  {Melconian}}]{Shidling2018}%
  \BibitemOpen
  \bibfield  {author} {\bibinfo {author} {\bibfnamefont {P.~D.}\ \bibnamefont
  {Shidling}}, \bibinfo {author} {\bibfnamefont {R.~S.}\ \bibnamefont
  {Behling}}, \bibinfo {author} {\bibfnamefont {B.}~\bibnamefont {Fenker}},
  \bibinfo {author} {\bibfnamefont {J.~C.}\ \bibnamefont {Hardy}}, \bibinfo
  {author} {\bibfnamefont {V.~E.}\ \bibnamefont {Iacob}}, \bibinfo {author}
  {\bibfnamefont {M.}~\bibnamefont {Mehlman}}, \bibinfo {author} {\bibfnamefont
  {H.~I.}\ \bibnamefont {Park}}, \bibinfo {author} {\bibfnamefont {B.~T.}\
  \bibnamefont {Roeder}},\ and\ \bibinfo {author} {\bibfnamefont
  {D.}~\bibnamefont {Melconian}},\ }\href
  {https://doi.org/10.1103/PhysRevC.98.015502} {\bibfield  {journal} {\bibinfo
  {journal} {Phys. Rev. C}\ }\textbf {\bibinfo {volume} {98}},\ \bibinfo
  {pages} {015502} (\bibinfo {year} {2018})}\BibitemShut {NoStop}%
\bibitem [{\citenamefont {Iacob}\ \emph {et~al.}(2022)\citenamefont {Iacob}
  \emph {et~al.}}]{Iacob-InPrep}%
  \BibitemOpen
  \bibfield  {author} {\bibinfo {author} {\bibfnamefont {I.}~\bibnamefont
  {Iacob}} \emph {et~al.},\ }\href@noop {} {} (\bibinfo {year} {2022}),\
  \bibinfo {note} {in preparation}\BibitemShut {NoStop}%
\bibitem [{\citenamefont {Ozmetin}\ \emph {et~al.}(2022)\citenamefont {Ozmetin}
  \emph {et~al.}}]{Ozmetin-InPrep}%
  \BibitemOpen
  \bibfield  {author} {\bibinfo {author} {\bibfnamefont {A.}~\bibnamefont
  {Ozmetin}} \emph {et~al.},\ }\href@noop {} {} (\bibinfo {year} {2022}),\
  \bibinfo {note} {in preparation}\BibitemShut {NoStop}%
\bibitem [{\citenamefont {Valverde}\ \emph {et~al.}(2018)\citenamefont
  {Valverde}, \citenamefont {Brodeur}, \citenamefont {Ahn}, \citenamefont
  {Allen}, \citenamefont {Bardayan}, \citenamefont {Becchetti}, \citenamefont
  {Blankstein}, \citenamefont {Brown}, \citenamefont {Burdette}, \citenamefont
  {Frentz}, \citenamefont {Gilardy}, \citenamefont {Hall}, \citenamefont
  {King}, \citenamefont {Kolata}, \citenamefont {Long}, \citenamefont {Macon},
  \citenamefont {Nelson}, \citenamefont {O'Malley}, \citenamefont {Skulski},
  \citenamefont {Strauss},\ and\ \citenamefont {Vande~Kolk}}]{Valverde2018}%
  \BibitemOpen
  \bibfield  {author} {\bibinfo {author} {\bibfnamefont {A.~A.}\ \bibnamefont
  {Valverde}}, \bibinfo {author} {\bibfnamefont {M.}~\bibnamefont {Brodeur}},
  \bibinfo {author} {\bibfnamefont {T.}~\bibnamefont {Ahn}}, \bibinfo {author}
  {\bibfnamefont {J.}~\bibnamefont {Allen}}, \bibinfo {author} {\bibfnamefont
  {D.~W.}\ \bibnamefont {Bardayan}}, \bibinfo {author} {\bibfnamefont {F.~D.}\
  \bibnamefont {Becchetti}}, \bibinfo {author} {\bibfnamefont {D.}~\bibnamefont
  {Blankstein}}, \bibinfo {author} {\bibfnamefont {G.}~\bibnamefont {Brown}},
  \bibinfo {author} {\bibfnamefont {D.~P.}\ \bibnamefont {Burdette}}, \bibinfo
  {author} {\bibfnamefont {B.}~\bibnamefont {Frentz}}, \bibinfo {author}
  {\bibfnamefont {G.}~\bibnamefont {Gilardy}}, \bibinfo {author} {\bibfnamefont
  {M.~R.}\ \bibnamefont {Hall}}, \bibinfo {author} {\bibfnamefont
  {S.}~\bibnamefont {King}}, \bibinfo {author} {\bibfnamefont {J.~J.}\
  \bibnamefont {Kolata}}, \bibinfo {author} {\bibfnamefont {J.}~\bibnamefont
  {Long}}, \bibinfo {author} {\bibfnamefont {K.~T.}\ \bibnamefont {Macon}},
  \bibinfo {author} {\bibfnamefont {A.}~\bibnamefont {Nelson}}, \bibinfo
  {author} {\bibfnamefont {P.~D.}\ \bibnamefont {O'Malley}}, \bibinfo {author}
  {\bibfnamefont {M.}~\bibnamefont {Skulski}}, \bibinfo {author} {\bibfnamefont
  {S.~Y.}\ \bibnamefont {Strauss}},\ and\ \bibinfo {author} {\bibfnamefont
  {B.}~\bibnamefont {Vande~Kolk}},\ }\href
  {https://doi.org/10.1103/PhysRevC.97.035503} {\bibfield  {journal} {\bibinfo
  {journal} {Phys. Rev. C}\ }\textbf {\bibinfo {volume} {97}},\ \bibinfo
  {pages} {035503} (\bibinfo {year} {2018})}\BibitemShut {NoStop}%
\bibitem [{\citenamefont {Long}\ \emph {et~al.}(2022)\citenamefont {Long},
  \citenamefont {Nicoloff}, \citenamefont {Bardayan}, \citenamefont
  {Becchetti}, \citenamefont {Blankstein}, \citenamefont {Boomershine},
  \citenamefont {Burdette}, \citenamefont {Caprio}, \citenamefont {Caves},
  \citenamefont {Fasano}, \citenamefont {Frentz}, \citenamefont {Henderson},
  \citenamefont {Kelly}, \citenamefont {Kolata}, \citenamefont {Liu},
  \citenamefont {O'Malley}, \citenamefont {Strauss}, \citenamefont {Zite},\
  and\ \citenamefont {Brodeur}}]{Long2022}%
  \BibitemOpen
  \bibfield  {author} {\bibinfo {author} {\bibfnamefont {J.}~\bibnamefont
  {Long}}, \bibinfo {author} {\bibfnamefont {C.~R.}\ \bibnamefont {Nicoloff}},
  \bibinfo {author} {\bibfnamefont {D.~W.}\ \bibnamefont {Bardayan}}, \bibinfo
  {author} {\bibfnamefont {F.~D.}\ \bibnamefont {Becchetti}}, \bibinfo {author}
  {\bibfnamefont {D.}~\bibnamefont {Blankstein}}, \bibinfo {author}
  {\bibfnamefont {C.}~\bibnamefont {Boomershine}}, \bibinfo {author}
  {\bibfnamefont {D.~P.}\ \bibnamefont {Burdette}}, \bibinfo {author}
  {\bibfnamefont {M.~A.}\ \bibnamefont {Caprio}}, \bibinfo {author}
  {\bibfnamefont {L.}~\bibnamefont {Caves}}, \bibinfo {author} {\bibfnamefont
  {P.~J.}\ \bibnamefont {Fasano}}, \bibinfo {author} {\bibfnamefont
  {B.}~\bibnamefont {Frentz}}, \bibinfo {author} {\bibfnamefont {S.~L.}\
  \bibnamefont {Henderson}}, \bibinfo {author} {\bibfnamefont {J.~M.}\
  \bibnamefont {Kelly}}, \bibinfo {author} {\bibfnamefont {J.~J.}\ \bibnamefont
  {Kolata}}, \bibinfo {author} {\bibfnamefont {B.}~\bibnamefont {Liu}},
  \bibinfo {author} {\bibfnamefont {P.~D.}\ \bibnamefont {O'Malley}}, \bibinfo
  {author} {\bibfnamefont {S.~Y.}\ \bibnamefont {Strauss}}, \bibinfo {author}
  {\bibfnamefont {R.}~\bibnamefont {Zite}},\ and\ \bibinfo {author}
  {\bibfnamefont {M.}~\bibnamefont {Brodeur}},\ }\href
  {https://doi.org/10.1103/PhysRevC.106.045501} {\bibfield  {journal} {\bibinfo
   {journal} {Phys. Rev. C}\ }\textbf {\bibinfo {volume} {106}},\ \bibinfo
  {pages} {045501} (\bibinfo {year} {2022})}\BibitemShut {NoStop}%
\bibitem [{\citenamefont {Burdette}\ \emph {et~al.}(2020)\citenamefont
  {Burdette}, \citenamefont {Brodeur}, \citenamefont {Bardayan}, \citenamefont
  {Becchetti}, \citenamefont {Blankstein}, \citenamefont {Boomershine},
  \citenamefont {Caves}, \citenamefont {Henderson}, \citenamefont {Kolata},
  \citenamefont {Liu}, \citenamefont {Long}, \citenamefont {O'Malley},\ and\
  \citenamefont {Strauss}}]{Burdette2020}%
  \BibitemOpen
  \bibfield  {author} {\bibinfo {author} {\bibfnamefont {D.~P.}\ \bibnamefont
  {Burdette}}, \bibinfo {author} {\bibfnamefont {M.}~\bibnamefont {Brodeur}},
  \bibinfo {author} {\bibfnamefont {D.~W.}\ \bibnamefont {Bardayan}}, \bibinfo
  {author} {\bibfnamefont {F.~D.}\ \bibnamefont {Becchetti}}, \bibinfo {author}
  {\bibfnamefont {D.}~\bibnamefont {Blankstein}}, \bibinfo {author}
  {\bibfnamefont {C.}~\bibnamefont {Boomershine}}, \bibinfo {author}
  {\bibfnamefont {L.}~\bibnamefont {Caves}}, \bibinfo {author} {\bibfnamefont
  {S.~L.}\ \bibnamefont {Henderson}}, \bibinfo {author} {\bibfnamefont {J.~J.}\
  \bibnamefont {Kolata}}, \bibinfo {author} {\bibfnamefont {B.}~\bibnamefont
  {Liu}}, \bibinfo {author} {\bibfnamefont {J.}~\bibnamefont {Long}}, \bibinfo
  {author} {\bibfnamefont {P.~D.}\ \bibnamefont {O'Malley}},\ and\ \bibinfo
  {author} {\bibfnamefont {S.~Y.}\ \bibnamefont {Strauss}},\ }\href
  {https://doi.org/10.1103/PhysRevC.101.055504} {\bibfield  {journal} {\bibinfo
   {journal} {Phys. Rev. C}\ }\textbf {\bibinfo {volume} {101}},\ \bibinfo
  {pages} {055504} (\bibinfo {year} {2020})}\BibitemShut {NoStop}%
\bibitem [{\citenamefont {Long}\ \emph {et~al.}(2017)\citenamefont {Long},
  \citenamefont {Ahn}, \citenamefont {Allen}, \citenamefont {Bardayan},
  \citenamefont {Becchetti}, \citenamefont {Blankstein}, \citenamefont
  {Brodeur}, \citenamefont {Burdette}, \citenamefont {Frentz}, \citenamefont
  {Hall}, \citenamefont {Kelly}, \citenamefont {Kolata}, \citenamefont
  {O'Malley}, \citenamefont {Schultz}, \citenamefont {Strauss},\ and\
  \citenamefont {Valverde}}]{Long2017}%
  \BibitemOpen
  \bibfield  {author} {\bibinfo {author} {\bibfnamefont {J.}~\bibnamefont
  {Long}}, \bibinfo {author} {\bibfnamefont {T.}~\bibnamefont {Ahn}}, \bibinfo
  {author} {\bibfnamefont {J.}~\bibnamefont {Allen}}, \bibinfo {author}
  {\bibfnamefont {D.~W.}\ \bibnamefont {Bardayan}}, \bibinfo {author}
  {\bibfnamefont {F.~D.}\ \bibnamefont {Becchetti}}, \bibinfo {author}
  {\bibfnamefont {D.}~\bibnamefont {Blankstein}}, \bibinfo {author}
  {\bibfnamefont {M.}~\bibnamefont {Brodeur}}, \bibinfo {author} {\bibfnamefont
  {D.}~\bibnamefont {Burdette}}, \bibinfo {author} {\bibfnamefont
  {B.}~\bibnamefont {Frentz}}, \bibinfo {author} {\bibfnamefont {M.~R.}\
  \bibnamefont {Hall}}, \bibinfo {author} {\bibfnamefont {J.~M.}\ \bibnamefont
  {Kelly}}, \bibinfo {author} {\bibfnamefont {J.~J.}\ \bibnamefont {Kolata}},
  \bibinfo {author} {\bibfnamefont {P.~D.}\ \bibnamefont {O'Malley}}, \bibinfo
  {author} {\bibfnamefont {B.~E.}\ \bibnamefont {Schultz}}, \bibinfo {author}
  {\bibfnamefont {S.~Y.}\ \bibnamefont {Strauss}},\ and\ \bibinfo {author}
  {\bibfnamefont {A.~A.}\ \bibnamefont {Valverde}},\ }\href
  {https://doi.org/10.1103/PhysRevC.96.015502} {\bibfield  {journal} {\bibinfo
  {journal} {Phys. Rev. C}\ }\textbf {\bibinfo {volume} {96}},\ \bibinfo
  {pages} {015502} (\bibinfo {year} {2017})}\BibitemShut {NoStop}%
\bibitem [{\citenamefont {Long}\ \emph {et~al.}(2020)\citenamefont {Long},
  \citenamefont {Brodeur}, \citenamefont {Baines}, \citenamefont {Bardayan},
  \citenamefont {Becchetti}, \citenamefont {Blankstein}, \citenamefont
  {Boomershine}, \citenamefont {Burdette}, \citenamefont {Clark}, \citenamefont
  {Frentz}, \citenamefont {Henderson}, \citenamefont {Kelly}, \citenamefont
  {Kolata}, \citenamefont {Liu}, \citenamefont {Macon}, \citenamefont
  {O'Malley}, \citenamefont {Pardo}, \citenamefont {Seymour}, \citenamefont
  {Strauss},\ and\ \citenamefont {Vande~Kolk}}]{Long2020}%
  \BibitemOpen
  \bibfield  {author} {\bibinfo {author} {\bibfnamefont {J.}~\bibnamefont
  {Long}}, \bibinfo {author} {\bibfnamefont {M.}~\bibnamefont {Brodeur}},
  \bibinfo {author} {\bibfnamefont {M.}~\bibnamefont {Baines}}, \bibinfo
  {author} {\bibfnamefont {D.~W.}\ \bibnamefont {Bardayan}}, \bibinfo {author}
  {\bibfnamefont {F.~D.}\ \bibnamefont {Becchetti}}, \bibinfo {author}
  {\bibfnamefont {D.}~\bibnamefont {Blankstein}}, \bibinfo {author}
  {\bibfnamefont {C.}~\bibnamefont {Boomershine}}, \bibinfo {author}
  {\bibfnamefont {D.~P.}\ \bibnamefont {Burdette}}, \bibinfo {author}
  {\bibfnamefont {A.~M.}\ \bibnamefont {Clark}}, \bibinfo {author}
  {\bibfnamefont {B.}~\bibnamefont {Frentz}}, \bibinfo {author} {\bibfnamefont
  {S.~L.}\ \bibnamefont {Henderson}}, \bibinfo {author} {\bibfnamefont {J.~M.}\
  \bibnamefont {Kelly}}, \bibinfo {author} {\bibfnamefont {J.~J.}\ \bibnamefont
  {Kolata}}, \bibinfo {author} {\bibfnamefont {B.}~\bibnamefont {Liu}},
  \bibinfo {author} {\bibfnamefont {K.~T.}\ \bibnamefont {Macon}}, \bibinfo
  {author} {\bibfnamefont {P.~D.}\ \bibnamefont {O'Malley}}, \bibinfo {author}
  {\bibfnamefont {A.}~\bibnamefont {Pardo}}, \bibinfo {author} {\bibfnamefont
  {C.}~\bibnamefont {Seymour}}, \bibinfo {author} {\bibfnamefont {S.~Y.}\
  \bibnamefont {Strauss}},\ and\ \bibinfo {author} {\bibfnamefont
  {B.}~\bibnamefont {Vande~Kolk}},\ }\href
  {https://doi.org/10.1103/PhysRevC.101.015501} {\bibfield  {journal} {\bibinfo
   {journal} {Phys. Rev. C}\ }\textbf {\bibinfo {volume} {101}},\ \bibinfo
  {pages} {015501} (\bibinfo {year} {2020})}\BibitemShut {NoStop}%
\bibitem [{\citenamefont {Gulyuz}\ \emph {et~al.}(2016)\citenamefont {Gulyuz},
  \citenamefont {Bollen}, \citenamefont {Brodeur}, \citenamefont {Bryce},
  \citenamefont {Cooper}, \citenamefont {Eibach}, \citenamefont {Izzo},
  \citenamefont {Kwan}, \citenamefont {Manukyan}, \citenamefont {Morrissey},
  \citenamefont {Naviliat-Cuncic}, \citenamefont {Redshaw}, \citenamefont
  {Ringle}, \citenamefont {Sandler}, \citenamefont {Schwarz}, \citenamefont
  {Sumithrarachchi}, \citenamefont {Valverde},\ and\ \citenamefont
  {Villari}}]{Gulyuz2016}%
  \BibitemOpen
  \bibfield  {author} {\bibinfo {author} {\bibfnamefont {K.}~\bibnamefont
  {Gulyuz}}, \bibinfo {author} {\bibfnamefont {G.}~\bibnamefont {Bollen}},
  \bibinfo {author} {\bibfnamefont {M.}~\bibnamefont {Brodeur}}, \bibinfo
  {author} {\bibfnamefont {R.~A.}\ \bibnamefont {Bryce}}, \bibinfo {author}
  {\bibfnamefont {K.}~\bibnamefont {Cooper}}, \bibinfo {author} {\bibfnamefont
  {M.}~\bibnamefont {Eibach}}, \bibinfo {author} {\bibfnamefont
  {C.}~\bibnamefont {Izzo}}, \bibinfo {author} {\bibfnamefont {E.}~\bibnamefont
  {Kwan}}, \bibinfo {author} {\bibfnamefont {K.}~\bibnamefont {Manukyan}},
  \bibinfo {author} {\bibfnamefont {D.~J.}\ \bibnamefont {Morrissey}}, \bibinfo
  {author} {\bibfnamefont {O.}~\bibnamefont {Naviliat-Cuncic}}, \bibinfo
  {author} {\bibfnamefont {M.}~\bibnamefont {Redshaw}}, \bibinfo {author}
  {\bibfnamefont {R.}~\bibnamefont {Ringle}}, \bibinfo {author} {\bibfnamefont
  {R.}~\bibnamefont {Sandler}}, \bibinfo {author} {\bibfnamefont
  {S.}~\bibnamefont {Schwarz}}, \bibinfo {author} {\bibfnamefont {C.~S.}\
  \bibnamefont {Sumithrarachchi}}, \bibinfo {author} {\bibfnamefont {A.~A.}\
  \bibnamefont {Valverde}},\ and\ \bibinfo {author} {\bibfnamefont {A.~C.~C.}\
  \bibnamefont {Villari}},\ }\href
  {https://doi.org/10.1103/PhysRevLett.116.012501} {\bibfield  {journal}
  {\bibinfo  {journal} {Phys. Rev. Lett.}\ }\textbf {\bibinfo {volume} {116}},\
  \bibinfo {pages} {012501} (\bibinfo {year} {2016})}\BibitemShut {NoStop}%
\bibitem [{\citenamefont {Eibach}\ \emph {et~al.}(2015)\citenamefont {Eibach},
  \citenamefont {Bollen}, \citenamefont {Brodeur}, \citenamefont {Cooper},
  \citenamefont {Gulyuz}, \citenamefont {Izzo}, \citenamefont {Morrissey},
  \citenamefont {Redshaw}, \citenamefont {Ringle}, \citenamefont {Sandler},
  \citenamefont {Schwarz}, \citenamefont {Sumithrarachchi}, \citenamefont
  {Valverde},\ and\ \citenamefont {Villari}}]{Eibach2015}%
  \BibitemOpen
  \bibfield  {author} {\bibinfo {author} {\bibfnamefont {M.}~\bibnamefont
  {Eibach}}, \bibinfo {author} {\bibfnamefont {G.}~\bibnamefont {Bollen}},
  \bibinfo {author} {\bibfnamefont {M.}~\bibnamefont {Brodeur}}, \bibinfo
  {author} {\bibfnamefont {K.}~\bibnamefont {Cooper}}, \bibinfo {author}
  {\bibfnamefont {K.}~\bibnamefont {Gulyuz}}, \bibinfo {author} {\bibfnamefont
  {C.}~\bibnamefont {Izzo}}, \bibinfo {author} {\bibfnamefont {D.~J.}\
  \bibnamefont {Morrissey}}, \bibinfo {author} {\bibfnamefont {M.}~\bibnamefont
  {Redshaw}}, \bibinfo {author} {\bibfnamefont {R.}~\bibnamefont {Ringle}},
  \bibinfo {author} {\bibfnamefont {R.}~\bibnamefont {Sandler}}, \bibinfo
  {author} {\bibfnamefont {S.}~\bibnamefont {Schwarz}}, \bibinfo {author}
  {\bibfnamefont {C.~S.}\ \bibnamefont {Sumithrarachchi}}, \bibinfo {author}
  {\bibfnamefont {A.~A.}\ \bibnamefont {Valverde}},\ and\ \bibinfo {author}
  {\bibfnamefont {A.~C.~C.}\ \bibnamefont {Villari}},\ }\href
  {https://doi.org/10.1103/PhysRevC.92.045502} {\bibfield  {journal} {\bibinfo
  {journal} {Phys. Rev. C}\ }\textbf {\bibinfo {volume} {92}},\ \bibinfo
  {pages} {045502} (\bibinfo {year} {2015})}\BibitemShut {NoStop}%
\bibitem [{\citenamefont {Fenker}\ \emph {et~al.}(2016)\citenamefont {Fenker},
  \citenamefont {Behr}, \citenamefont {Melconian}, \citenamefont {Anderson},
  \citenamefont {Anholm}, \citenamefont {Ashery}, \citenamefont {Behling},
  \citenamefont {Cohen}, \citenamefont {Craiciu}, \citenamefont {Donohue},
  \citenamefont {Farfan}, \citenamefont {Friesen}, \citenamefont {Gorelov},
  \citenamefont {McNeil}, \citenamefont {Mehlman}, \citenamefont {Norton},
  \citenamefont {Olchanski}, \citenamefont {Smale}, \citenamefont
  {Th{\'{e}}riault}, \citenamefont {Vantyghem},\ and\ \citenamefont
  {Warner}}]{fenkerNJP}%
  \BibitemOpen
  \bibfield  {author} {\bibinfo {author} {\bibfnamefont {B.}~\bibnamefont
  {Fenker}}, \bibinfo {author} {\bibfnamefont {J.~A.}\ \bibnamefont {Behr}},
  \bibinfo {author} {\bibfnamefont {D.}~\bibnamefont {Melconian}}, \bibinfo
  {author} {\bibfnamefont {R.~M.~A.}\ \bibnamefont {Anderson}}, \bibinfo
  {author} {\bibfnamefont {M.}~\bibnamefont {Anholm}}, \bibinfo {author}
  {\bibfnamefont {D.}~\bibnamefont {Ashery}}, \bibinfo {author} {\bibfnamefont
  {R.~S.}\ \bibnamefont {Behling}}, \bibinfo {author} {\bibfnamefont
  {I.}~\bibnamefont {Cohen}}, \bibinfo {author} {\bibfnamefont
  {I.}~\bibnamefont {Craiciu}}, \bibinfo {author} {\bibfnamefont {J.~M.}\
  \bibnamefont {Donohue}}, \bibinfo {author} {\bibfnamefont {C.}~\bibnamefont
  {Farfan}}, \bibinfo {author} {\bibfnamefont {D.}~\bibnamefont {Friesen}},
  \bibinfo {author} {\bibfnamefont {A.}~\bibnamefont {Gorelov}}, \bibinfo
  {author} {\bibfnamefont {J.}~\bibnamefont {McNeil}}, \bibinfo {author}
  {\bibfnamefont {M.}~\bibnamefont {Mehlman}}, \bibinfo {author} {\bibfnamefont
  {H.}~\bibnamefont {Norton}}, \bibinfo {author} {\bibfnamefont
  {K.}~\bibnamefont {Olchanski}}, \bibinfo {author} {\bibfnamefont
  {S.}~\bibnamefont {Smale}}, \bibinfo {author} {\bibfnamefont
  {O.}~\bibnamefont {Th{\'{e}}riault}}, \bibinfo {author} {\bibfnamefont
  {A.~N.}\ \bibnamefont {Vantyghem}},\ and\ \bibinfo {author} {\bibfnamefont
  {C.~L.}\ \bibnamefont {Warner}},\ }\href
  {https://doi.org/10.1088/1367-2630/18/7/073028} {\bibfield  {journal}
  {\bibinfo  {journal} {New Journal of Physics}\ }\textbf {\bibinfo {volume}
  {18}},\ \bibinfo {pages} {073028} (\bibinfo {year} {2016})}\BibitemShut
  {NoStop}%
\bibitem [{\citenamefont {Fenker}\ \emph {et~al.}(2018)\citenamefont {Fenker},
  \citenamefont {Gorelov}, \citenamefont {Melconian}, \citenamefont {Behr},
  \citenamefont {Anholm}, \citenamefont {Ashery}, \citenamefont {Behling},
  \citenamefont {Cohen}, \citenamefont {Craiciu}, \citenamefont {Gwinner},
  \citenamefont {McNeil}, \citenamefont {Mehlman}, \citenamefont {Olchanski},
  \citenamefont {Shidling}, \citenamefont {Smale},\ and\ \citenamefont
  {Warner}}]{Fenker2018}%
  \BibitemOpen
  \bibfield  {author} {\bibinfo {author} {\bibfnamefont {B.}~\bibnamefont
  {Fenker}}, \bibinfo {author} {\bibfnamefont {A.}~\bibnamefont {Gorelov}},
  \bibinfo {author} {\bibfnamefont {D.}~\bibnamefont {Melconian}}, \bibinfo
  {author} {\bibfnamefont {J.~A.}\ \bibnamefont {Behr}}, \bibinfo {author}
  {\bibfnamefont {M.}~\bibnamefont {Anholm}}, \bibinfo {author} {\bibfnamefont
  {D.}~\bibnamefont {Ashery}}, \bibinfo {author} {\bibfnamefont {R.~S.}\
  \bibnamefont {Behling}}, \bibinfo {author} {\bibfnamefont {I.}~\bibnamefont
  {Cohen}}, \bibinfo {author} {\bibfnamefont {I.}~\bibnamefont {Craiciu}},
  \bibinfo {author} {\bibfnamefont {G.}~\bibnamefont {Gwinner}}, \bibinfo
  {author} {\bibfnamefont {J.}~\bibnamefont {McNeil}}, \bibinfo {author}
  {\bibfnamefont {M.}~\bibnamefont {Mehlman}}, \bibinfo {author} {\bibfnamefont
  {K.}~\bibnamefont {Olchanski}}, \bibinfo {author} {\bibfnamefont {P.~D.}\
  \bibnamefont {Shidling}}, \bibinfo {author} {\bibfnamefont {S.}~\bibnamefont
  {Smale}},\ and\ \bibinfo {author} {\bibfnamefont {C.~L.}\ \bibnamefont
  {Warner}},\ }\href {https://doi.org/10.1103/PhysRevLett.120.062502}
  {\bibfield  {journal} {\bibinfo  {journal} {Phys. Rev. Lett.}\ }\textbf
  {\bibinfo {volume} {120}},\ \bibinfo {pages} {062502} (\bibinfo {year}
  {2018})}\BibitemShut {NoStop}%
\bibitem [{\citenamefont {Seng}\ and\ \citenamefont
  {Gorchtein}(2022{\natexlab{b}})}]{Seng:2022epj}%
  \BibitemOpen
  \bibfield  {author} {\bibinfo {author} {\bibfnamefont {C.-Y.}\ \bibnamefont
  {Seng}}\ and\ \bibinfo {author} {\bibfnamefont {M.}~\bibnamefont
  {Gorchtein}},\ }\href@noop {} {\  (\bibinfo {year} {2022}{\natexlab{b}})},\
  \Eprint {https://arxiv.org/abs/2208.03037} {arXiv:2208.03037 [nucl-th]}
  \BibitemShut {NoStop}%
\bibitem [{\citenamefont {Seng}(2022)}]{Seng:2022inj}%
  \BibitemOpen
  \bibfield  {author} {\bibinfo {author} {\bibfnamefont {C.-Y.}\ \bibnamefont
  {Seng}},\ }\href@noop {} {\  (\bibinfo {year} {2022})},\ \Eprint
  {https://arxiv.org/abs/2212.02681} {arXiv:2212.02681 [nucl-th]} \BibitemShut
  {NoStop}%
\bibitem [{\citenamefont {Shidling}\ \emph {et~al.}(2021)\citenamefont
  {Shidling} \emph {et~al.}}]{shidling2021}%
  \BibitemOpen
  \bibfield  {author} {\bibinfo {author} {\bibfnamefont {P.}~\bibnamefont
  {Shidling}} \emph {et~al.},\ }\href@noop {} {\bibfield  {journal} {\bibinfo
  {journal} {Int. J. Mass Spectr.}\ }\textbf {\bibinfo {volume} {468}},\
  \bibinfo {pages} {116636} (\bibinfo {year} {2021})}\BibitemShut {NoStop}%
\bibitem [{\citenamefont {Naviliat-Cuncic}\ and\ \citenamefont
  {Severijns}(2009)}]{Naviliat2009}%
  \BibitemOpen
  \bibfield  {author} {\bibinfo {author} {\bibfnamefont {O.}~\bibnamefont
  {Naviliat-Cuncic}}\ and\ \bibinfo {author} {\bibfnamefont {N.}~\bibnamefont
  {Severijns}},\ }\href {https://doi.org/10.1103/PhysRevLett.102.142302}
  {\bibfield  {journal} {\bibinfo  {journal} {Phys. Rev. Lett.}\ }\textbf
  {\bibinfo {volume} {102}},\ \bibinfo {pages} {142302} (\bibinfo {year}
  {2009})}\BibitemShut {NoStop}%
\bibitem [{\citenamefont {Hayen}\ and\ \citenamefont
  {Young}(2020)}]{Hayen2020}%
  \BibitemOpen
  \bibfield  {author} {\bibinfo {author} {\bibfnamefont {L.}~\bibnamefont
  {Hayen}}\ and\ \bibinfo {author} {\bibfnamefont {A.~R.}\ \bibnamefont
  {Young}},\ }\href@noop {} {\bibinfo {title} {Consistent description of
  angular correlations in $\beta$ decay for beyond standard model physics
  searches}} (\bibinfo {year} {2020}),\ \Eprint
  {https://arxiv.org/abs/2009.11364} {arXiv:2009.11364 [nucl-th]} \BibitemShut
  {NoStop}%
\bibitem [{\citenamefont {Brodeur}\ \emph {et~al.}(2016)\citenamefont
  {Brodeur}, \citenamefont {Kelly}, \citenamefont {Long}, \citenamefont
  {Nicoloff},\ and\ \citenamefont {Schultz}}]{Brodeur2016}%
  \BibitemOpen
  \bibfield  {author} {\bibinfo {author} {\bibfnamefont {M.}~\bibnamefont
  {Brodeur}}, \bibinfo {author} {\bibfnamefont {J.}~\bibnamefont {Kelly}},
  \bibinfo {author} {\bibfnamefont {J.}~\bibnamefont {Long}}, \bibinfo {author}
  {\bibfnamefont {C.}~\bibnamefont {Nicoloff}},\ and\ \bibinfo {author}
  {\bibfnamefont {B.}~\bibnamefont {Schultz}},\ }\href
  {https://doi.org/https://doi.org/10.1016/j.nimb.2015.12.038} {\bibfield
  {journal} {\bibinfo  {journal} {Nuclear Instruments and Methods in Physics
  Research Section B: Beam Interactions with Materials and Atoms}\ }\textbf
  {\bibinfo {volume} {376}},\ \bibinfo {pages} {281 } (\bibinfo {year}
  {2016})},\ \bibinfo {note} {proceedings of the XVIIth International
  Conference on Electromagnetic Isotope Separators and Related Topics
  (EMIS2015), Grand Rapids, MI, U.S.A., 11-15 May 2015}\BibitemShut {NoStop}%
\bibitem [{\citenamefont {O'Malley}\ \emph {et~al.}(2020)\citenamefont
  {O'Malley}, \citenamefont {Brodeur}, \citenamefont {Burdette}, \citenamefont
  {Klimes}, \citenamefont {Valverde}, \citenamefont {Clark}, \citenamefont
  {Savard}, \citenamefont {Ringle},\ and\ \citenamefont
  {Varentsov}}]{OMalley2020}%
  \BibitemOpen
  \bibfield  {author} {\bibinfo {author} {\bibfnamefont {P.}~\bibnamefont
  {O'Malley}}, \bibinfo {author} {\bibfnamefont {M.}~\bibnamefont {Brodeur}},
  \bibinfo {author} {\bibfnamefont {D.}~\bibnamefont {Burdette}}, \bibinfo
  {author} {\bibfnamefont {J.}~\bibnamefont {Klimes}}, \bibinfo {author}
  {\bibfnamefont {A.}~\bibnamefont {Valverde}}, \bibinfo {author}
  {\bibfnamefont {J.}~\bibnamefont {Clark}}, \bibinfo {author} {\bibfnamefont
  {G.}~\bibnamefont {Savard}}, \bibinfo {author} {\bibfnamefont
  {R.}~\bibnamefont {Ringle}},\ and\ \bibinfo {author} {\bibfnamefont
  {V.}~\bibnamefont {Varentsov}},\ }\href
  {https://doi.org/https://doi.org/10.1016/j.nimb.2019.04.017} {\bibfield
  {journal} {\bibinfo  {journal} {Nuclear Instruments and Methods in Physics
  Research Section B: Beam Interactions with Materials and Atoms}\ }\textbf
  {\bibinfo {volume} {463}},\ \bibinfo {pages} {488 } (\bibinfo {year}
  {2020})}\BibitemShut {NoStop}%
\bibitem [{\citenamefont {Fretwell}\ \emph {et~al.}(2020)\citenamefont
  {Fretwell}, \citenamefont {Leach}, \citenamefont {Bray}, \citenamefont {Kim},
  \citenamefont {Dilling}, \citenamefont {Lennarz}, \citenamefont {Mougeot},
  \citenamefont {Ponce}, \citenamefont {Ruiz}, \citenamefont {Stackhouse},\
  and\ \citenamefont {Friedrich}}]{Fretwell2020}%
  \BibitemOpen
  \bibfield  {author} {\bibinfo {author} {\bibfnamefont {S.}~\bibnamefont
  {Fretwell}}, \bibinfo {author} {\bibfnamefont {K.~G.}\ \bibnamefont {Leach}},
  \bibinfo {author} {\bibfnamefont {C.}~\bibnamefont {Bray}}, \bibinfo {author}
  {\bibfnamefont {G.~B.}\ \bibnamefont {Kim}}, \bibinfo {author} {\bibfnamefont
  {J.}~\bibnamefont {Dilling}}, \bibinfo {author} {\bibfnamefont
  {A.}~\bibnamefont {Lennarz}}, \bibinfo {author} {\bibfnamefont
  {X.}~\bibnamefont {Mougeot}}, \bibinfo {author} {\bibfnamefont
  {F.}~\bibnamefont {Ponce}}, \bibinfo {author} {\bibfnamefont
  {C.}~\bibnamefont {Ruiz}}, \bibinfo {author} {\bibfnamefont {J.}~\bibnamefont
  {Stackhouse}},\ and\ \bibinfo {author} {\bibfnamefont {S.}~\bibnamefont
  {Friedrich}},\ }\href {https://doi.org/10.1103/PhysRevLett.125.032701}
  {\bibfield  {journal} {\bibinfo  {journal} {Physical Review Letters}\
  }\textbf {\bibinfo {volume} {125}},\ \bibinfo {pages} {32701} (\bibinfo
  {year} {2020})}\BibitemShut {NoStop}%
\bibitem [{\citenamefont {Friedrich}\ \emph {et~al.}(2021)\citenamefont
  {Friedrich}, \citenamefont {Kim}, \citenamefont {Bray}, \citenamefont
  {Cantor}, \citenamefont {Dilling}, \citenamefont {Fretwell}, \citenamefont
  {Hall}, \citenamefont {Lennarz}, \citenamefont {Lordi}, \citenamefont
  {Machule}, \citenamefont {McKeen}, \citenamefont {Mougeot}, \citenamefont
  {Ponce}, \citenamefont {Ruiz}, \citenamefont {Samanta}, \citenamefont
  {Warburton},\ and\ \citenamefont {Leach}}]{Friedrich2021}%
  \BibitemOpen
  \bibfield  {author} {\bibinfo {author} {\bibfnamefont {S.}~\bibnamefont
  {Friedrich}}, \bibinfo {author} {\bibfnamefont {G.~B.}\ \bibnamefont {Kim}},
  \bibinfo {author} {\bibfnamefont {C.}~\bibnamefont {Bray}}, \bibinfo {author}
  {\bibfnamefont {R.}~\bibnamefont {Cantor}}, \bibinfo {author} {\bibfnamefont
  {J.}~\bibnamefont {Dilling}}, \bibinfo {author} {\bibfnamefont
  {S.}~\bibnamefont {Fretwell}}, \bibinfo {author} {\bibfnamefont {J.~A.}\
  \bibnamefont {Hall}}, \bibinfo {author} {\bibfnamefont {A.}~\bibnamefont
  {Lennarz}}, \bibinfo {author} {\bibfnamefont {V.}~\bibnamefont {Lordi}},
  \bibinfo {author} {\bibfnamefont {P.}~\bibnamefont {Machule}}, \bibinfo
  {author} {\bibfnamefont {D.}~\bibnamefont {McKeen}}, \bibinfo {author}
  {\bibfnamefont {X.}~\bibnamefont {Mougeot}}, \bibinfo {author} {\bibfnamefont
  {F.}~\bibnamefont {Ponce}}, \bibinfo {author} {\bibfnamefont
  {C.}~\bibnamefont {Ruiz}}, \bibinfo {author} {\bibfnamefont {A.}~\bibnamefont
  {Samanta}}, \bibinfo {author} {\bibfnamefont {W.~K.}\ \bibnamefont
  {Warburton}},\ and\ \bibinfo {author} {\bibfnamefont {K.~G.}\ \bibnamefont
  {Leach}},\ }\href {https://doi.org/10.1103/PhysRevLett.126.021803} {\bibfield
   {journal} {\bibinfo  {journal} {Physical Review Letters}\ }\textbf {\bibinfo
  {volume} {126}},\ \bibinfo {pages} {021803} (\bibinfo {year} {2021})},\
  \Eprint {https://arxiv.org/abs/2010.09603} {arXiv:2010.09603} \BibitemShut
  {NoStop}%
\bibitem [{\citenamefont {Glick-Magid}\ \emph {et~al.}(2017)\citenamefont
  {Glick-Magid}, \citenamefont {Mishnayot}, \citenamefont {Mukul},
  \citenamefont {Hass}, \citenamefont {Vaintraub}, \citenamefont {Ron},\ and\
  \citenamefont {Gazit}}]{Glick-Magid2017}%
  \BibitemOpen
  \bibfield  {author} {\bibinfo {author} {\bibfnamefont {A.}~\bibnamefont
  {Glick-Magid}}, \bibinfo {author} {\bibfnamefont {Y.}~\bibnamefont
  {Mishnayot}}, \bibinfo {author} {\bibfnamefont {I.}~\bibnamefont {Mukul}},
  \bibinfo {author} {\bibfnamefont {M.}~\bibnamefont {Hass}}, \bibinfo {author}
  {\bibfnamefont {S.}~\bibnamefont {Vaintraub}}, \bibinfo {author}
  {\bibfnamefont {G.}~\bibnamefont {Ron}},\ and\ \bibinfo {author}
  {\bibfnamefont {D.}~\bibnamefont {Gazit}},\ }\href
  {https://doi.org/https://doi.org/10.1016/j.physletb.2017.02.023} {\bibfield
  {journal} {\bibinfo  {journal} {Physics Letters B}\ }\textbf {\bibinfo
  {volume} {767}},\ \bibinfo {pages} {285} (\bibinfo {year}
  {2017})}\BibitemShut {NoStop}%
\bibitem [{\citenamefont {Sternberg}\ \emph {et~al.}(2015)\citenamefont
  {Sternberg}, \citenamefont {Segel}, \citenamefont {Scielzo}, \citenamefont
  {Savard}, \citenamefont {Clark}, \citenamefont {Bertone}, \citenamefont
  {Buchinger}, \citenamefont {Burkey}, \citenamefont {Caldwell}, \citenamefont
  {Chaudhuri}, \citenamefont {Crawford}, \citenamefont {Deibel}, \citenamefont
  {Greene}, \citenamefont {Gulick}, \citenamefont {Lascar}, \citenamefont
  {Levand}, \citenamefont {Li}, \citenamefont {P\'erez~Galv\'an}, \citenamefont
  {Sharma}, \citenamefont {Van~Schelt}, \citenamefont {Yee},\ and\
  \citenamefont {Zabransky}}]{sternberg2015}%
  \BibitemOpen
  \bibfield  {author} {\bibinfo {author} {\bibfnamefont {M.~G.}\ \bibnamefont
  {Sternberg}}, \bibinfo {author} {\bibfnamefont {R.}~\bibnamefont {Segel}},
  \bibinfo {author} {\bibfnamefont {N.~D.}\ \bibnamefont {Scielzo}}, \bibinfo
  {author} {\bibfnamefont {G.}~\bibnamefont {Savard}}, \bibinfo {author}
  {\bibfnamefont {J.~A.}\ \bibnamefont {Clark}}, \bibinfo {author}
  {\bibfnamefont {P.~F.}\ \bibnamefont {Bertone}}, \bibinfo {author}
  {\bibfnamefont {F.}~\bibnamefont {Buchinger}}, \bibinfo {author}
  {\bibfnamefont {M.}~\bibnamefont {Burkey}}, \bibinfo {author} {\bibfnamefont
  {S.}~\bibnamefont {Caldwell}}, \bibinfo {author} {\bibfnamefont
  {A.}~\bibnamefont {Chaudhuri}}, \bibinfo {author} {\bibfnamefont {J.~E.}\
  \bibnamefont {Crawford}}, \bibinfo {author} {\bibfnamefont {C.~M.}\
  \bibnamefont {Deibel}}, \bibinfo {author} {\bibfnamefont {J.}~\bibnamefont
  {Greene}}, \bibinfo {author} {\bibfnamefont {S.}~\bibnamefont {Gulick}},
  \bibinfo {author} {\bibfnamefont {D.}~\bibnamefont {Lascar}}, \bibinfo
  {author} {\bibfnamefont {A.~F.}\ \bibnamefont {Levand}}, \bibinfo {author}
  {\bibfnamefont {G.}~\bibnamefont {Li}}, \bibinfo {author} {\bibfnamefont
  {A.}~\bibnamefont {P\'erez~Galv\'an}}, \bibinfo {author} {\bibfnamefont
  {K.~S.}\ \bibnamefont {Sharma}}, \bibinfo {author} {\bibfnamefont
  {J.}~\bibnamefont {Van~Schelt}}, \bibinfo {author} {\bibfnamefont {R.~M.}\
  \bibnamefont {Yee}},\ and\ \bibinfo {author} {\bibfnamefont {B.~J.}\
  \bibnamefont {Zabransky}},\ }\href
  {https://doi.org/10.1103/PhysRevLett.115.182501} {\bibfield  {journal}
  {\bibinfo  {journal} {Phys. Rev. Lett.}\ }\textbf {\bibinfo {volume} {115}},\
  \bibinfo {pages} {182501} (\bibinfo {year} {2015})}\BibitemShut {NoStop}%
\bibitem [{\citenamefont {Gallant}\ \emph {et~al.}(2022)\citenamefont {Gallant}
  \emph {et~al.}}]{gallant2022}%
  \BibitemOpen
  \bibfield  {author} {\bibinfo {author} {\bibfnamefont {A.~T.}\ \bibnamefont
  {Gallant}} \emph {et~al.}} (\bibinfo {year} {2022}),\ \bibinfo {note}
  {submitted to Phys. Rev. Lett.}\BibitemShut {Stop}%
\bibitem [{\citenamefont {M\"uller}\ \emph {et~al.}(2022)\citenamefont
  {M\"uller} \emph {et~al.}}]{mueller2022}%
  \BibitemOpen
  \bibfield  {author} {\bibinfo {author} {\bibfnamefont {P.}~\bibnamefont
  {M\"uller}} \emph {et~al.},\ }\href@noop {} {\bibfield  {journal} {\bibinfo
  {journal} {Phys.\ Rev.\ Lett.}\ }\textbf {\bibinfo {volume} {129}},\ \bibinfo
  {pages} {182502} (\bibinfo {year} {2022})}\BibitemShut {NoStop}%
\bibitem [{\citenamefont {Byron}\ \emph {et~al.}(2022)\citenamefont {Byron}
  \emph {et~al.}}]{CRES-arXiv}%
  \BibitemOpen
  \bibfield  {author} {\bibinfo {author} {\bibfnamefont {W.}~\bibnamefont
  {Byron}} \emph {et~al.},\ }\href@noop {} {\bibinfo {title} {First observation
  of cyclotron radiation from mev-scale $e^\pm$ following nuclear beta decay}}
  (\bibinfo {year} {2022}),\ \bibinfo {note} {arXiv:2209.02870}\BibitemShut
  {NoStop}%
\bibitem [{\citenamefont {Hayen}\ \emph {et~al.}(2018)\citenamefont {Hayen},
  \citenamefont {Severijns}, \citenamefont {Bodek}, \citenamefont {Rozpedzik},\
  and\ \citenamefont {Mougeot}}]{Hayen2018}%
  \BibitemOpen
  \bibfield  {author} {\bibinfo {author} {\bibfnamefont {L.}~\bibnamefont
  {Hayen}}, \bibinfo {author} {\bibfnamefont {N.}~\bibnamefont {Severijns}},
  \bibinfo {author} {\bibfnamefont {K.}~\bibnamefont {Bodek}}, \bibinfo
  {author} {\bibfnamefont {D.}~\bibnamefont {Rozpedzik}},\ and\ \bibinfo
  {author} {\bibfnamefont {X.}~\bibnamefont {Mougeot}},\ }\href
  {https://doi.org/10.1103/RevModPhys.90.015008} {\bibfield  {journal}
  {\bibinfo  {journal} {Reviews of Modern Physics}\ }\textbf {\bibinfo {volume}
  {90}},\ \bibinfo {pages} {015008} (\bibinfo {year} {2018})},\ \Eprint
  {https://arxiv.org/abs/1709.07530} {arXiv:1709.07530} \BibitemShut {NoStop}%
\bibitem [{\citenamefont {Glick-Magid}\ and\ \citenamefont
  {Gazit}(2022)}]{Glick-Magid2022formalism}%
  \BibitemOpen
  \bibfield  {author} {\bibinfo {author} {\bibfnamefont {A.}~\bibnamefont
  {Glick-Magid}}\ and\ \bibinfo {author} {\bibfnamefont {D.}~\bibnamefont
  {Gazit}},\ }\href@noop {} {\bibfield  {journal} {\bibinfo  {journal} {Journal
  of Physics G: Nuclear and Particle Physics}\ }\textbf {\bibinfo {volume}
  {49}},\ \bibinfo {pages} {105105} (\bibinfo {year} {2022})}\BibitemShut
  {NoStop}%
\bibitem [{\citenamefont {King}\ \emph {et~al.}(2022)\citenamefont {King},
  \citenamefont {Baroni}, \citenamefont {Cirigliano}, \citenamefont {Gandolfi},
  \citenamefont {Hayen}, \citenamefont {Mereghetti}, \citenamefont {Pastore},\
  and\ \citenamefont {Piarulli}}]{King2022}%
  \BibitemOpen
  \bibfield  {author} {\bibinfo {author} {\bibfnamefont {G.~B.}\ \bibnamefont
  {King}}, \bibinfo {author} {\bibfnamefont {A.}~\bibnamefont {Baroni}},
  \bibinfo {author} {\bibfnamefont {V.}~\bibnamefont {Cirigliano}}, \bibinfo
  {author} {\bibfnamefont {S.}~\bibnamefont {Gandolfi}}, \bibinfo {author}
  {\bibfnamefont {L.}~\bibnamefont {Hayen}}, \bibinfo {author} {\bibfnamefont
  {E.}~\bibnamefont {Mereghetti}}, \bibinfo {author} {\bibfnamefont
  {S.}~\bibnamefont {Pastore}},\ and\ \bibinfo {author} {\bibfnamefont
  {M.}~\bibnamefont {Piarulli}},\ }\href {http://arxiv.org/abs/2207.11179} {\
  (\bibinfo {year} {2022})},\ \Eprint {https://arxiv.org/abs/2207.11179}
  {arXiv:2207.11179} \BibitemShut {NoStop}%
\bibitem [{\citenamefont {Glick-Magid}\ \emph {et~al.}(2022)\citenamefont
  {Glick-Magid}, \citenamefont {Forss{\'{e}}n}, \citenamefont {Gazda},
  \citenamefont {Gazit}, \citenamefont {Gysbers},\ and\ \citenamefont
  {Navr{\'{a}}til}}]{Glick-Magid2022}%
  \BibitemOpen
  \bibfield  {author} {\bibinfo {author} {\bibfnamefont {A.}~\bibnamefont
  {Glick-Magid}}, \bibinfo {author} {\bibfnamefont {C.}~\bibnamefont
  {Forss{\'{e}}n}}, \bibinfo {author} {\bibfnamefont {D.}~\bibnamefont
  {Gazda}}, \bibinfo {author} {\bibfnamefont {D.}~\bibnamefont {Gazit}},
  \bibinfo {author} {\bibfnamefont {P.}~\bibnamefont {Gysbers}},\ and\ \bibinfo
  {author} {\bibfnamefont {P.}~\bibnamefont {Navr{\'{a}}til}},\ }\href
  {https://doi.org/10.1016/j.physletb.2022.137259} {\bibfield  {journal}
  {\bibinfo  {journal} {Physics Letters B}\ }\textbf {\bibinfo {volume}
  {832}},\ \bibinfo {pages} {137259} (\bibinfo {year} {2022})},\ \Eprint
  {https://arxiv.org/abs/2107.10212} {arXiv:2107.10212} \BibitemShut {NoStop}%
\bibitem [{\citenamefont {Sargsyan}\ \emph {et~al.}(2022)\citenamefont
  {Sargsyan}, \citenamefont {Launey}, \citenamefont {Burkey}, \citenamefont
  {Gallant}, \citenamefont {Scielzo}, \citenamefont {Savard}, \citenamefont
  {Mercenne}, \citenamefont {Dytrych}, \citenamefont {Langr}, \citenamefont
  {Varriano}, \citenamefont {Longfellow}, \citenamefont {Hirsh},\ and\
  \citenamefont {Draayer}}]{Sargsyan2022}%
  \BibitemOpen
  \bibfield  {author} {\bibinfo {author} {\bibfnamefont {G.~H.}\ \bibnamefont
  {Sargsyan}}, \bibinfo {author} {\bibfnamefont {K.~D.}\ \bibnamefont
  {Launey}}, \bibinfo {author} {\bibfnamefont {M.~T.}\ \bibnamefont {Burkey}},
  \bibinfo {author} {\bibfnamefont {A.~T.}\ \bibnamefont {Gallant}}, \bibinfo
  {author} {\bibfnamefont {N.~D.}\ \bibnamefont {Scielzo}}, \bibinfo {author}
  {\bibfnamefont {G.}~\bibnamefont {Savard}}, \bibinfo {author} {\bibfnamefont
  {A.}~\bibnamefont {Mercenne}}, \bibinfo {author} {\bibfnamefont
  {T.}~\bibnamefont {Dytrych}}, \bibinfo {author} {\bibfnamefont
  {D.}~\bibnamefont {Langr}}, \bibinfo {author} {\bibfnamefont
  {L.}~\bibnamefont {Varriano}}, \bibinfo {author} {\bibfnamefont
  {B.}~\bibnamefont {Longfellow}}, \bibinfo {author} {\bibfnamefont {T.~Y.}\
  \bibnamefont {Hirsh}},\ and\ \bibinfo {author} {\bibfnamefont {J.~P.}\
  \bibnamefont {Draayer}},\ }\href
  {https://doi.org/10.1103/PhysRevLett.128.202503} {\bibfield  {journal}
  {\bibinfo  {journal} {Phys. Rev. Lett.}\ }\textbf {\bibinfo {volume} {128}},\
  \bibinfo {pages} {202503} (\bibinfo {year} {2022})}\BibitemShut {NoStop}%
\bibitem [{\citenamefont {Zelevinsky}\ \emph {et~al.}(2017)\citenamefont
  {Zelevinsky}, \citenamefont {Auerbach},\ and\ \citenamefont
  {Loc}}]{Zelevinsky2017}%
  \BibitemOpen
  \bibfield  {author} {\bibinfo {author} {\bibfnamefont {V.}~\bibnamefont
  {Zelevinsky}}, \bibinfo {author} {\bibfnamefont {N.}~\bibnamefont
  {Auerbach}},\ and\ \bibinfo {author} {\bibfnamefont {B.~M.}\ \bibnamefont
  {Loc}},\ }\href {https://doi.org/10.1103/PhysRevC.96.044319} {\bibfield
  {journal} {\bibinfo  {journal} {Phys. Rev. C}\ }\textbf {\bibinfo {volume}
  {96}},\ \bibinfo {pages} {044319} (\bibinfo {year} {2017})}\BibitemShut
  {NoStop}%
\bibitem [{\citenamefont {Falkowski}\ \emph {et~al.}(2021)\citenamefont
  {Falkowski}, \citenamefont {Gonzalez-Alonso},\ and\ \citenamefont
  {Naviliat-Cuncic}}]{Falkowski2021}%
  \BibitemOpen
  \bibfield  {author} {\bibinfo {author} {\bibfnamefont {A.}~\bibnamefont
  {Falkowski}}, \bibinfo {author} {\bibfnamefont {M.}~\bibnamefont
  {Gonzalez-Alonso}},\ and\ \bibinfo {author} {\bibfnamefont {O.}~\bibnamefont
  {Naviliat-Cuncic}},\ }\href {https://doi.org/10.1007/JHEP04(2021)126}
  {\bibfield  {journal} {\bibinfo  {journal} {J. High Energ. Phys.}\ }\textbf
  {\bibinfo {volume} {04}},\ \bibinfo {pages} {126}}\BibitemShut {NoStop}%
\bibitem [{\citenamefont {Fukuda}\ \emph {et~al.}(1998)\citenamefont {Fukuda},
  \citenamefont {Hayakawa}, \citenamefont {Ichihara}, \citenamefont {Inoue},
  \citenamefont {Ishihara}, \citenamefont {Ishino}, \citenamefont {Itow},
  \citenamefont {Kajita}, \citenamefont {Kameda}, \citenamefont {Kasuga},
  \citenamefont {Kobayashi}, \citenamefont {Kobayashi}, \citenamefont {Koshio},
  \citenamefont {Miura}, \citenamefont {Nakahata}, \citenamefont {Nakayama},
  \citenamefont {Okada}, \citenamefont {Okumura}, \citenamefont {Sakurai},
  \citenamefont {Shiozawa}, \citenamefont {Suzuki}, \citenamefont {Takeuchi},
  \citenamefont {Totsuka}, \citenamefont {Yamada}, \citenamefont {Earl},
  \citenamefont {Habig}, \citenamefont {Kearns}, \citenamefont {Messier},
  \citenamefont {Scholberg}, \citenamefont {Stone}, \citenamefont {Sulak},
  \citenamefont {Walter}, \citenamefont {Goldhaber}, \citenamefont
  {Barszczxak}, \citenamefont {Casper}, \citenamefont {Gajewski}, \citenamefont
  {Halverson}, \citenamefont {Hsu}, \citenamefont {Kropp}, \citenamefont
  {Price}, \citenamefont {Reines}, \citenamefont {Smy}, \citenamefont {Sobel},
  \citenamefont {Vagins}, \citenamefont {Ganezer}, \citenamefont {Keig},
  \citenamefont {Ellsworth}, \citenamefont {Tasaka}, \citenamefont {Flanagan},
  \citenamefont {Kibayashi}, \citenamefont {Learned}, \citenamefont {Matsuno},
  \citenamefont {Stenger}, \citenamefont {Takemori}, \citenamefont {Ishii},
  \citenamefont {Kanzaki}, \citenamefont {Kobayashi}, \citenamefont {Mine},
  \citenamefont {Nakamura}, \citenamefont {Nishikawa}, \citenamefont {Oyama},
  \citenamefont {Sakai}, \citenamefont {Sakuda}, \citenamefont {Sasaki},
  \citenamefont {Echigo}, \citenamefont {Kohama}, \citenamefont {Suzuki},
  \citenamefont {Haines}, \citenamefont {Blaufuss}, \citenamefont {Kim},
  \citenamefont {Sanford}, \citenamefont {Svoboda}, \citenamefont {Chen},
  \citenamefont {Conner}, \citenamefont {Goodman}, \citenamefont {Sullivan},
  \citenamefont {Hill}, \citenamefont {Jung}, \citenamefont {Martens},
  \citenamefont {Mauger}, \citenamefont {McGrew}, \citenamefont {Sharkey},
  \citenamefont {Viren}, \citenamefont {Yanagisawa}, \citenamefont {Doki},
  \citenamefont {Miyano}, \citenamefont {Okazawa}, \citenamefont {Saji},
  \citenamefont {Takahata}, \citenamefont {Nagashima}, \citenamefont {Takita},
  \citenamefont {Yamaguchi}, \citenamefont {Yoshida}, \citenamefont {Kim},
  \citenamefont {Etoh}, \citenamefont {Fujita}, \citenamefont {Hasegawa},
  \citenamefont {Hasegawa}, \citenamefont {Hatakeyama}, \citenamefont
  {Iwamoto}, \citenamefont {Koga}, \citenamefont {Maruyama}, \citenamefont
  {Ogawa}, \citenamefont {Shirai}, \citenamefont {Suzuki}, \citenamefont
  {Tsushima}, \citenamefont {Koshiba}, \citenamefont {Nemoto}, \citenamefont
  {Nishijima}, \citenamefont {Futagami}, \citenamefont {Hayato}, \citenamefont
  {Kanaya}, \citenamefont {Kaneyuki}, \citenamefont {Watanabe}, \citenamefont
  {Kielczewska}, \citenamefont {Doyle}, \citenamefont {George}, \citenamefont
  {Stachyra}, \citenamefont {Wai}, \citenamefont {Wilkes},\ and\ \citenamefont
  {Young}}]{Fuk98}%
  \BibitemOpen
  \bibfield  {author} {\bibinfo {author} {\bibfnamefont {Y.}~\bibnamefont
  {Fukuda}}, \bibinfo {author} {\bibfnamefont {T.}~\bibnamefont {Hayakawa}},
  \bibinfo {author} {\bibfnamefont {E.}~\bibnamefont {Ichihara}}, \bibinfo
  {author} {\bibfnamefont {K.}~\bibnamefont {Inoue}}, \bibinfo {author}
  {\bibfnamefont {K.}~\bibnamefont {Ishihara}}, \bibinfo {author}
  {\bibfnamefont {H.}~\bibnamefont {Ishino}}, \bibinfo {author} {\bibfnamefont
  {Y.}~\bibnamefont {Itow}}, \bibinfo {author} {\bibfnamefont {T.}~\bibnamefont
  {Kajita}}, \bibinfo {author} {\bibfnamefont {J.}~\bibnamefont {Kameda}},
  \bibinfo {author} {\bibfnamefont {S.}~\bibnamefont {Kasuga}}, \bibinfo
  {author} {\bibfnamefont {K.}~\bibnamefont {Kobayashi}}, \bibinfo {author}
  {\bibfnamefont {Y.}~\bibnamefont {Kobayashi}}, \bibinfo {author}
  {\bibfnamefont {Y.}~\bibnamefont {Koshio}}, \bibinfo {author} {\bibfnamefont
  {M.}~\bibnamefont {Miura}}, \bibinfo {author} {\bibfnamefont
  {M.}~\bibnamefont {Nakahata}}, \bibinfo {author} {\bibfnamefont
  {S.}~\bibnamefont {Nakayama}}, \bibinfo {author} {\bibfnamefont
  {A.}~\bibnamefont {Okada}}, \bibinfo {author} {\bibfnamefont
  {K.}~\bibnamefont {Okumura}}, \bibinfo {author} {\bibfnamefont
  {N.}~\bibnamefont {Sakurai}}, \bibinfo {author} {\bibfnamefont
  {M.}~\bibnamefont {Shiozawa}}, \bibinfo {author} {\bibfnamefont
  {Y.}~\bibnamefont {Suzuki}}, \bibinfo {author} {\bibfnamefont
  {Y.}~\bibnamefont {Takeuchi}}, \bibinfo {author} {\bibfnamefont
  {Y.}~\bibnamefont {Totsuka}}, \bibinfo {author} {\bibfnamefont
  {S.}~\bibnamefont {Yamada}}, \bibinfo {author} {\bibfnamefont
  {M.}~\bibnamefont {Earl}}, \bibinfo {author} {\bibfnamefont {A.}~\bibnamefont
  {Habig}}, \bibinfo {author} {\bibfnamefont {E.}~\bibnamefont {Kearns}},
  \bibinfo {author} {\bibfnamefont {M.~D.}\ \bibnamefont {Messier}}, \bibinfo
  {author} {\bibfnamefont {K.}~\bibnamefont {Scholberg}}, \bibinfo {author}
  {\bibfnamefont {J.~L.}\ \bibnamefont {Stone}}, \bibinfo {author}
  {\bibfnamefont {L.~R.}\ \bibnamefont {Sulak}}, \bibinfo {author}
  {\bibfnamefont {C.~W.}\ \bibnamefont {Walter}}, \bibinfo {author}
  {\bibfnamefont {M.}~\bibnamefont {Goldhaber}}, \bibinfo {author}
  {\bibfnamefont {T.}~\bibnamefont {Barszczxak}}, \bibinfo {author}
  {\bibfnamefont {D.}~\bibnamefont {Casper}}, \bibinfo {author} {\bibfnamefont
  {W.}~\bibnamefont {Gajewski}}, \bibinfo {author} {\bibfnamefont {P.~G.}\
  \bibnamefont {Halverson}}, \bibinfo {author} {\bibfnamefont {J.}~\bibnamefont
  {Hsu}}, \bibinfo {author} {\bibfnamefont {W.~R.}\ \bibnamefont {Kropp}},
  \bibinfo {author} {\bibfnamefont {L.~R.}\ \bibnamefont {Price}}, \bibinfo
  {author} {\bibfnamefont {F.}~\bibnamefont {Reines}}, \bibinfo {author}
  {\bibfnamefont {M.}~\bibnamefont {Smy}}, \bibinfo {author} {\bibfnamefont
  {H.~W.}\ \bibnamefont {Sobel}}, \bibinfo {author} {\bibfnamefont {M.~R.}\
  \bibnamefont {Vagins}}, \bibinfo {author} {\bibfnamefont {K.~S.}\
  \bibnamefont {Ganezer}}, \bibinfo {author} {\bibfnamefont {W.~E.}\
  \bibnamefont {Keig}}, \bibinfo {author} {\bibfnamefont {R.~W.}\ \bibnamefont
  {Ellsworth}}, \bibinfo {author} {\bibfnamefont {S.}~\bibnamefont {Tasaka}},
  \bibinfo {author} {\bibfnamefont {J.~W.}\ \bibnamefont {Flanagan}}, \bibinfo
  {author} {\bibfnamefont {A.}~\bibnamefont {Kibayashi}}, \bibinfo {author}
  {\bibfnamefont {J.~G.}\ \bibnamefont {Learned}}, \bibinfo {author}
  {\bibfnamefont {S.}~\bibnamefont {Matsuno}}, \bibinfo {author} {\bibfnamefont
  {V.~J.}\ \bibnamefont {Stenger}}, \bibinfo {author} {\bibfnamefont
  {D.}~\bibnamefont {Takemori}}, \bibinfo {author} {\bibfnamefont
  {T.}~\bibnamefont {Ishii}}, \bibinfo {author} {\bibfnamefont
  {J.}~\bibnamefont {Kanzaki}}, \bibinfo {author} {\bibfnamefont
  {T.}~\bibnamefont {Kobayashi}}, \bibinfo {author} {\bibfnamefont
  {S.}~\bibnamefont {Mine}}, \bibinfo {author} {\bibfnamefont {K.}~\bibnamefont
  {Nakamura}}, \bibinfo {author} {\bibfnamefont {K.}~\bibnamefont {Nishikawa}},
  \bibinfo {author} {\bibfnamefont {Y.}~\bibnamefont {Oyama}}, \bibinfo
  {author} {\bibfnamefont {A.}~\bibnamefont {Sakai}}, \bibinfo {author}
  {\bibfnamefont {M.}~\bibnamefont {Sakuda}}, \bibinfo {author} {\bibfnamefont
  {O.}~\bibnamefont {Sasaki}}, \bibinfo {author} {\bibfnamefont
  {S.}~\bibnamefont {Echigo}}, \bibinfo {author} {\bibfnamefont
  {M.}~\bibnamefont {Kohama}}, \bibinfo {author} {\bibfnamefont {A.~T.}\
  \bibnamefont {Suzuki}}, \bibinfo {author} {\bibfnamefont {T.~J.}\
  \bibnamefont {Haines}}, \bibinfo {author} {\bibfnamefont {E.}~\bibnamefont
  {Blaufuss}}, \bibinfo {author} {\bibfnamefont {B.~K.}\ \bibnamefont {Kim}},
  \bibinfo {author} {\bibfnamefont {R.}~\bibnamefont {Sanford}}, \bibinfo
  {author} {\bibfnamefont {R.}~\bibnamefont {Svoboda}}, \bibinfo {author}
  {\bibfnamefont {M.~L.}\ \bibnamefont {Chen}}, \bibinfo {author}
  {\bibfnamefont {Z.}~\bibnamefont {Conner}}, \bibinfo {author} {\bibfnamefont
  {J.~A.}\ \bibnamefont {Goodman}}, \bibinfo {author} {\bibfnamefont {G.~W.}\
  \bibnamefont {Sullivan}}, \bibinfo {author} {\bibfnamefont {J.}~\bibnamefont
  {Hill}}, \bibinfo {author} {\bibfnamefont {C.~K.}\ \bibnamefont {Jung}},
  \bibinfo {author} {\bibfnamefont {K.}~\bibnamefont {Martens}}, \bibinfo
  {author} {\bibfnamefont {C.}~\bibnamefont {Mauger}}, \bibinfo {author}
  {\bibfnamefont {C.}~\bibnamefont {McGrew}}, \bibinfo {author} {\bibfnamefont
  {E.}~\bibnamefont {Sharkey}}, \bibinfo {author} {\bibfnamefont
  {B.}~\bibnamefont {Viren}}, \bibinfo {author} {\bibfnamefont
  {C.}~\bibnamefont {Yanagisawa}}, \bibinfo {author} {\bibfnamefont
  {W.}~\bibnamefont {Doki}}, \bibinfo {author} {\bibfnamefont {K.}~\bibnamefont
  {Miyano}}, \bibinfo {author} {\bibfnamefont {H.}~\bibnamefont {Okazawa}},
  \bibinfo {author} {\bibfnamefont {C.}~\bibnamefont {Saji}}, \bibinfo {author}
  {\bibfnamefont {M.}~\bibnamefont {Takahata}}, \bibinfo {author}
  {\bibfnamefont {Y.}~\bibnamefont {Nagashima}}, \bibinfo {author}
  {\bibfnamefont {M.}~\bibnamefont {Takita}}, \bibinfo {author} {\bibfnamefont
  {T.}~\bibnamefont {Yamaguchi}}, \bibinfo {author} {\bibfnamefont
  {M.}~\bibnamefont {Yoshida}}, \bibinfo {author} {\bibfnamefont {S.~B.}\
  \bibnamefont {Kim}}, \bibinfo {author} {\bibfnamefont {M.}~\bibnamefont
  {Etoh}}, \bibinfo {author} {\bibfnamefont {K.}~\bibnamefont {Fujita}},
  \bibinfo {author} {\bibfnamefont {A.}~\bibnamefont {Hasegawa}}, \bibinfo
  {author} {\bibfnamefont {T.}~\bibnamefont {Hasegawa}}, \bibinfo {author}
  {\bibfnamefont {S.}~\bibnamefont {Hatakeyama}}, \bibinfo {author}
  {\bibfnamefont {T.}~\bibnamefont {Iwamoto}}, \bibinfo {author} {\bibfnamefont
  {M.}~\bibnamefont {Koga}}, \bibinfo {author} {\bibfnamefont {T.}~\bibnamefont
  {Maruyama}}, \bibinfo {author} {\bibfnamefont {H.}~\bibnamefont {Ogawa}},
  \bibinfo {author} {\bibfnamefont {J.}~\bibnamefont {Shirai}}, \bibinfo
  {author} {\bibfnamefont {A.}~\bibnamefont {Suzuki}}, \bibinfo {author}
  {\bibfnamefont {F.}~\bibnamefont {Tsushima}}, \bibinfo {author}
  {\bibfnamefont {M.}~\bibnamefont {Koshiba}}, \bibinfo {author} {\bibfnamefont
  {M.}~\bibnamefont {Nemoto}}, \bibinfo {author} {\bibfnamefont
  {K.}~\bibnamefont {Nishijima}}, \bibinfo {author} {\bibfnamefont
  {T.}~\bibnamefont {Futagami}}, \bibinfo {author} {\bibfnamefont
  {Y.}~\bibnamefont {Hayato}}, \bibinfo {author} {\bibfnamefont
  {Y.}~\bibnamefont {Kanaya}}, \bibinfo {author} {\bibfnamefont
  {K.}~\bibnamefont {Kaneyuki}}, \bibinfo {author} {\bibfnamefont
  {Y.}~\bibnamefont {Watanabe}}, \bibinfo {author} {\bibfnamefont
  {D.}~\bibnamefont {Kielczewska}}, \bibinfo {author} {\bibfnamefont {R.~A.}\
  \bibnamefont {Doyle}}, \bibinfo {author} {\bibfnamefont {J.~S.}\ \bibnamefont
  {George}}, \bibinfo {author} {\bibfnamefont {A.~L.}\ \bibnamefont
  {Stachyra}}, \bibinfo {author} {\bibfnamefont {L.~L.}\ \bibnamefont {Wai}},
  \bibinfo {author} {\bibfnamefont {R.~J.}\ \bibnamefont {Wilkes}},\ and\
  \bibinfo {author} {\bibfnamefont {K.~K.}\ \bibnamefont {Young}} (\bibinfo
  {collaboration} {Super-Kamiokande Collaboration}),\ }\href
  {https://doi.org/10.1103/PhysRevLett.81.1562} {\bibfield  {journal} {\bibinfo
   {journal} {Phys. Rev. Lett.}\ }\textbf {\bibinfo {volume} {81}},\ \bibinfo
  {pages} {1562} (\bibinfo {year} {1998})}\BibitemShut {NoStop}%
\bibitem [{\citenamefont {Ahmad}\ \emph {et~al.}(2001)\citenamefont {Ahmad},
  \citenamefont {Allen}, \citenamefont {Andersen}, \citenamefont {Anglin},
  \citenamefont {B\"uhler}, \citenamefont {Barton}, \citenamefont {Beier},
  \citenamefont {Bercovitch}, \citenamefont {Bigu}, \citenamefont {Biller},
  \citenamefont {Black}, \citenamefont {Blevis}, \citenamefont {Boardman},
  \citenamefont {Boger}, \citenamefont {Bonvin}, \citenamefont {Boulay},
  \citenamefont {Bowler}, \citenamefont {Bowles}, \citenamefont {Brice},
  \citenamefont {Browne}, \citenamefont {Bullard}, \citenamefont {Burritt},
  \citenamefont {Cameron}, \citenamefont {Cameron}, \citenamefont {Chan},
  \citenamefont {Chen}, \citenamefont {Chen}, \citenamefont {Chen},
  \citenamefont {Chon}, \citenamefont {Cleveland}, \citenamefont {Clifford},
  \citenamefont {Cowan}, \citenamefont {Cowen}, \citenamefont {Cox},
  \citenamefont {Dai}, \citenamefont {Dai}, \citenamefont {Dalnoki-Veress},
  \citenamefont {Davidson}, \citenamefont {Doe}, \citenamefont {Doucas},
  \citenamefont {Dragowsky}, \citenamefont {Duba}, \citenamefont {Duncan},
  \citenamefont {Dunmore}, \citenamefont {Earle}, \citenamefont {Elliott},
  \citenamefont {Evans}, \citenamefont {Ewan}, \citenamefont {Farine},
  \citenamefont {Fergani}, \citenamefont {Ferraris}, \citenamefont {Ford},
  \citenamefont {Fowler}, \citenamefont {Frame}, \citenamefont {Frank},
  \citenamefont {Frati}, \citenamefont {Germani}, \citenamefont {Gil},
  \citenamefont {Goldschmidt}, \citenamefont {Grant}, \citenamefont {Hahn},
  \citenamefont {Hallin}, \citenamefont {Hallman}, \citenamefont {Hamer},
  \citenamefont {Hamian}, \citenamefont {Haq}, \citenamefont {Hargrove},
  \citenamefont {Harvey}, \citenamefont {Hazama}, \citenamefont {Heaton},
  \citenamefont {Heeger}, \citenamefont {Heintzelman}, \citenamefont {Heise},
  \citenamefont {Helmer}, \citenamefont {Hepburn}, \citenamefont {Heron},
  \citenamefont {Hewett}, \citenamefont {Hime}, \citenamefont {Howe},
  \citenamefont {Hykawy}, \citenamefont {Isaac}, \citenamefont {Jagam},
  \citenamefont {Jelley}, \citenamefont {Jillings}, \citenamefont {Jonkmans},
  \citenamefont {Karn}, \citenamefont {Keener}, \citenamefont {Kirch},
  \citenamefont {Klein}, \citenamefont {Knox}, \citenamefont {Komar},
  \citenamefont {Kouzes}, \citenamefont {Kutter}, \citenamefont {Kyba},
  \citenamefont {Law}, \citenamefont {Lawson}, \citenamefont {Lay},
  \citenamefont {Lee}, \citenamefont {Lesko}, \citenamefont {Leslie},
  \citenamefont {Levine}, \citenamefont {Locke}, \citenamefont {Lowry},
  \citenamefont {Luoma}, \citenamefont {Lyon}, \citenamefont {Majerus},
  \citenamefont {Mak}, \citenamefont {Marino}, \citenamefont {McCauley},
  \citenamefont {McDonald}, \citenamefont {McDonald}, \citenamefont
  {McFarlane}, \citenamefont {McGregor}, \citenamefont {McLatchie},
  \citenamefont {Drees}, \citenamefont {Mes}, \citenamefont {Mifflin},
  \citenamefont {Miller}, \citenamefont {Milton}, \citenamefont {Moffat},
  \citenamefont {Moorhead}, \citenamefont {Nally}, \citenamefont {Neubauer},
  \citenamefont {Newcomer}, \citenamefont {Ng}, \citenamefont {Noble},
  \citenamefont {Norman}, \citenamefont {Novikov}, \citenamefont {O'Neill},
  \citenamefont {Okada}, \citenamefont {Ollerhead}, \citenamefont {Omori},
  \citenamefont {Orrell}, \citenamefont {Oser}, \citenamefont {Poon},
  \citenamefont {Radcliffe}, \citenamefont {Roberge}, \citenamefont
  {Robertson}, \citenamefont {Robertson}, \citenamefont {Rowley}, \citenamefont
  {Rusu}, \citenamefont {Saettler}, \citenamefont {Schaffer}, \citenamefont
  {Schuelke}, \citenamefont {Schwendener}, \citenamefont {Seifert},
  \citenamefont {Shatkay}, \citenamefont {Simpson}, \citenamefont {Sinclair},
  \citenamefont {Skensved}, \citenamefont {Smith}, \citenamefont {Smith},
  \citenamefont {Starinsky}, \citenamefont {Steiger}, \citenamefont {Stokstad},
  \citenamefont {Storey}, \citenamefont {Sur}, \citenamefont {Tafirout},
  \citenamefont {Tagg}, \citenamefont {Tanner}, \citenamefont {Taplin},
  \citenamefont {Thorman}, \citenamefont {Thornewell}, \citenamefont {Trent},
  \citenamefont {Tserkovnyak}, \citenamefont {Van~Berg}, \citenamefont {Van~de
  Water}, \citenamefont {Virtue}, \citenamefont {Waltham}, \citenamefont
  {Wang}, \citenamefont {Wark}, \citenamefont {West}, \citenamefont {Wilhelmy},
  \citenamefont {Wilkerson}, \citenamefont {Wilson}, \citenamefont {Wittich},
  \citenamefont {Wouters},\ and\ \citenamefont {Yeh}}]{Ahm01}%
  \BibitemOpen
  \bibfield  {author} {\bibinfo {author} {\bibfnamefont {Q.~R.}\ \bibnamefont
  {Ahmad}}, \bibinfo {author} {\bibfnamefont {R.~C.}\ \bibnamefont {Allen}},
  \bibinfo {author} {\bibfnamefont {T.~C.}\ \bibnamefont {Andersen}}, \bibinfo
  {author} {\bibfnamefont {J.~D.}\ \bibnamefont {Anglin}}, \bibinfo {author}
  {\bibfnamefont {G.}~\bibnamefont {B\"uhler}}, \bibinfo {author}
  {\bibfnamefont {J.~C.}\ \bibnamefont {Barton}}, \bibinfo {author}
  {\bibfnamefont {E.~W.}\ \bibnamefont {Beier}}, \bibinfo {author}
  {\bibfnamefont {M.}~\bibnamefont {Bercovitch}}, \bibinfo {author}
  {\bibfnamefont {J.}~\bibnamefont {Bigu}}, \bibinfo {author} {\bibfnamefont
  {S.}~\bibnamefont {Biller}}, \bibinfo {author} {\bibfnamefont {R.~A.}\
  \bibnamefont {Black}}, \bibinfo {author} {\bibfnamefont {I.}~\bibnamefont
  {Blevis}}, \bibinfo {author} {\bibfnamefont {R.~J.}\ \bibnamefont
  {Boardman}}, \bibinfo {author} {\bibfnamefont {J.}~\bibnamefont {Boger}},
  \bibinfo {author} {\bibfnamefont {E.}~\bibnamefont {Bonvin}}, \bibinfo
  {author} {\bibfnamefont {M.~G.}\ \bibnamefont {Boulay}}, \bibinfo {author}
  {\bibfnamefont {M.~G.}\ \bibnamefont {Bowler}}, \bibinfo {author}
  {\bibfnamefont {T.~J.}\ \bibnamefont {Bowles}}, \bibinfo {author}
  {\bibfnamefont {S.~J.}\ \bibnamefont {Brice}}, \bibinfo {author}
  {\bibfnamefont {M.~C.}\ \bibnamefont {Browne}}, \bibinfo {author}
  {\bibfnamefont {T.~V.}\ \bibnamefont {Bullard}}, \bibinfo {author}
  {\bibfnamefont {T.~H.}\ \bibnamefont {Burritt}}, \bibinfo {author}
  {\bibfnamefont {K.}~\bibnamefont {Cameron}}, \bibinfo {author} {\bibfnamefont
  {J.}~\bibnamefont {Cameron}}, \bibinfo {author} {\bibfnamefont {Y.~D.}\
  \bibnamefont {Chan}}, \bibinfo {author} {\bibfnamefont {M.}~\bibnamefont
  {Chen}}, \bibinfo {author} {\bibfnamefont {H.~H.}\ \bibnamefont {Chen}},
  \bibinfo {author} {\bibfnamefont {X.}~\bibnamefont {Chen}}, \bibinfo {author}
  {\bibfnamefont {M.~C.}\ \bibnamefont {Chon}}, \bibinfo {author}
  {\bibfnamefont {B.~T.}\ \bibnamefont {Cleveland}}, \bibinfo {author}
  {\bibfnamefont {E.~T.~H.}\ \bibnamefont {Clifford}}, \bibinfo {author}
  {\bibfnamefont {J.~H.~M.}\ \bibnamefont {Cowan}}, \bibinfo {author}
  {\bibfnamefont {D.~F.}\ \bibnamefont {Cowen}}, \bibinfo {author}
  {\bibfnamefont {G.~A.}\ \bibnamefont {Cox}}, \bibinfo {author} {\bibfnamefont
  {Y.}~\bibnamefont {Dai}}, \bibinfo {author} {\bibfnamefont {X.}~\bibnamefont
  {Dai}}, \bibinfo {author} {\bibfnamefont {F.}~\bibnamefont {Dalnoki-Veress}},
  \bibinfo {author} {\bibfnamefont {W.~F.}\ \bibnamefont {Davidson}}, \bibinfo
  {author} {\bibfnamefont {P.~J.}\ \bibnamefont {Doe}}, \bibinfo {author}
  {\bibfnamefont {G.}~\bibnamefont {Doucas}}, \bibinfo {author} {\bibfnamefont
  {M.~R.}\ \bibnamefont {Dragowsky}}, \bibinfo {author} {\bibfnamefont {C.~A.}\
  \bibnamefont {Duba}}, \bibinfo {author} {\bibfnamefont {F.~A.}\ \bibnamefont
  {Duncan}}, \bibinfo {author} {\bibfnamefont {J.}~\bibnamefont {Dunmore}},
  \bibinfo {author} {\bibfnamefont {E.~D.}\ \bibnamefont {Earle}}, \bibinfo
  {author} {\bibfnamefont {S.~R.}\ \bibnamefont {Elliott}}, \bibinfo {author}
  {\bibfnamefont {H.~C.}\ \bibnamefont {Evans}}, \bibinfo {author}
  {\bibfnamefont {G.~T.}\ \bibnamefont {Ewan}}, \bibinfo {author}
  {\bibfnamefont {J.}~\bibnamefont {Farine}}, \bibinfo {author} {\bibfnamefont
  {H.}~\bibnamefont {Fergani}}, \bibinfo {author} {\bibfnamefont {A.~P.}\
  \bibnamefont {Ferraris}}, \bibinfo {author} {\bibfnamefont {R.~J.}\
  \bibnamefont {Ford}}, \bibinfo {author} {\bibfnamefont {M.~M.}\ \bibnamefont
  {Fowler}}, \bibinfo {author} {\bibfnamefont {K.}~\bibnamefont {Frame}},
  \bibinfo {author} {\bibfnamefont {E.~D.}\ \bibnamefont {Frank}}, \bibinfo
  {author} {\bibfnamefont {W.}~\bibnamefont {Frati}}, \bibinfo {author}
  {\bibfnamefont {J.~V.}\ \bibnamefont {Germani}}, \bibinfo {author}
  {\bibfnamefont {S.}~\bibnamefont {Gil}}, \bibinfo {author} {\bibfnamefont
  {A.}~\bibnamefont {Goldschmidt}}, \bibinfo {author} {\bibfnamefont {D.~R.}\
  \bibnamefont {Grant}}, \bibinfo {author} {\bibfnamefont {R.~L.}\ \bibnamefont
  {Hahn}}, \bibinfo {author} {\bibfnamefont {A.~L.}\ \bibnamefont {Hallin}},
  \bibinfo {author} {\bibfnamefont {E.~D.}\ \bibnamefont {Hallman}}, \bibinfo
  {author} {\bibfnamefont {A.}~\bibnamefont {Hamer}}, \bibinfo {author}
  {\bibfnamefont {A.~A.}\ \bibnamefont {Hamian}}, \bibinfo {author}
  {\bibfnamefont {R.~U.}\ \bibnamefont {Haq}}, \bibinfo {author} {\bibfnamefont
  {C.~K.}\ \bibnamefont {Hargrove}}, \bibinfo {author} {\bibfnamefont {P.~J.}\
  \bibnamefont {Harvey}}, \bibinfo {author} {\bibfnamefont {R.}~\bibnamefont
  {Hazama}}, \bibinfo {author} {\bibfnamefont {R.}~\bibnamefont {Heaton}},
  \bibinfo {author} {\bibfnamefont {K.~M.}\ \bibnamefont {Heeger}}, \bibinfo
  {author} {\bibfnamefont {W.~J.}\ \bibnamefont {Heintzelman}}, \bibinfo
  {author} {\bibfnamefont {J.}~\bibnamefont {Heise}}, \bibinfo {author}
  {\bibfnamefont {R.~L.}\ \bibnamefont {Helmer}}, \bibinfo {author}
  {\bibfnamefont {J.~D.}\ \bibnamefont {Hepburn}}, \bibinfo {author}
  {\bibfnamefont {H.}~\bibnamefont {Heron}}, \bibinfo {author} {\bibfnamefont
  {J.}~\bibnamefont {Hewett}}, \bibinfo {author} {\bibfnamefont
  {A.}~\bibnamefont {Hime}}, \bibinfo {author} {\bibfnamefont {M.}~\bibnamefont
  {Howe}}, \bibinfo {author} {\bibfnamefont {J.~G.}\ \bibnamefont {Hykawy}},
  \bibinfo {author} {\bibfnamefont {M.~C.~P.}\ \bibnamefont {Isaac}}, \bibinfo
  {author} {\bibfnamefont {P.}~\bibnamefont {Jagam}}, \bibinfo {author}
  {\bibfnamefont {N.~A.}\ \bibnamefont {Jelley}}, \bibinfo {author}
  {\bibfnamefont {C.}~\bibnamefont {Jillings}}, \bibinfo {author}
  {\bibfnamefont {G.}~\bibnamefont {Jonkmans}}, \bibinfo {author}
  {\bibfnamefont {J.}~\bibnamefont {Karn}}, \bibinfo {author} {\bibfnamefont
  {P.~T.}\ \bibnamefont {Keener}}, \bibinfo {author} {\bibfnamefont
  {K.}~\bibnamefont {Kirch}}, \bibinfo {author} {\bibfnamefont {J.~R.}\
  \bibnamefont {Klein}}, \bibinfo {author} {\bibfnamefont {A.~B.}\ \bibnamefont
  {Knox}}, \bibinfo {author} {\bibfnamefont {R.~J.}\ \bibnamefont {Komar}},
  \bibinfo {author} {\bibfnamefont {R.}~\bibnamefont {Kouzes}}, \bibinfo
  {author} {\bibfnamefont {T.}~\bibnamefont {Kutter}}, \bibinfo {author}
  {\bibfnamefont {C.~C.~M.}\ \bibnamefont {Kyba}}, \bibinfo {author}
  {\bibfnamefont {J.}~\bibnamefont {Law}}, \bibinfo {author} {\bibfnamefont
  {I.~T.}\ \bibnamefont {Lawson}}, \bibinfo {author} {\bibfnamefont
  {M.}~\bibnamefont {Lay}}, \bibinfo {author} {\bibfnamefont {H.~W.}\
  \bibnamefont {Lee}}, \bibinfo {author} {\bibfnamefont {K.~T.}\ \bibnamefont
  {Lesko}}, \bibinfo {author} {\bibfnamefont {J.~R.}\ \bibnamefont {Leslie}},
  \bibinfo {author} {\bibfnamefont {I.}~\bibnamefont {Levine}}, \bibinfo
  {author} {\bibfnamefont {W.}~\bibnamefont {Locke}}, \bibinfo {author}
  {\bibfnamefont {M.~M.}\ \bibnamefont {Lowry}}, \bibinfo {author}
  {\bibfnamefont {S.}~\bibnamefont {Luoma}}, \bibinfo {author} {\bibfnamefont
  {J.}~\bibnamefont {Lyon}}, \bibinfo {author} {\bibfnamefont {S.}~\bibnamefont
  {Majerus}}, \bibinfo {author} {\bibfnamefont {H.~B.}\ \bibnamefont {Mak}},
  \bibinfo {author} {\bibfnamefont {A.~D.}\ \bibnamefont {Marino}}, \bibinfo
  {author} {\bibfnamefont {N.}~\bibnamefont {McCauley}}, \bibinfo {author}
  {\bibfnamefont {A.~B.}\ \bibnamefont {McDonald}}, \bibinfo {author}
  {\bibfnamefont {D.~S.}\ \bibnamefont {McDonald}}, \bibinfo {author}
  {\bibfnamefont {K.}~\bibnamefont {McFarlane}}, \bibinfo {author}
  {\bibfnamefont {G.}~\bibnamefont {McGregor}}, \bibinfo {author}
  {\bibfnamefont {W.}~\bibnamefont {McLatchie}}, \bibinfo {author}
  {\bibfnamefont {R.~M.}\ \bibnamefont {Drees}}, \bibinfo {author}
  {\bibfnamefont {H.}~\bibnamefont {Mes}}, \bibinfo {author} {\bibfnamefont
  {C.}~\bibnamefont {Mifflin}}, \bibinfo {author} {\bibfnamefont {G.~G.}\
  \bibnamefont {Miller}}, \bibinfo {author} {\bibfnamefont {G.}~\bibnamefont
  {Milton}}, \bibinfo {author} {\bibfnamefont {B.~A.}\ \bibnamefont {Moffat}},
  \bibinfo {author} {\bibfnamefont {M.}~\bibnamefont {Moorhead}}, \bibinfo
  {author} {\bibfnamefont {C.~W.}\ \bibnamefont {Nally}}, \bibinfo {author}
  {\bibfnamefont {M.~S.}\ \bibnamefont {Neubauer}}, \bibinfo {author}
  {\bibfnamefont {F.~M.}\ \bibnamefont {Newcomer}}, \bibinfo {author}
  {\bibfnamefont {H.~S.}\ \bibnamefont {Ng}}, \bibinfo {author} {\bibfnamefont
  {A.~J.}\ \bibnamefont {Noble}}, \bibinfo {author} {\bibfnamefont {E.~B.}\
  \bibnamefont {Norman}}, \bibinfo {author} {\bibfnamefont {V.~M.}\
  \bibnamefont {Novikov}}, \bibinfo {author} {\bibfnamefont {M.}~\bibnamefont
  {O'Neill}}, \bibinfo {author} {\bibfnamefont {C.~E.}\ \bibnamefont {Okada}},
  \bibinfo {author} {\bibfnamefont {R.~W.}\ \bibnamefont {Ollerhead}}, \bibinfo
  {author} {\bibfnamefont {M.}~\bibnamefont {Omori}}, \bibinfo {author}
  {\bibfnamefont {J.~L.}\ \bibnamefont {Orrell}}, \bibinfo {author}
  {\bibfnamefont {S.~M.}\ \bibnamefont {Oser}}, \bibinfo {author}
  {\bibfnamefont {A.~W.~P.}\ \bibnamefont {Poon}}, \bibinfo {author}
  {\bibfnamefont {T.~J.}\ \bibnamefont {Radcliffe}}, \bibinfo {author}
  {\bibfnamefont {A.}~\bibnamefont {Roberge}}, \bibinfo {author} {\bibfnamefont
  {B.~C.}\ \bibnamefont {Robertson}}, \bibinfo {author} {\bibfnamefont
  {R.~G.~H.}\ \bibnamefont {Robertson}}, \bibinfo {author} {\bibfnamefont
  {J.~K.}\ \bibnamefont {Rowley}}, \bibinfo {author} {\bibfnamefont {V.~L.}\
  \bibnamefont {Rusu}}, \bibinfo {author} {\bibfnamefont {E.}~\bibnamefont
  {Saettler}}, \bibinfo {author} {\bibfnamefont {K.~K.}\ \bibnamefont
  {Schaffer}}, \bibinfo {author} {\bibfnamefont {A.}~\bibnamefont {Schuelke}},
  \bibinfo {author} {\bibfnamefont {M.~H.}\ \bibnamefont {Schwendener}},
  \bibinfo {author} {\bibfnamefont {H.}~\bibnamefont {Seifert}}, \bibinfo
  {author} {\bibfnamefont {M.}~\bibnamefont {Shatkay}}, \bibinfo {author}
  {\bibfnamefont {J.~J.}\ \bibnamefont {Simpson}}, \bibinfo {author}
  {\bibfnamefont {D.}~\bibnamefont {Sinclair}}, \bibinfo {author}
  {\bibfnamefont {P.}~\bibnamefont {Skensved}}, \bibinfo {author}
  {\bibfnamefont {A.~R.}\ \bibnamefont {Smith}}, \bibinfo {author}
  {\bibfnamefont {M.~W.~E.}\ \bibnamefont {Smith}}, \bibinfo {author}
  {\bibfnamefont {N.}~\bibnamefont {Starinsky}}, \bibinfo {author}
  {\bibfnamefont {T.~D.}\ \bibnamefont {Steiger}}, \bibinfo {author}
  {\bibfnamefont {R.~G.}\ \bibnamefont {Stokstad}}, \bibinfo {author}
  {\bibfnamefont {R.~S.}\ \bibnamefont {Storey}}, \bibinfo {author}
  {\bibfnamefont {B.}~\bibnamefont {Sur}}, \bibinfo {author} {\bibfnamefont
  {R.}~\bibnamefont {Tafirout}}, \bibinfo {author} {\bibfnamefont
  {N.}~\bibnamefont {Tagg}}, \bibinfo {author} {\bibfnamefont {N.~W.}\
  \bibnamefont {Tanner}}, \bibinfo {author} {\bibfnamefont {R.~K.}\
  \bibnamefont {Taplin}}, \bibinfo {author} {\bibfnamefont {M.}~\bibnamefont
  {Thorman}}, \bibinfo {author} {\bibfnamefont {P.}~\bibnamefont {Thornewell}},
  \bibinfo {author} {\bibfnamefont {P.~T.}\ \bibnamefont {Trent}}, \bibinfo
  {author} {\bibfnamefont {Y.~I.}\ \bibnamefont {Tserkovnyak}}, \bibinfo
  {author} {\bibfnamefont {R.}~\bibnamefont {Van~Berg}}, \bibinfo {author}
  {\bibfnamefont {R.~G.}\ \bibnamefont {Van~de Water}}, \bibinfo {author}
  {\bibfnamefont {C.~J.}\ \bibnamefont {Virtue}}, \bibinfo {author}
  {\bibfnamefont {C.~E.}\ \bibnamefont {Waltham}}, \bibinfo {author}
  {\bibfnamefont {J.-X.}\ \bibnamefont {Wang}}, \bibinfo {author}
  {\bibfnamefont {D.~L.}\ \bibnamefont {Wark}}, \bibinfo {author}
  {\bibfnamefont {N.}~\bibnamefont {West}}, \bibinfo {author} {\bibfnamefont
  {J.~B.}\ \bibnamefont {Wilhelmy}}, \bibinfo {author} {\bibfnamefont {J.~F.}\
  \bibnamefont {Wilkerson}}, \bibinfo {author} {\bibfnamefont {J.}~\bibnamefont
  {Wilson}}, \bibinfo {author} {\bibfnamefont {P.}~\bibnamefont {Wittich}},
  \bibinfo {author} {\bibfnamefont {J.~M.}\ \bibnamefont {Wouters}},\ and\
  \bibinfo {author} {\bibfnamefont {M.}~\bibnamefont {Yeh}} (\bibinfo
  {collaboration} {SNO Collaboration}),\ }\href
  {https://doi.org/10.1103/PhysRevLett.87.071301} {\bibfield  {journal}
  {\bibinfo  {journal} {Phys. Rev. Lett.}\ }\textbf {\bibinfo {volume} {87}},\
  \bibinfo {pages} {071301} (\bibinfo {year} {2001})}\BibitemShut {NoStop}%
\bibitem [{\citenamefont {Abi}\ \emph {et~al.}(2021)\citenamefont {Abi},
  \citenamefont {Albahri}, \citenamefont {Al-Kilani}, \citenamefont {Allspach},
  \citenamefont {Alonzi}, \citenamefont {Anastasi}, \citenamefont {Anisenkov},
  \citenamefont {Azfar}, \citenamefont {Badgley}, \citenamefont {Bae\ss{}ler},
  \citenamefont {Bailey}, \citenamefont {Baranov}, \citenamefont
  {Barlas-Yucel}, \citenamefont {Barrett}, \citenamefont {Barzi}, \citenamefont
  {Basti}, \citenamefont {Bedeschi}, \citenamefont {Behnke}, \citenamefont
  {Berz}, \citenamefont {Bhattacharya}, \citenamefont {Binney}, \citenamefont
  {Bjorkquist}, \citenamefont {Bloom}, \citenamefont {Bono}, \citenamefont
  {Bottalico}, \citenamefont {Bowcock}, \citenamefont {Boyden}, \citenamefont
  {Cantatore}, \citenamefont {Carey}, \citenamefont {Carroll}, \citenamefont
  {Casey}, \citenamefont {Cauz}, \citenamefont {Ceravolo}, \citenamefont
  {Chakraborty}, \citenamefont {Chang}, \citenamefont {Chapelain},
  \citenamefont {Chappa}, \citenamefont {Charity}, \citenamefont {Chislett},
  \citenamefont {Choi}, \citenamefont {Chu}, \citenamefont {Chupp},
  \citenamefont {Convery}, \citenamefont {Conway}, \citenamefont {Corradi},
  \citenamefont {Corrodi}, \citenamefont {Cotrozzi}, \citenamefont {Crnkovic},
  \citenamefont {Dabagov}, \citenamefont {De~Lurgio}, \citenamefont {Debevec},
  \citenamefont {Di~Falco}, \citenamefont {Di~Meo}, \citenamefont
  {Di~Sciascio}, \citenamefont {Di~Stefano}, \citenamefont {Drendel},
  \citenamefont {Driutti}, \citenamefont {Duginov}, \citenamefont {Eads},
  \citenamefont {Eggert}, \citenamefont {Epps}, \citenamefont {Esquivel},
  \citenamefont {Farooq}, \citenamefont {Fatemi}, \citenamefont {Ferrari},
  \citenamefont {Fertl}, \citenamefont {Fiedler}, \citenamefont {Fienberg},
  \citenamefont {Fioretti}, \citenamefont {Flay}, \citenamefont {Foster},
  \citenamefont {Friedsam}, \citenamefont {Frle\ifmmode~\check{z}\else
  \v{z}\fi{}}, \citenamefont {Froemming}, \citenamefont {Fry}, \citenamefont
  {Fu}, \citenamefont {Gabbanini}, \citenamefont {Galati}, \citenamefont
  {Ganguly}, \citenamefont {Garcia}, \citenamefont {Gastler}, \citenamefont
  {George}, \citenamefont {Gibbons}, \citenamefont {Gioiosa}, \citenamefont
  {Giovanetti}, \citenamefont {Girotti}, \citenamefont {Gohn}, \citenamefont
  {Gorringe}, \citenamefont {Grange}, \citenamefont {Grant}, \citenamefont
  {Gray}, \citenamefont {Haciomeroglu}, \citenamefont {Hahn}, \citenamefont
  {Halewood-Leagas}, \citenamefont {Hampai}, \citenamefont {Han}, \citenamefont
  {Hazen}, \citenamefont {Hempstead}, \citenamefont {Henry}, \citenamefont
  {Herrod}, \citenamefont {Hertzog}, \citenamefont {Hesketh}, \citenamefont
  {Hibbert}, \citenamefont {Hodge}, \citenamefont {Holzbauer}, \citenamefont
  {Hong}, \citenamefont {Hong}, \citenamefont {Iacovacci}, \citenamefont
  {Incagli}, \citenamefont {Johnstone}, \citenamefont {Johnstone},
  \citenamefont {Kammel}, \citenamefont {Kargiantoulakis}, \citenamefont
  {Karuza}, \citenamefont {Kaspar}, \citenamefont {Kawall}, \citenamefont
  {Kelton}, \citenamefont {Keshavarzi}, \citenamefont {Kessler}, \citenamefont
  {Khaw}, \citenamefont {Khechadoorian}, \citenamefont {Khomutov},
  \citenamefont {Kiburg}, \citenamefont {Kiburg}, \citenamefont {Kim},
  \citenamefont {Kim}, \citenamefont {Kim}, \citenamefont {King}, \citenamefont
  {Kinnaird}, \citenamefont {Korostelev}, \citenamefont {Kourbanis},
  \citenamefont {Kraegeloh}, \citenamefont {Krylov}, \citenamefont
  {Kuchibhotla}, \citenamefont {Kuchinskiy}, \citenamefont {Labe},
  \citenamefont {LaBounty}, \citenamefont {Lancaster}, \citenamefont {Lee},
  \citenamefont {Lee}, \citenamefont {Leo}, \citenamefont {Li}, \citenamefont
  {Li}, \citenamefont {Li}, \citenamefont {Logashenko}, \citenamefont
  {Lorente~Campos}, \citenamefont {Luc\`a}, \citenamefont {Lukicov},
  \citenamefont {Luo}, \citenamefont {Lusiani}, \citenamefont {Lyon},
  \citenamefont {MacCoy}, \citenamefont {Madrak}, \citenamefont {Makino},
  \citenamefont {Marignetti}, \citenamefont {Mastroianni}, \citenamefont
  {Maxfield}, \citenamefont {McEvoy}, \citenamefont {Merritt}, \citenamefont
  {Mikhailichenko}, \citenamefont {Miller}, \citenamefont {Miozzi},
  \citenamefont {Morgan}, \citenamefont {Morse}, \citenamefont {Mott},
  \citenamefont {Motuk}, \citenamefont {Nath}, \citenamefont {Newton},
  \citenamefont {Nguyen}, \citenamefont {Oberling}, \citenamefont {Osofsky},
  \citenamefont {Ostiguy}, \citenamefont {Park}, \citenamefont {Pauletta},
  \citenamefont {Piacentino}, \citenamefont {Pilato}, \citenamefont {Pitts},
  \citenamefont {Plaster}, \citenamefont {Po\ifmmode \check{c}\else
  \v{c}\fi{}ani\ifmmode~\acute{c}\else \'{c}\fi{}}, \citenamefont {Pohlman},
  \citenamefont {Polly}, \citenamefont {Popovic}, \citenamefont {Price},
  \citenamefont {Quinn}, \citenamefont {Raha}, \citenamefont {Ramachandran},
  \citenamefont {Ramberg}, \citenamefont {Rider}, \citenamefont {Ritchie},
  \citenamefont {Roberts}, \citenamefont {Rubin}, \citenamefont {Santi},
  \citenamefont {Sathyan}, \citenamefont {Schellman}, \citenamefont
  {Schlesier}, \citenamefont {Schreckenberger}, \citenamefont {Semertzidis},
  \citenamefont {Shatunov}, \citenamefont {Shemyakin}, \citenamefont {Shenk},
  \citenamefont {Sim}, \citenamefont {Smith}, \citenamefont {Smith},
  \citenamefont {Soha}, \citenamefont {Sorbara}, \citenamefont {St\"ockinger},
  \citenamefont {Stapleton}, \citenamefont {Still}, \citenamefont {Stoughton},
  \citenamefont {Stratakis}, \citenamefont {Strohman}, \citenamefont
  {Stuttard}, \citenamefont {Swanson}, \citenamefont {Sweetmore}, \citenamefont
  {Sweigart}, \citenamefont {Syphers}, \citenamefont {Tarazona}, \citenamefont
  {Teubner}, \citenamefont {Tewsley-Booth}, \citenamefont {Thomson},
  \citenamefont {Tishchenko}, \citenamefont {Tran}, \citenamefont {Turner},
  \citenamefont {Valetov}, \citenamefont {Vasilkova}, \citenamefont
  {Venanzoni}, \citenamefont {Volnykh}, \citenamefont {Walton}, \citenamefont
  {Warren}, \citenamefont {Weisskopf}, \citenamefont {Welty-Rieger},
  \citenamefont {Whitley}, \citenamefont {Winter}, \citenamefont {Wolski},
  \citenamefont {Wormald}, \citenamefont {Wu},\ and\ \citenamefont
  {Yoshikawa}}]{PhysRevLett.126.141801}%
  \BibitemOpen
  \bibfield  {author} {\bibinfo {author} {\bibfnamefont {B.}~\bibnamefont
  {Abi}}, \bibinfo {author} {\bibfnamefont {T.}~\bibnamefont {Albahri}},
  \bibinfo {author} {\bibfnamefont {S.}~\bibnamefont {Al-Kilani}}, \bibinfo
  {author} {\bibfnamefont {D.}~\bibnamefont {Allspach}}, \bibinfo {author}
  {\bibfnamefont {L.~P.}\ \bibnamefont {Alonzi}}, \bibinfo {author}
  {\bibfnamefont {A.}~\bibnamefont {Anastasi}}, \bibinfo {author}
  {\bibfnamefont {A.}~\bibnamefont {Anisenkov}}, \bibinfo {author}
  {\bibfnamefont {F.}~\bibnamefont {Azfar}}, \bibinfo {author} {\bibfnamefont
  {K.}~\bibnamefont {Badgley}}, \bibinfo {author} {\bibfnamefont
  {S.}~\bibnamefont {Bae\ss{}ler}}, \bibinfo {author} {\bibfnamefont
  {I.}~\bibnamefont {Bailey}}, \bibinfo {author} {\bibfnamefont {V.~A.}\
  \bibnamefont {Baranov}}, \bibinfo {author} {\bibfnamefont {E.}~\bibnamefont
  {Barlas-Yucel}}, \bibinfo {author} {\bibfnamefont {T.}~\bibnamefont
  {Barrett}}, \bibinfo {author} {\bibfnamefont {E.}~\bibnamefont {Barzi}},
  \bibinfo {author} {\bibfnamefont {A.}~\bibnamefont {Basti}}, \bibinfo
  {author} {\bibfnamefont {F.}~\bibnamefont {Bedeschi}}, \bibinfo {author}
  {\bibfnamefont {A.}~\bibnamefont {Behnke}}, \bibinfo {author} {\bibfnamefont
  {M.}~\bibnamefont {Berz}}, \bibinfo {author} {\bibfnamefont {M.}~\bibnamefont
  {Bhattacharya}}, \bibinfo {author} {\bibfnamefont {H.~P.}\ \bibnamefont
  {Binney}}, \bibinfo {author} {\bibfnamefont {R.}~\bibnamefont {Bjorkquist}},
  \bibinfo {author} {\bibfnamefont {P.}~\bibnamefont {Bloom}}, \bibinfo
  {author} {\bibfnamefont {J.}~\bibnamefont {Bono}}, \bibinfo {author}
  {\bibfnamefont {E.}~\bibnamefont {Bottalico}}, \bibinfo {author}
  {\bibfnamefont {T.}~\bibnamefont {Bowcock}}, \bibinfo {author} {\bibfnamefont
  {D.}~\bibnamefont {Boyden}}, \bibinfo {author} {\bibfnamefont
  {G.}~\bibnamefont {Cantatore}}, \bibinfo {author} {\bibfnamefont {R.~M.}\
  \bibnamefont {Carey}}, \bibinfo {author} {\bibfnamefont {J.}~\bibnamefont
  {Carroll}}, \bibinfo {author} {\bibfnamefont {B.~C.~K.}\ \bibnamefont
  {Casey}}, \bibinfo {author} {\bibfnamefont {D.}~\bibnamefont {Cauz}},
  \bibinfo {author} {\bibfnamefont {S.}~\bibnamefont {Ceravolo}}, \bibinfo
  {author} {\bibfnamefont {R.}~\bibnamefont {Chakraborty}}, \bibinfo {author}
  {\bibfnamefont {S.~P.}\ \bibnamefont {Chang}}, \bibinfo {author}
  {\bibfnamefont {A.}~\bibnamefont {Chapelain}}, \bibinfo {author}
  {\bibfnamefont {S.}~\bibnamefont {Chappa}}, \bibinfo {author} {\bibfnamefont
  {S.}~\bibnamefont {Charity}}, \bibinfo {author} {\bibfnamefont
  {R.}~\bibnamefont {Chislett}}, \bibinfo {author} {\bibfnamefont
  {J.}~\bibnamefont {Choi}}, \bibinfo {author} {\bibfnamefont {Z.}~\bibnamefont
  {Chu}}, \bibinfo {author} {\bibfnamefont {T.~E.}\ \bibnamefont {Chupp}},
  \bibinfo {author} {\bibfnamefont {M.~E.}\ \bibnamefont {Convery}}, \bibinfo
  {author} {\bibfnamefont {A.}~\bibnamefont {Conway}}, \bibinfo {author}
  {\bibfnamefont {G.}~\bibnamefont {Corradi}}, \bibinfo {author} {\bibfnamefont
  {S.}~\bibnamefont {Corrodi}}, \bibinfo {author} {\bibfnamefont
  {L.}~\bibnamefont {Cotrozzi}}, \bibinfo {author} {\bibfnamefont {J.~D.}\
  \bibnamefont {Crnkovic}}, \bibinfo {author} {\bibfnamefont {S.}~\bibnamefont
  {Dabagov}}, \bibinfo {author} {\bibfnamefont {P.~M.}\ \bibnamefont
  {De~Lurgio}}, \bibinfo {author} {\bibfnamefont {P.~T.}\ \bibnamefont
  {Debevec}}, \bibinfo {author} {\bibfnamefont {S.}~\bibnamefont {Di~Falco}},
  \bibinfo {author} {\bibfnamefont {P.}~\bibnamefont {Di~Meo}}, \bibinfo
  {author} {\bibfnamefont {G.}~\bibnamefont {Di~Sciascio}}, \bibinfo {author}
  {\bibfnamefont {R.}~\bibnamefont {Di~Stefano}}, \bibinfo {author}
  {\bibfnamefont {B.}~\bibnamefont {Drendel}}, \bibinfo {author} {\bibfnamefont
  {A.}~\bibnamefont {Driutti}}, \bibinfo {author} {\bibfnamefont {V.~N.}\
  \bibnamefont {Duginov}}, \bibinfo {author} {\bibfnamefont {M.}~\bibnamefont
  {Eads}}, \bibinfo {author} {\bibfnamefont {N.}~\bibnamefont {Eggert}},
  \bibinfo {author} {\bibfnamefont {A.}~\bibnamefont {Epps}}, \bibinfo {author}
  {\bibfnamefont {J.}~\bibnamefont {Esquivel}}, \bibinfo {author}
  {\bibfnamefont {M.}~\bibnamefont {Farooq}}, \bibinfo {author} {\bibfnamefont
  {R.}~\bibnamefont {Fatemi}}, \bibinfo {author} {\bibfnamefont
  {C.}~\bibnamefont {Ferrari}}, \bibinfo {author} {\bibfnamefont
  {M.}~\bibnamefont {Fertl}}, \bibinfo {author} {\bibfnamefont
  {A.}~\bibnamefont {Fiedler}}, \bibinfo {author} {\bibfnamefont {A.~T.}\
  \bibnamefont {Fienberg}}, \bibinfo {author} {\bibfnamefont {A.}~\bibnamefont
  {Fioretti}}, \bibinfo {author} {\bibfnamefont {D.}~\bibnamefont {Flay}},
  \bibinfo {author} {\bibfnamefont {S.~B.}\ \bibnamefont {Foster}}, \bibinfo
  {author} {\bibfnamefont {H.}~\bibnamefont {Friedsam}}, \bibinfo {author}
  {\bibfnamefont {E.}~\bibnamefont {Frle\ifmmode~\check{z}\else \v{z}\fi{}}},
  \bibinfo {author} {\bibfnamefont {N.~S.}\ \bibnamefont {Froemming}}, \bibinfo
  {author} {\bibfnamefont {J.}~\bibnamefont {Fry}}, \bibinfo {author}
  {\bibfnamefont {C.}~\bibnamefont {Fu}}, \bibinfo {author} {\bibfnamefont
  {C.}~\bibnamefont {Gabbanini}}, \bibinfo {author} {\bibfnamefont {M.~D.}\
  \bibnamefont {Galati}}, \bibinfo {author} {\bibfnamefont {S.}~\bibnamefont
  {Ganguly}}, \bibinfo {author} {\bibfnamefont {A.}~\bibnamefont {Garcia}},
  \bibinfo {author} {\bibfnamefont {D.~E.}\ \bibnamefont {Gastler}}, \bibinfo
  {author} {\bibfnamefont {J.}~\bibnamefont {George}}, \bibinfo {author}
  {\bibfnamefont {L.~K.}\ \bibnamefont {Gibbons}}, \bibinfo {author}
  {\bibfnamefont {A.}~\bibnamefont {Gioiosa}}, \bibinfo {author} {\bibfnamefont
  {K.~L.}\ \bibnamefont {Giovanetti}}, \bibinfo {author} {\bibfnamefont
  {P.}~\bibnamefont {Girotti}}, \bibinfo {author} {\bibfnamefont
  {W.}~\bibnamefont {Gohn}}, \bibinfo {author} {\bibfnamefont {T.}~\bibnamefont
  {Gorringe}}, \bibinfo {author} {\bibfnamefont {J.}~\bibnamefont {Grange}},
  \bibinfo {author} {\bibfnamefont {S.}~\bibnamefont {Grant}}, \bibinfo
  {author} {\bibfnamefont {F.}~\bibnamefont {Gray}}, \bibinfo {author}
  {\bibfnamefont {S.}~\bibnamefont {Haciomeroglu}}, \bibinfo {author}
  {\bibfnamefont {D.}~\bibnamefont {Hahn}}, \bibinfo {author} {\bibfnamefont
  {T.}~\bibnamefont {Halewood-Leagas}}, \bibinfo {author} {\bibfnamefont
  {D.}~\bibnamefont {Hampai}}, \bibinfo {author} {\bibfnamefont
  {F.}~\bibnamefont {Han}}, \bibinfo {author} {\bibfnamefont {E.}~\bibnamefont
  {Hazen}}, \bibinfo {author} {\bibfnamefont {J.}~\bibnamefont {Hempstead}},
  \bibinfo {author} {\bibfnamefont {S.}~\bibnamefont {Henry}}, \bibinfo
  {author} {\bibfnamefont {A.~T.}\ \bibnamefont {Herrod}}, \bibinfo {author}
  {\bibfnamefont {D.~W.}\ \bibnamefont {Hertzog}}, \bibinfo {author}
  {\bibfnamefont {G.}~\bibnamefont {Hesketh}}, \bibinfo {author} {\bibfnamefont
  {A.}~\bibnamefont {Hibbert}}, \bibinfo {author} {\bibfnamefont
  {Z.}~\bibnamefont {Hodge}}, \bibinfo {author} {\bibfnamefont {J.~L.}\
  \bibnamefont {Holzbauer}}, \bibinfo {author} {\bibfnamefont {K.~W.}\
  \bibnamefont {Hong}}, \bibinfo {author} {\bibfnamefont {R.}~\bibnamefont
  {Hong}}, \bibinfo {author} {\bibfnamefont {M.}~\bibnamefont {Iacovacci}},
  \bibinfo {author} {\bibfnamefont {M.}~\bibnamefont {Incagli}}, \bibinfo
  {author} {\bibfnamefont {C.}~\bibnamefont {Johnstone}}, \bibinfo {author}
  {\bibfnamefont {J.~A.}\ \bibnamefont {Johnstone}}, \bibinfo {author}
  {\bibfnamefont {P.}~\bibnamefont {Kammel}}, \bibinfo {author} {\bibfnamefont
  {M.}~\bibnamefont {Kargiantoulakis}}, \bibinfo {author} {\bibfnamefont
  {M.}~\bibnamefont {Karuza}}, \bibinfo {author} {\bibfnamefont
  {J.}~\bibnamefont {Kaspar}}, \bibinfo {author} {\bibfnamefont
  {D.}~\bibnamefont {Kawall}}, \bibinfo {author} {\bibfnamefont
  {L.}~\bibnamefont {Kelton}}, \bibinfo {author} {\bibfnamefont
  {A.}~\bibnamefont {Keshavarzi}}, \bibinfo {author} {\bibfnamefont
  {D.}~\bibnamefont {Kessler}}, \bibinfo {author} {\bibfnamefont {K.~S.}\
  \bibnamefont {Khaw}}, \bibinfo {author} {\bibfnamefont {Z.}~\bibnamefont
  {Khechadoorian}}, \bibinfo {author} {\bibfnamefont {N.~V.}\ \bibnamefont
  {Khomutov}}, \bibinfo {author} {\bibfnamefont {B.}~\bibnamefont {Kiburg}},
  \bibinfo {author} {\bibfnamefont {M.}~\bibnamefont {Kiburg}}, \bibinfo
  {author} {\bibfnamefont {O.}~\bibnamefont {Kim}}, \bibinfo {author}
  {\bibfnamefont {S.~C.}\ \bibnamefont {Kim}}, \bibinfo {author} {\bibfnamefont
  {Y.~I.}\ \bibnamefont {Kim}}, \bibinfo {author} {\bibfnamefont
  {B.}~\bibnamefont {King}}, \bibinfo {author} {\bibfnamefont {N.}~\bibnamefont
  {Kinnaird}}, \bibinfo {author} {\bibfnamefont {M.}~\bibnamefont
  {Korostelev}}, \bibinfo {author} {\bibfnamefont {I.}~\bibnamefont
  {Kourbanis}}, \bibinfo {author} {\bibfnamefont {E.}~\bibnamefont
  {Kraegeloh}}, \bibinfo {author} {\bibfnamefont {V.~A.}\ \bibnamefont
  {Krylov}}, \bibinfo {author} {\bibfnamefont {A.}~\bibnamefont {Kuchibhotla}},
  \bibinfo {author} {\bibfnamefont {N.~A.}\ \bibnamefont {Kuchinskiy}},
  \bibinfo {author} {\bibfnamefont {K.~R.}\ \bibnamefont {Labe}}, \bibinfo
  {author} {\bibfnamefont {J.}~\bibnamefont {LaBounty}}, \bibinfo {author}
  {\bibfnamefont {M.}~\bibnamefont {Lancaster}}, \bibinfo {author}
  {\bibfnamefont {M.~J.}\ \bibnamefont {Lee}}, \bibinfo {author} {\bibfnamefont
  {S.}~\bibnamefont {Lee}}, \bibinfo {author} {\bibfnamefont {S.}~\bibnamefont
  {Leo}}, \bibinfo {author} {\bibfnamefont {B.}~\bibnamefont {Li}}, \bibinfo
  {author} {\bibfnamefont {D.}~\bibnamefont {Li}}, \bibinfo {author}
  {\bibfnamefont {L.}~\bibnamefont {Li}}, \bibinfo {author} {\bibfnamefont
  {I.}~\bibnamefont {Logashenko}}, \bibinfo {author} {\bibfnamefont
  {A.}~\bibnamefont {Lorente~Campos}}, \bibinfo {author} {\bibfnamefont
  {A.}~\bibnamefont {Luc\`a}}, \bibinfo {author} {\bibfnamefont
  {G.}~\bibnamefont {Lukicov}}, \bibinfo {author} {\bibfnamefont
  {G.}~\bibnamefont {Luo}}, \bibinfo {author} {\bibfnamefont {A.}~\bibnamefont
  {Lusiani}}, \bibinfo {author} {\bibfnamefont {A.~L.}\ \bibnamefont {Lyon}},
  \bibinfo {author} {\bibfnamefont {B.}~\bibnamefont {MacCoy}}, \bibinfo
  {author} {\bibfnamefont {R.}~\bibnamefont {Madrak}}, \bibinfo {author}
  {\bibfnamefont {K.}~\bibnamefont {Makino}}, \bibinfo {author} {\bibfnamefont
  {F.}~\bibnamefont {Marignetti}}, \bibinfo {author} {\bibfnamefont
  {S.}~\bibnamefont {Mastroianni}}, \bibinfo {author} {\bibfnamefont
  {S.}~\bibnamefont {Maxfield}}, \bibinfo {author} {\bibfnamefont
  {M.}~\bibnamefont {McEvoy}}, \bibinfo {author} {\bibfnamefont
  {W.}~\bibnamefont {Merritt}}, \bibinfo {author} {\bibfnamefont {A.~A.}\
  \bibnamefont {Mikhailichenko}}, \bibinfo {author} {\bibfnamefont {J.~P.}\
  \bibnamefont {Miller}}, \bibinfo {author} {\bibfnamefont {S.}~\bibnamefont
  {Miozzi}}, \bibinfo {author} {\bibfnamefont {J.~P.}\ \bibnamefont {Morgan}},
  \bibinfo {author} {\bibfnamefont {W.~M.}\ \bibnamefont {Morse}}, \bibinfo
  {author} {\bibfnamefont {J.}~\bibnamefont {Mott}}, \bibinfo {author}
  {\bibfnamefont {E.}~\bibnamefont {Motuk}}, \bibinfo {author} {\bibfnamefont
  {A.}~\bibnamefont {Nath}}, \bibinfo {author} {\bibfnamefont {D.}~\bibnamefont
  {Newton}}, \bibinfo {author} {\bibfnamefont {H.}~\bibnamefont {Nguyen}},
  \bibinfo {author} {\bibfnamefont {M.}~\bibnamefont {Oberling}}, \bibinfo
  {author} {\bibfnamefont {R.}~\bibnamefont {Osofsky}}, \bibinfo {author}
  {\bibfnamefont {J.-F.}\ \bibnamefont {Ostiguy}}, \bibinfo {author}
  {\bibfnamefont {S.}~\bibnamefont {Park}}, \bibinfo {author} {\bibfnamefont
  {G.}~\bibnamefont {Pauletta}}, \bibinfo {author} {\bibfnamefont {G.~M.}\
  \bibnamefont {Piacentino}}, \bibinfo {author} {\bibfnamefont {R.~N.}\
  \bibnamefont {Pilato}}, \bibinfo {author} {\bibfnamefont {K.~T.}\
  \bibnamefont {Pitts}}, \bibinfo {author} {\bibfnamefont {B.}~\bibnamefont
  {Plaster}}, \bibinfo {author} {\bibfnamefont {D.}~\bibnamefont {Po\ifmmode
  \check{c}\else \v{c}\fi{}ani\ifmmode~\acute{c}\else \'{c}\fi{}}}, \bibinfo
  {author} {\bibfnamefont {N.}~\bibnamefont {Pohlman}}, \bibinfo {author}
  {\bibfnamefont {C.~C.}\ \bibnamefont {Polly}}, \bibinfo {author}
  {\bibfnamefont {M.}~\bibnamefont {Popovic}}, \bibinfo {author} {\bibfnamefont
  {J.}~\bibnamefont {Price}}, \bibinfo {author} {\bibfnamefont
  {B.}~\bibnamefont {Quinn}}, \bibinfo {author} {\bibfnamefont
  {N.}~\bibnamefont {Raha}}, \bibinfo {author} {\bibfnamefont {S.}~\bibnamefont
  {Ramachandran}}, \bibinfo {author} {\bibfnamefont {E.}~\bibnamefont
  {Ramberg}}, \bibinfo {author} {\bibfnamefont {N.~T.}\ \bibnamefont {Rider}},
  \bibinfo {author} {\bibfnamefont {J.~L.}\ \bibnamefont {Ritchie}}, \bibinfo
  {author} {\bibfnamefont {B.~L.}\ \bibnamefont {Roberts}}, \bibinfo {author}
  {\bibfnamefont {D.~L.}\ \bibnamefont {Rubin}}, \bibinfo {author}
  {\bibfnamefont {L.}~\bibnamefont {Santi}}, \bibinfo {author} {\bibfnamefont
  {D.}~\bibnamefont {Sathyan}}, \bibinfo {author} {\bibfnamefont
  {H.}~\bibnamefont {Schellman}}, \bibinfo {author} {\bibfnamefont
  {C.}~\bibnamefont {Schlesier}}, \bibinfo {author} {\bibfnamefont
  {A.}~\bibnamefont {Schreckenberger}}, \bibinfo {author} {\bibfnamefont
  {Y.~K.}\ \bibnamefont {Semertzidis}}, \bibinfo {author} {\bibfnamefont
  {Y.~M.}\ \bibnamefont {Shatunov}}, \bibinfo {author} {\bibfnamefont
  {D.}~\bibnamefont {Shemyakin}}, \bibinfo {author} {\bibfnamefont
  {M.}~\bibnamefont {Shenk}}, \bibinfo {author} {\bibfnamefont
  {D.}~\bibnamefont {Sim}}, \bibinfo {author} {\bibfnamefont {M.~W.}\
  \bibnamefont {Smith}}, \bibinfo {author} {\bibfnamefont {A.}~\bibnamefont
  {Smith}}, \bibinfo {author} {\bibfnamefont {A.~K.}\ \bibnamefont {Soha}},
  \bibinfo {author} {\bibfnamefont {M.}~\bibnamefont {Sorbara}}, \bibinfo
  {author} {\bibfnamefont {D.}~\bibnamefont {St\"ockinger}}, \bibinfo {author}
  {\bibfnamefont {J.}~\bibnamefont {Stapleton}}, \bibinfo {author}
  {\bibfnamefont {D.}~\bibnamefont {Still}}, \bibinfo {author} {\bibfnamefont
  {C.}~\bibnamefont {Stoughton}}, \bibinfo {author} {\bibfnamefont
  {D.}~\bibnamefont {Stratakis}}, \bibinfo {author} {\bibfnamefont
  {C.}~\bibnamefont {Strohman}}, \bibinfo {author} {\bibfnamefont
  {T.}~\bibnamefont {Stuttard}}, \bibinfo {author} {\bibfnamefont {H.~E.}\
  \bibnamefont {Swanson}}, \bibinfo {author} {\bibfnamefont {G.}~\bibnamefont
  {Sweetmore}}, \bibinfo {author} {\bibfnamefont {D.~A.}\ \bibnamefont
  {Sweigart}}, \bibinfo {author} {\bibfnamefont {M.~J.}\ \bibnamefont
  {Syphers}}, \bibinfo {author} {\bibfnamefont {D.~A.}\ \bibnamefont
  {Tarazona}}, \bibinfo {author} {\bibfnamefont {T.}~\bibnamefont {Teubner}},
  \bibinfo {author} {\bibfnamefont {A.~E.}\ \bibnamefont {Tewsley-Booth}},
  \bibinfo {author} {\bibfnamefont {K.}~\bibnamefont {Thomson}}, \bibinfo
  {author} {\bibfnamefont {V.}~\bibnamefont {Tishchenko}}, \bibinfo {author}
  {\bibfnamefont {N.~H.}\ \bibnamefont {Tran}}, \bibinfo {author}
  {\bibfnamefont {W.}~\bibnamefont {Turner}}, \bibinfo {author} {\bibfnamefont
  {E.}~\bibnamefont {Valetov}}, \bibinfo {author} {\bibfnamefont
  {D.}~\bibnamefont {Vasilkova}}, \bibinfo {author} {\bibfnamefont
  {G.}~\bibnamefont {Venanzoni}}, \bibinfo {author} {\bibfnamefont {V.~P.}\
  \bibnamefont {Volnykh}}, \bibinfo {author} {\bibfnamefont {T.}~\bibnamefont
  {Walton}}, \bibinfo {author} {\bibfnamefont {M.}~\bibnamefont {Warren}},
  \bibinfo {author} {\bibfnamefont {A.}~\bibnamefont {Weisskopf}}, \bibinfo
  {author} {\bibfnamefont {L.}~\bibnamefont {Welty-Rieger}}, \bibinfo {author}
  {\bibfnamefont {M.}~\bibnamefont {Whitley}}, \bibinfo {author} {\bibfnamefont
  {P.}~\bibnamefont {Winter}}, \bibinfo {author} {\bibfnamefont
  {A.}~\bibnamefont {Wolski}}, \bibinfo {author} {\bibfnamefont
  {M.}~\bibnamefont {Wormald}}, \bibinfo {author} {\bibfnamefont
  {W.}~\bibnamefont {Wu}},\ and\ \bibinfo {author} {\bibfnamefont
  {C.}~\bibnamefont {Yoshikawa}} (\bibinfo {collaboration} {Muon
  $g\ensuremath{-}2$ Collaboration}),\ }\href
  {https://doi.org/10.1103/PhysRevLett.126.141801} {\bibfield  {journal}
  {\bibinfo  {journal} {Phys. Rev. Lett.}\ }\textbf {\bibinfo {volume} {126}},\
  \bibinfo {pages} {141801} (\bibinfo {year} {2021})}\BibitemShut {NoStop}%
\bibitem [{\citenamefont {Aker}\ \emph {et~al.}(2022)\citenamefont {Aker},
  \citenamefont {Beglarian}, \citenamefont {Behrens}, \citenamefont {Berlev},
  \citenamefont {Besserer}, \citenamefont {Bieringer}, \citenamefont {Block},
  \citenamefont {Bobien}, \citenamefont {B{\"o}ttcher}, \citenamefont
  {Bornschein}, \citenamefont {Bornschein}, \citenamefont {Brunst},
  \citenamefont {Caldwell}, \citenamefont {Carney}, \citenamefont {La~Cascio},
  \citenamefont {Chilingaryan}, \citenamefont {Choi}, \citenamefont {Debowski},
  \citenamefont {Deffert}, \citenamefont {Descher}, \citenamefont
  {D{\'i}az~Barrero}, \citenamefont {Doe}, \citenamefont {Dragoun},
  \citenamefont {Drexlin}, \citenamefont {Eitel}, \citenamefont {Ellinger},
  \citenamefont {Engel}, \citenamefont {Enomoto}, \citenamefont {Felden},
  \citenamefont {Formaggio}, \citenamefont {Fr{\"a}nkle}, \citenamefont
  {Franklin}, \citenamefont {Friedel}, \citenamefont {Fulst}, \citenamefont
  {Gauda}, \citenamefont {Gil}, \citenamefont {Gl{\"u}ck}, \citenamefont
  {Gr{\"o}ssle}, \citenamefont {Gumbsheimer}, \citenamefont {Gupta},
  \citenamefont {H{\"o}hn}, \citenamefont {Hannen}, \citenamefont
  {Hau{\ss}mann}, \citenamefont {Helbing}, \citenamefont {Hickford},
  \citenamefont {Hiller}, \citenamefont {Hillesheimer}, \citenamefont {Hinz},
  \citenamefont {Houdy}, \citenamefont {Huber}, \citenamefont {Jansen},
  \citenamefont {Karl}, \citenamefont {Kellerer}, \citenamefont {Kellerer},
  \citenamefont {Kleifges}, \citenamefont {Klein}, \citenamefont {K{\"o}hler},
  \citenamefont {K{\"o}llenberger}, \citenamefont {Kopmann}, \citenamefont
  {Korzeczek}, \citenamefont {Koval{\'i}k}, \citenamefont {Krasch},
  \citenamefont {Krause}, \citenamefont {Kunka}, \citenamefont {Lasserre},
  \citenamefont {Le}, \citenamefont {Lebeda}, \citenamefont {Lehnert},
  \citenamefont {Lokhov}, \citenamefont {Machatschek}, \citenamefont
  {Malcherek}, \citenamefont {Mark}, \citenamefont {Marsteller}, \citenamefont
  {Martin}, \citenamefont {Melzer}, \citenamefont {Menshikov}, \citenamefont
  {Mertens}, \citenamefont {Mostafa}, \citenamefont {M{\"u}ller}, \citenamefont
  {Neumann}, \citenamefont {Niemes}, \citenamefont {Oelpmann}, \citenamefont
  {Parno}, \citenamefont {Poon}, \citenamefont {Poyato}, \citenamefont
  {Priester}, \citenamefont {Ramachandran}, \citenamefont {Robertson},
  \citenamefont {Rodejohann}, \citenamefont {R{\"o}llig}, \citenamefont
  {R{\"o}ttele}, \citenamefont {Rodenbeck}, \citenamefont {Ry{\v{s}}av{\'y}},
  \citenamefont {Sack}, \citenamefont {Saenz}, \citenamefont {Sch{\"a}fer},
  \citenamefont {Schaller~n{\'e}e Pollithy}, \citenamefont {Schimpf},
  \citenamefont {Schl{\"o}sser}, \citenamefont {Schl{\"o}sser}, \citenamefont
  {Schl{\"u}ter}, \citenamefont {Schneidewind}, \citenamefont {Schrank},
  \citenamefont {Schulz}, \citenamefont {Schwemmer}, \citenamefont
  {{\v{S}}ef{\v{c}}{\'i}k}, \citenamefont {Sibille}, \citenamefont {Siegmann},
  \citenamefont {Slez{\'a}k}, \citenamefont {Spanier}, \citenamefont {Steidl},
  \citenamefont {Sturm}, \citenamefont {Sun}, \citenamefont {Tcherniakhovski},
  \citenamefont {Telle}, \citenamefont {Thorne}, \citenamefont {Th{\"u}mmler},
  \citenamefont {Titov}, \citenamefont {Tkachev}, \citenamefont {Urban},
  \citenamefont {Valerius}, \citenamefont {V{\'e}nos}, \citenamefont
  {Vizcaya~Hern{\'a}ndez}, \citenamefont {Weinheimer}, \citenamefont {Welte},
  \citenamefont {Wendel}, \citenamefont {Wilkerson}, \citenamefont {Wolf},
  \citenamefont {W{\"u}stling}, \citenamefont {Wydra}, \citenamefont {Xu},
  \citenamefont {Yen}, \citenamefont {Zadoroghny}, \citenamefont {Zeller},\
  and\ \citenamefont {Collaboration}}]{Aker2022}%
  \BibitemOpen
  \bibfield  {author} {\bibinfo {author} {\bibfnamefont {M.}~\bibnamefont
  {Aker}}, \bibinfo {author} {\bibfnamefont {A.}~\bibnamefont {Beglarian}},
  \bibinfo {author} {\bibfnamefont {J.}~\bibnamefont {Behrens}}, \bibinfo
  {author} {\bibfnamefont {A.}~\bibnamefont {Berlev}}, \bibinfo {author}
  {\bibfnamefont {U.}~\bibnamefont {Besserer}}, \bibinfo {author}
  {\bibfnamefont {B.}~\bibnamefont {Bieringer}}, \bibinfo {author}
  {\bibfnamefont {F.}~\bibnamefont {Block}}, \bibinfo {author} {\bibfnamefont
  {S.}~\bibnamefont {Bobien}}, \bibinfo {author} {\bibfnamefont
  {M.}~\bibnamefont {B{\"o}ttcher}}, \bibinfo {author} {\bibfnamefont
  {B.}~\bibnamefont {Bornschein}}, \bibinfo {author} {\bibfnamefont
  {L.}~\bibnamefont {Bornschein}}, \bibinfo {author} {\bibfnamefont
  {T.}~\bibnamefont {Brunst}}, \bibinfo {author} {\bibfnamefont {T.~S.}\
  \bibnamefont {Caldwell}}, \bibinfo {author} {\bibfnamefont {R.~M.~D.}\
  \bibnamefont {Carney}}, \bibinfo {author} {\bibfnamefont {L.}~\bibnamefont
  {La~Cascio}}, \bibinfo {author} {\bibfnamefont {S.}~\bibnamefont
  {Chilingaryan}}, \bibinfo {author} {\bibfnamefont {W.}~\bibnamefont {Choi}},
  \bibinfo {author} {\bibfnamefont {K.}~\bibnamefont {Debowski}}, \bibinfo
  {author} {\bibfnamefont {M.}~\bibnamefont {Deffert}}, \bibinfo {author}
  {\bibfnamefont {M.}~\bibnamefont {Descher}}, \bibinfo {author} {\bibfnamefont
  {D.}~\bibnamefont {D{\'i}az~Barrero}}, \bibinfo {author} {\bibfnamefont
  {P.~J.}\ \bibnamefont {Doe}}, \bibinfo {author} {\bibfnamefont
  {O.}~\bibnamefont {Dragoun}}, \bibinfo {author} {\bibfnamefont
  {G.}~\bibnamefont {Drexlin}}, \bibinfo {author} {\bibfnamefont
  {K.}~\bibnamefont {Eitel}}, \bibinfo {author} {\bibfnamefont
  {E.}~\bibnamefont {Ellinger}}, \bibinfo {author} {\bibfnamefont
  {R.}~\bibnamefont {Engel}}, \bibinfo {author} {\bibfnamefont
  {S.}~\bibnamefont {Enomoto}}, \bibinfo {author} {\bibfnamefont
  {A.}~\bibnamefont {Felden}}, \bibinfo {author} {\bibfnamefont {J.~A.}\
  \bibnamefont {Formaggio}}, \bibinfo {author} {\bibfnamefont {F.~M.}\
  \bibnamefont {Fr{\"a}nkle}}, \bibinfo {author} {\bibfnamefont {G.~B.}\
  \bibnamefont {Franklin}}, \bibinfo {author} {\bibfnamefont {F.}~\bibnamefont
  {Friedel}}, \bibinfo {author} {\bibfnamefont {A.}~\bibnamefont {Fulst}},
  \bibinfo {author} {\bibfnamefont {K.}~\bibnamefont {Gauda}}, \bibinfo
  {author} {\bibfnamefont {W.}~\bibnamefont {Gil}}, \bibinfo {author}
  {\bibfnamefont {F.}~\bibnamefont {Gl{\"u}ck}}, \bibinfo {author}
  {\bibfnamefont {R.}~\bibnamefont {Gr{\"o}ssle}}, \bibinfo {author}
  {\bibfnamefont {R.}~\bibnamefont {Gumbsheimer}}, \bibinfo {author}
  {\bibfnamefont {V.}~\bibnamefont {Gupta}}, \bibinfo {author} {\bibfnamefont
  {T.}~\bibnamefont {H{\"o}hn}}, \bibinfo {author} {\bibfnamefont
  {V.}~\bibnamefont {Hannen}}, \bibinfo {author} {\bibfnamefont
  {N.}~\bibnamefont {Hau{\ss}mann}}, \bibinfo {author} {\bibfnamefont
  {K.}~\bibnamefont {Helbing}}, \bibinfo {author} {\bibfnamefont
  {S.}~\bibnamefont {Hickford}}, \bibinfo {author} {\bibfnamefont
  {R.}~\bibnamefont {Hiller}}, \bibinfo {author} {\bibfnamefont
  {D.}~\bibnamefont {Hillesheimer}}, \bibinfo {author} {\bibfnamefont
  {D.}~\bibnamefont {Hinz}}, \bibinfo {author} {\bibfnamefont {T.}~\bibnamefont
  {Houdy}}, \bibinfo {author} {\bibfnamefont {A.}~\bibnamefont {Huber}},
  \bibinfo {author} {\bibfnamefont {A.}~\bibnamefont {Jansen}}, \bibinfo
  {author} {\bibfnamefont {C.}~\bibnamefont {Karl}}, \bibinfo {author}
  {\bibfnamefont {F.}~\bibnamefont {Kellerer}}, \bibinfo {author}
  {\bibfnamefont {J.}~\bibnamefont {Kellerer}}, \bibinfo {author}
  {\bibfnamefont {M.}~\bibnamefont {Kleifges}}, \bibinfo {author}
  {\bibfnamefont {M.}~\bibnamefont {Klein}}, \bibinfo {author} {\bibfnamefont
  {C.}~\bibnamefont {K{\"o}hler}}, \bibinfo {author} {\bibfnamefont
  {L.}~\bibnamefont {K{\"o}llenberger}}, \bibinfo {author} {\bibfnamefont
  {A.}~\bibnamefont {Kopmann}}, \bibinfo {author} {\bibfnamefont
  {M.}~\bibnamefont {Korzeczek}}, \bibinfo {author} {\bibfnamefont
  {A.}~\bibnamefont {Koval{\'i}k}}, \bibinfo {author} {\bibfnamefont
  {B.}~\bibnamefont {Krasch}}, \bibinfo {author} {\bibfnamefont
  {H.}~\bibnamefont {Krause}}, \bibinfo {author} {\bibfnamefont
  {N.}~\bibnamefont {Kunka}}, \bibinfo {author} {\bibfnamefont
  {T.}~\bibnamefont {Lasserre}}, \bibinfo {author} {\bibfnamefont {T.~L.}\
  \bibnamefont {Le}}, \bibinfo {author} {\bibfnamefont {O.}~\bibnamefont
  {Lebeda}}, \bibinfo {author} {\bibfnamefont {B.}~\bibnamefont {Lehnert}},
  \bibinfo {author} {\bibfnamefont {A.}~\bibnamefont {Lokhov}}, \bibinfo
  {author} {\bibfnamefont {M.}~\bibnamefont {Machatschek}}, \bibinfo {author}
  {\bibfnamefont {E.}~\bibnamefont {Malcherek}}, \bibinfo {author}
  {\bibfnamefont {M.}~\bibnamefont {Mark}}, \bibinfo {author} {\bibfnamefont
  {A.}~\bibnamefont {Marsteller}}, \bibinfo {author} {\bibfnamefont {E.~L.}\
  \bibnamefont {Martin}}, \bibinfo {author} {\bibfnamefont {C.}~\bibnamefont
  {Melzer}}, \bibinfo {author} {\bibfnamefont {A.}~\bibnamefont {Menshikov}},
  \bibinfo {author} {\bibfnamefont {S.}~\bibnamefont {Mertens}}, \bibinfo
  {author} {\bibfnamefont {J.}~\bibnamefont {Mostafa}}, \bibinfo {author}
  {\bibfnamefont {K.}~\bibnamefont {M{\"u}ller}}, \bibinfo {author}
  {\bibfnamefont {H.}~\bibnamefont {Neumann}}, \bibinfo {author} {\bibfnamefont
  {S.}~\bibnamefont {Niemes}}, \bibinfo {author} {\bibfnamefont
  {P.}~\bibnamefont {Oelpmann}}, \bibinfo {author} {\bibfnamefont {D.~S.}\
  \bibnamefont {Parno}}, \bibinfo {author} {\bibfnamefont {A.~W.~P.}\
  \bibnamefont {Poon}}, \bibinfo {author} {\bibfnamefont {J.~M.~L.}\
  \bibnamefont {Poyato}}, \bibinfo {author} {\bibfnamefont {F.}~\bibnamefont
  {Priester}}, \bibinfo {author} {\bibfnamefont {S.}~\bibnamefont
  {Ramachandran}}, \bibinfo {author} {\bibfnamefont {R.~G.~H.}\ \bibnamefont
  {Robertson}}, \bibinfo {author} {\bibfnamefont {W.}~\bibnamefont
  {Rodejohann}}, \bibinfo {author} {\bibfnamefont {M.}~\bibnamefont
  {R{\"o}llig}}, \bibinfo {author} {\bibfnamefont {C.}~\bibnamefont
  {R{\"o}ttele}}, \bibinfo {author} {\bibfnamefont {C.}~\bibnamefont
  {Rodenbeck}}, \bibinfo {author} {\bibfnamefont {M.}~\bibnamefont
  {Ry{\v{s}}av{\'y}}}, \bibinfo {author} {\bibfnamefont {R.}~\bibnamefont
  {Sack}}, \bibinfo {author} {\bibfnamefont {A.}~\bibnamefont {Saenz}},
  \bibinfo {author} {\bibfnamefont {P.}~\bibnamefont {Sch{\"a}fer}}, \bibinfo
  {author} {\bibfnamefont {A.}~\bibnamefont {Schaller~n{\'e}e Pollithy}},
  \bibinfo {author} {\bibfnamefont {L.}~\bibnamefont {Schimpf}}, \bibinfo
  {author} {\bibfnamefont {K.}~\bibnamefont {Schl{\"o}sser}}, \bibinfo {author}
  {\bibfnamefont {M.}~\bibnamefont {Schl{\"o}sser}}, \bibinfo {author}
  {\bibfnamefont {L.}~\bibnamefont {Schl{\"u}ter}}, \bibinfo {author}
  {\bibfnamefont {S.}~\bibnamefont {Schneidewind}}, \bibinfo {author}
  {\bibfnamefont {M.}~\bibnamefont {Schrank}}, \bibinfo {author} {\bibfnamefont
  {B.}~\bibnamefont {Schulz}}, \bibinfo {author} {\bibfnamefont
  {A.}~\bibnamefont {Schwemmer}}, \bibinfo {author} {\bibfnamefont
  {M.}~\bibnamefont {{\v{S}}ef{\v{c}}{\'i}k}}, \bibinfo {author} {\bibfnamefont
  {V.}~\bibnamefont {Sibille}}, \bibinfo {author} {\bibfnamefont
  {D.}~\bibnamefont {Siegmann}}, \bibinfo {author} {\bibfnamefont
  {M.}~\bibnamefont {Slez{\'a}k}}, \bibinfo {author} {\bibfnamefont
  {F.}~\bibnamefont {Spanier}}, \bibinfo {author} {\bibfnamefont
  {M.}~\bibnamefont {Steidl}}, \bibinfo {author} {\bibfnamefont
  {M.}~\bibnamefont {Sturm}}, \bibinfo {author} {\bibfnamefont
  {M.}~\bibnamefont {Sun}}, \bibinfo {author} {\bibfnamefont {D.}~\bibnamefont
  {Tcherniakhovski}}, \bibinfo {author} {\bibfnamefont {H.~H.}\ \bibnamefont
  {Telle}}, \bibinfo {author} {\bibfnamefont {L.~A.}\ \bibnamefont {Thorne}},
  \bibinfo {author} {\bibfnamefont {T.}~\bibnamefont {Th{\"u}mmler}}, \bibinfo
  {author} {\bibfnamefont {N.}~\bibnamefont {Titov}}, \bibinfo {author}
  {\bibfnamefont {I.}~\bibnamefont {Tkachev}}, \bibinfo {author} {\bibfnamefont
  {K.}~\bibnamefont {Urban}}, \bibinfo {author} {\bibfnamefont
  {K.}~\bibnamefont {Valerius}}, \bibinfo {author} {\bibfnamefont
  {D.}~\bibnamefont {V{\'e}nos}}, \bibinfo {author} {\bibfnamefont {A.~P.}\
  \bibnamefont {Vizcaya~Hern{\'a}ndez}}, \bibinfo {author} {\bibfnamefont
  {C.}~\bibnamefont {Weinheimer}}, \bibinfo {author} {\bibfnamefont
  {S.}~\bibnamefont {Welte}}, \bibinfo {author} {\bibfnamefont
  {J.}~\bibnamefont {Wendel}}, \bibinfo {author} {\bibfnamefont {J.~F.}\
  \bibnamefont {Wilkerson}}, \bibinfo {author} {\bibfnamefont {J.}~\bibnamefont
  {Wolf}}, \bibinfo {author} {\bibfnamefont {S.}~\bibnamefont {W{\"u}stling}},
  \bibinfo {author} {\bibfnamefont {J.}~\bibnamefont {Wydra}}, \bibinfo
  {author} {\bibfnamefont {W.}~\bibnamefont {Xu}}, \bibinfo {author}
  {\bibfnamefont {Y.-R.}\ \bibnamefont {Yen}}, \bibinfo {author} {\bibfnamefont
  {S.}~\bibnamefont {Zadoroghny}}, \bibinfo {author} {\bibfnamefont
  {G.}~\bibnamefont {Zeller}},\ and\ \bibinfo {author} {\bibfnamefont {T.~K.}\
  \bibnamefont {Collaboration}},\ }\href
  {https://doi.org/10.1038/s41567-021-01463-1} {\bibfield  {journal} {\bibinfo
  {journal} {Nature Physics}\ }\textbf {\bibinfo {volume} {18}},\ \bibinfo
  {pages} {160} (\bibinfo {year} {2022})}\BibitemShut {NoStop}%
\bibitem [{\citenamefont {Esfahani}\ \emph {et~al.}(2022)\citenamefont
  {Esfahani} \emph {et~al.}}]{Project8:2022wqh}%
  \BibitemOpen
  \bibfield  {author} {\bibinfo {author} {\bibfnamefont {A.~A.}\ \bibnamefont
  {Esfahani}} \emph {et~al.} (\bibinfo {collaboration} {Project 8}),\ }in\
  \href@noop {} {\emph {\bibinfo {booktitle} {{2022 Snowmass Summer Study}}}}\
  (\bibinfo {year} {2022})\ \Eprint {https://arxiv.org/abs/2203.07349}
  {arXiv:2203.07349 [nucl-ex]} \BibitemShut {NoStop}%
\bibitem [{\citenamefont {Cattadori}\ \emph {et~al.}(2007)\citenamefont
  {Cattadori}, \citenamefont {De~Deo}, \citenamefont {Laubenstein},
  \citenamefont {Pandola},\ and\ \citenamefont
  {Tretyak}}]{Cattadori2007_115In}%
  \BibitemOpen
  \bibfield  {author} {\bibinfo {author} {\bibfnamefont {C.~M.}\ \bibnamefont
  {Cattadori}}, \bibinfo {author} {\bibfnamefont {M.}~\bibnamefont {De~Deo}},
  \bibinfo {author} {\bibfnamefont {M.}~\bibnamefont {Laubenstein}}, \bibinfo
  {author} {\bibfnamefont {L.}~\bibnamefont {Pandola}},\ and\ \bibinfo {author}
  {\bibfnamefont {V.~I.}\ \bibnamefont {Tretyak}},\ }\href@noop {} {\bibfield
  {journal} {\bibinfo  {journal} {Physics of Atomic Nuclei}\ }\textbf {\bibinfo
  {volume} {70}},\ \bibinfo {pages} {127} (\bibinfo {year} {2007})}\BibitemShut
  {NoStop}%
\bibitem [{\citenamefont {Mustonen}\ and\ \citenamefont
  {Suhonen}(2010{\natexlab{a}})}]{Mustonen2010_ULQCalcs}%
  \BibitemOpen
  \bibfield  {author} {\bibinfo {author} {\bibfnamefont {M.~T.}\ \bibnamefont
  {Mustonen}}\ and\ \bibinfo {author} {\bibfnamefont {J.}~\bibnamefont
  {Suhonen}},\ }\href {https://doi.org/10.1088/0954-3899/37/6/064008}
  {\bibfield  {journal} {\bibinfo  {journal} {J. Phys. G: Nucl. Part. Phys.}\
  }\textbf {\bibinfo {volume} {37}},\ \bibinfo {pages} {064008} (\bibinfo
  {year} {2010}{\natexlab{a}})}\BibitemShut {NoStop}%
\bibitem [{\citenamefont {Mustonen}\ and\ \citenamefont
  {Suhonen}(2010{\natexlab{b}})}]{Mustonen2010_ULQs}%
  \BibitemOpen
  \bibfield  {author} {\bibinfo {author} {\bibfnamefont {M.~T.}\ \bibnamefont
  {Mustonen}}\ and\ \bibinfo {author} {\bibfnamefont {J.}~\bibnamefont
  {Suhonen}},\ }\href {https://doi.org/10.1063/1.3527233} {\bibfield  {journal}
  {\bibinfo  {journal} {AIP Conf. Proc.}\ }\textbf {\bibinfo {volume} {1304}},\
  \bibinfo {pages} {401} (\bibinfo {year} {2010}{\natexlab{b}})}\BibitemShut
  {NoStop}%
\bibitem [{\citenamefont {Mustonen}\ and\ \citenamefont
  {Suhonen}(2011)}]{Mustonen2011_135Cs}%
  \BibitemOpen
  \bibfield  {author} {\bibinfo {author} {\bibfnamefont {M.}~\bibnamefont
  {Mustonen}}\ and\ \bibinfo {author} {\bibfnamefont {J.}~\bibnamefont
  {Suhonen}},\ }\href
  {https://doi.org/https://doi.org/10.1016/j.physletb.2011.07.088} {\bibfield
  {journal} {\bibinfo  {journal} {Physics Letters B}\ }\textbf {\bibinfo
  {volume} {703}},\ \bibinfo {pages} {370 } (\bibinfo {year}
  {2011})}\BibitemShut {NoStop}%
\bibitem [{\citenamefont {Haaranen}(2013)}]{Haaranen2013_115Cd}%
  \BibitemOpen
  \bibfield  {author} {\bibinfo {author} {\bibfnamefont {J.}~\bibnamefont
  {Haaranen}, \bibfnamefont {M.and~Suhonen}},\ }\href
  {https://doi.org/10.1140/epja/i2013-13093-8} {\bibfield  {journal} {\bibinfo
  {journal} {The European Physical Journal A}\ }\textbf {\bibinfo {volume}
  {49}},\ \bibinfo {pages} {93} (\bibinfo {year} {2013})}\BibitemShut {NoStop}%
\bibitem [{\citenamefont {Suhonen}(2014)}]{Suhonen2014_ULQs}%
  \BibitemOpen
  \bibfield  {author} {\bibinfo {author} {\bibfnamefont {J.}~\bibnamefont
  {Suhonen}},\ }\href {https://doi.org/10.1088/0031-8949/89/5/054032}
  {\bibfield  {journal} {\bibinfo  {journal} {Physica Scripta}\ }\textbf
  {\bibinfo {volume} {89}},\ \bibinfo {pages} {054032} (\bibinfo {year}
  {2014})}\BibitemShut {NoStop}%
\bibitem [{\citenamefont {Gamage}\ \emph {et~al.}(2019)\citenamefont {Gamage},
  \citenamefont {Bhandari}, \citenamefont {Gamage}, \citenamefont {Sandler},\
  and\ \citenamefont {Redshaw}}]{Gamage2019_ULQs}%
  \BibitemOpen
  \bibfield  {author} {\bibinfo {author} {\bibfnamefont {N.~D.}\ \bibnamefont
  {Gamage}}, \bibinfo {author} {\bibfnamefont {R.}~\bibnamefont {Bhandari}},
  \bibinfo {author} {\bibfnamefont {M.~H.}\ \bibnamefont {Gamage}}, \bibinfo
  {author} {\bibfnamefont {R.}~\bibnamefont {Sandler}},\ and\ \bibinfo {author}
  {\bibfnamefont {M.}~\bibnamefont {Redshaw}},\ }\href
  {hhttps://doi.org/10.1007/s10751-019-1588-51} {\bibfield  {journal} {\bibinfo
   {journal} {Hyp. Int.}\ }\textbf {\bibinfo {volume} {240}},\ \bibinfo {pages}
  {43} (\bibinfo {year} {2019})}\BibitemShut {NoStop}%
\bibitem [{\citenamefont {Keblbeck}\ \emph {et~al.}(2022)\citenamefont
  {Keblbeck}, \citenamefont {Bhandari}, \citenamefont {Gamage}, \citenamefont
  {Gamage},\ and\ \citenamefont {Redshaw}}]{Keblbeck2022_ULQs}%
  \BibitemOpen
  \bibfield  {author} {\bibinfo {author} {\bibfnamefont {D.~K.}\ \bibnamefont
  {Keblbeck}}, \bibinfo {author} {\bibfnamefont {R.}~\bibnamefont {Bhandari}},
  \bibinfo {author} {\bibfnamefont {N.~D.}\ \bibnamefont {Gamage}}, \bibinfo
  {author} {\bibfnamefont {M.~H.}\ \bibnamefont {Gamage}},\ and\ \bibinfo
  {author} {\bibfnamefont {M.}~\bibnamefont {Redshaw}},\ }\href@noop {}
  {\bibfield  {journal} {\bibinfo  {journal} {arXiv:2201.08790}\ } (\bibinfo
  {year} {2022})}\BibitemShut {NoStop}%
\bibitem [{\citenamefont {Horana~Gamage}\ \emph {et~al.}(2022)\citenamefont
  {Horana~Gamage}, \citenamefont {Bhandari}, \citenamefont {Bollen},
  \citenamefont {Gamage}, \citenamefont {Hamaker}, \citenamefont {Puentes},
  \citenamefont {Redshaw}, \citenamefont {Ringle}, \citenamefont {Schwarz},
  \citenamefont {Sumithrarachchi},\ and\ \citenamefont
  {Yandow}}]{Horana2022_75As}%
  \BibitemOpen
  \bibfield  {author} {\bibinfo {author} {\bibfnamefont {M.}~\bibnamefont
  {Horana~Gamage}}, \bibinfo {author} {\bibfnamefont {R.}~\bibnamefont
  {Bhandari}}, \bibinfo {author} {\bibfnamefont {G.}~\bibnamefont {Bollen}},
  \bibinfo {author} {\bibfnamefont {N.~D.}\ \bibnamefont {Gamage}}, \bibinfo
  {author} {\bibfnamefont {A.}~\bibnamefont {Hamaker}}, \bibinfo {author}
  {\bibfnamefont {D.}~\bibnamefont {Puentes}}, \bibinfo {author} {\bibfnamefont
  {M.}~\bibnamefont {Redshaw}}, \bibinfo {author} {\bibfnamefont
  {R.}~\bibnamefont {Ringle}}, \bibinfo {author} {\bibfnamefont
  {S.}~\bibnamefont {Schwarz}}, \bibinfo {author} {\bibfnamefont {C.~S.}\
  \bibnamefont {Sumithrarachchi}},\ and\ \bibinfo {author} {\bibfnamefont
  {I.}~\bibnamefont {Yandow}},\ }\href
  {https://doi.org/10.48550/arXiv.2208.11182} {\bibfield  {journal} {\bibinfo
  {journal} {arXiv:2208.11182}\ } (\bibinfo {year} {2022})}\BibitemShut
  {NoStop}%
\bibitem [{\citenamefont {Ramalho}\ \emph
  {et~al.}(2022{\natexlab{a}})\citenamefont {Ramalho}, \citenamefont {Ge},
  \citenamefont {Eronen}, \citenamefont {Nesterenko}, \citenamefont {Jaatinen},
  \citenamefont {Jokinen}, \citenamefont {Kankainen}, \citenamefont
  {Kostensalo}, \citenamefont {Kotila}, \citenamefont {Krivoruchenko},
  \citenamefont {Suhonen}, \citenamefont {Tyrin},\ and\ \citenamefont
  {Virtanen}}]{Ramalho2022_75As}%
  \BibitemOpen
  \bibfield  {author} {\bibinfo {author} {\bibfnamefont {M.}~\bibnamefont
  {Ramalho}}, \bibinfo {author} {\bibfnamefont {Z.}~\bibnamefont {Ge}},
  \bibinfo {author} {\bibfnamefont {T.}~\bibnamefont {Eronen}}, \bibinfo
  {author} {\bibfnamefont {D.~A.}\ \bibnamefont {Nesterenko}}, \bibinfo
  {author} {\bibfnamefont {J.}~\bibnamefont {Jaatinen}}, \bibinfo {author}
  {\bibfnamefont {A.}~\bibnamefont {Jokinen}}, \bibinfo {author} {\bibfnamefont
  {A.}~\bibnamefont {Kankainen}}, \bibinfo {author} {\bibfnamefont
  {J.}~\bibnamefont {Kostensalo}}, \bibinfo {author} {\bibfnamefont
  {J.}~\bibnamefont {Kotila}}, \bibinfo {author} {\bibfnamefont {M.~I.}\
  \bibnamefont {Krivoruchenko}}, \bibinfo {author} {\bibfnamefont
  {J.}~\bibnamefont {Suhonen}}, \bibinfo {author} {\bibfnamefont {K.~S.}\
  \bibnamefont {Tyrin}},\ and\ \bibinfo {author} {\bibfnamefont
  {V.}~\bibnamefont {Virtanen}},\ }\href
  {https://doi.org/10.1103/PhysRevC.106.015501} {\bibfield  {journal} {\bibinfo
   {journal} {Phys. Rev. C}\ }\textbf {\bibinfo {volume} {106}},\ \bibinfo
  {pages} {015501} (\bibinfo {year} {2022}{\natexlab{a}})}\BibitemShut
  {NoStop}%
\bibitem [{\citenamefont {Ge}\ \emph {et~al.}(2022)\citenamefont {Ge},
  \citenamefont {Eronen}, \citenamefont {{de Roubin}}, \citenamefont {Tyrin},
  \citenamefont {Canete}, \citenamefont {Geldhof}, \citenamefont {Jokinen},
  \citenamefont {Kankainen}, \citenamefont {Kostensalo}, \citenamefont
  {Kotila}, \citenamefont {Krivoruchenko}, \citenamefont {Moore}, \citenamefont
  {Nesterenko}, \citenamefont {Suhonen},\ and\ \citenamefont
  {Vil{\'e}n}}]{Ge2022_111In}%
  \BibitemOpen
  \bibfield  {author} {\bibinfo {author} {\bibfnamefont {Z.}~\bibnamefont
  {Ge}}, \bibinfo {author} {\bibfnamefont {T.}~\bibnamefont {Eronen}}, \bibinfo
  {author} {\bibfnamefont {A.}~\bibnamefont {{de Roubin}}}, \bibinfo {author}
  {\bibfnamefont {K.}~\bibnamefont {Tyrin}}, \bibinfo {author} {\bibfnamefont
  {L.}~\bibnamefont {Canete}}, \bibinfo {author} {\bibfnamefont
  {S.}~\bibnamefont {Geldhof}}, \bibinfo {author} {\bibfnamefont
  {A.}~\bibnamefont {Jokinen}}, \bibinfo {author} {\bibfnamefont
  {A.}~\bibnamefont {Kankainen}}, \bibinfo {author} {\bibfnamefont
  {J.}~\bibnamefont {Kostensalo}}, \bibinfo {author} {\bibfnamefont
  {J.}~\bibnamefont {Kotila}}, \bibinfo {author} {\bibfnamefont
  {M.}~\bibnamefont {Krivoruchenko}}, \bibinfo {author} {\bibfnamefont
  {I.}~\bibnamefont {Moore}}, \bibinfo {author} {\bibfnamefont
  {D.}~\bibnamefont {Nesterenko}}, \bibinfo {author} {\bibfnamefont
  {J.}~\bibnamefont {Suhonen}},\ and\ \bibinfo {author} {\bibfnamefont
  {M.}~\bibnamefont {Vil{\'e}n}},\ }\href@noop {} {\bibfield  {journal}
  {\bibinfo  {journal} {Physics Letters B}\ }\textbf {\bibinfo {volume}
  {832}},\ \bibinfo {pages} {137226} (\bibinfo {year} {2022})}\BibitemShut
  {NoStop}%
\bibitem [{\citenamefont {Eronen}\ \emph {et~al.}(2022)\citenamefont {Eronen},
  \citenamefont {Ge}, \citenamefont {{de Roubin}}, \citenamefont {Ramalho},
  \citenamefont {Kostensalo}, \citenamefont {Kotila}, \citenamefont
  {Beliushkina}, \citenamefont {Delafosse}, \citenamefont {Geldhof},
  \citenamefont {Gins}, \citenamefont {Hukkanen}, \citenamefont {Jokinen},
  \citenamefont {Kankainen}, \citenamefont {Moore}, \citenamefont {Nesterenko},
  \citenamefont {Stryjczyk},\ and\ \citenamefont {Suhonen}}]{Eronen2022_131I}%
  \BibitemOpen
  \bibfield  {author} {\bibinfo {author} {\bibfnamefont {T.}~\bibnamefont
  {Eronen}}, \bibinfo {author} {\bibfnamefont {Z.}~\bibnamefont {Ge}}, \bibinfo
  {author} {\bibfnamefont {A.}~\bibnamefont {{de Roubin}}}, \bibinfo {author}
  {\bibfnamefont {M.}~\bibnamefont {Ramalho}}, \bibinfo {author} {\bibfnamefont
  {J.}~\bibnamefont {Kostensalo}}, \bibinfo {author} {\bibfnamefont
  {J.}~\bibnamefont {Kotila}}, \bibinfo {author} {\bibfnamefont
  {O.}~\bibnamefont {Beliushkina}}, \bibinfo {author} {\bibfnamefont
  {C.}~\bibnamefont {Delafosse}}, \bibinfo {author} {\bibfnamefont
  {S.}~\bibnamefont {Geldhof}}, \bibinfo {author} {\bibfnamefont
  {W.}~\bibnamefont {Gins}}, \bibinfo {author} {\bibfnamefont {M.}~\bibnamefont
  {Hukkanen}}, \bibinfo {author} {\bibfnamefont {A.}~\bibnamefont {Jokinen}},
  \bibinfo {author} {\bibfnamefont {A.}~\bibnamefont {Kankainen}}, \bibinfo
  {author} {\bibfnamefont {I.}~\bibnamefont {Moore}}, \bibinfo {author}
  {\bibfnamefont {D.}~\bibnamefont {Nesterenko}}, \bibinfo {author}
  {\bibfnamefont {M.}~\bibnamefont {Stryjczyk}},\ and\ \bibinfo {author}
  {\bibfnamefont {J.}~\bibnamefont {Suhonen}},\ }\href@noop {} {\bibfield
  {journal} {\bibinfo  {journal} {Physics Letters B}\ }\textbf {\bibinfo
  {volume} {830}},\ \bibinfo {pages} {137135} (\bibinfo {year}
  {2022})}\BibitemShut {NoStop}%
\bibitem [{\citenamefont {Ge}\ \emph {et~al.}(2021)\citenamefont {Ge},
  \citenamefont {Eronen}, \citenamefont {Tyrin}, \citenamefont {Kotila},
  \citenamefont {Kostensalo}, \citenamefont {Nesterenko}, \citenamefont
  {Beliuskina}, \citenamefont {de~Groote}, \citenamefont {de~Roubin},
  \citenamefont {Geldhof}, \citenamefont {Gins}, \citenamefont {Hukkanen},
  \citenamefont {Jokinen}, \citenamefont {Kankainen}, \citenamefont
  {Koszor\'us}, \citenamefont {Krivoruchenko}, \citenamefont {Kujanp\"a\"a},
  \citenamefont {Moore}, \citenamefont {Raggio}, \citenamefont {Rinta-Antila},
  \citenamefont {Suhonen}, \citenamefont {Virtanen}, \citenamefont {Weaver},\
  and\ \citenamefont {Zadvornaya}}]{Ge2021_159Dy}%
  \BibitemOpen
  \bibfield  {author} {\bibinfo {author} {\bibfnamefont {Z.}~\bibnamefont
  {Ge}}, \bibinfo {author} {\bibfnamefont {T.}~\bibnamefont {Eronen}}, \bibinfo
  {author} {\bibfnamefont {K.~S.}\ \bibnamefont {Tyrin}}, \bibinfo {author}
  {\bibfnamefont {J.}~\bibnamefont {Kotila}}, \bibinfo {author} {\bibfnamefont
  {J.}~\bibnamefont {Kostensalo}}, \bibinfo {author} {\bibfnamefont {D.~A.}\
  \bibnamefont {Nesterenko}}, \bibinfo {author} {\bibfnamefont
  {O.}~\bibnamefont {Beliuskina}}, \bibinfo {author} {\bibfnamefont
  {R.}~\bibnamefont {de~Groote}}, \bibinfo {author} {\bibfnamefont
  {A.}~\bibnamefont {de~Roubin}}, \bibinfo {author} {\bibfnamefont
  {S.}~\bibnamefont {Geldhof}}, \bibinfo {author} {\bibfnamefont
  {W.}~\bibnamefont {Gins}}, \bibinfo {author} {\bibfnamefont {M.}~\bibnamefont
  {Hukkanen}}, \bibinfo {author} {\bibfnamefont {A.}~\bibnamefont {Jokinen}},
  \bibinfo {author} {\bibfnamefont {A.}~\bibnamefont {Kankainen}}, \bibinfo
  {author} {\bibfnamefont {A.}~\bibnamefont {Koszor\'us}}, \bibinfo {author}
  {\bibfnamefont {M.~I.}\ \bibnamefont {Krivoruchenko}}, \bibinfo {author}
  {\bibfnamefont {S.}~\bibnamefont {Kujanp\"a\"a}}, \bibinfo {author}
  {\bibfnamefont {I.~D.}\ \bibnamefont {Moore}}, \bibinfo {author}
  {\bibfnamefont {A.}~\bibnamefont {Raggio}}, \bibinfo {author} {\bibfnamefont
  {S.}~\bibnamefont {Rinta-Antila}}, \bibinfo {author} {\bibfnamefont
  {J.}~\bibnamefont {Suhonen}}, \bibinfo {author} {\bibfnamefont
  {V.}~\bibnamefont {Virtanen}}, \bibinfo {author} {\bibfnamefont {A.~P.}\
  \bibnamefont {Weaver}},\ and\ \bibinfo {author} {\bibfnamefont
  {A.}~\bibnamefont {Zadvornaya}},\ }\href
  {https://doi.org/10.1103/PhysRevLett.127.272301} {\bibfield  {journal}
  {\bibinfo  {journal} {Phys. Rev. Lett.}\ }\textbf {\bibinfo {volume} {127}},\
  \bibinfo {pages} {272301} (\bibinfo {year} {2021})}\BibitemShut {NoStop}%
\bibitem [{\citenamefont {de~Roubin}\ \emph {et~al.}(2020)\citenamefont
  {de~Roubin}, \citenamefont {Kostensalo}, \citenamefont {Eronen},
  \citenamefont {Canete}, \citenamefont {de~Groote}, \citenamefont {Jokinen},
  \citenamefont {Kankainen}, \citenamefont {Nesterenko}, \citenamefont {Moore},
  \citenamefont {Rinta-Antila}, \citenamefont {Suhonen},\ and\ \citenamefont
  {Vil\'en}}]{deRoubin2020_135Cs}%
  \BibitemOpen
  \bibfield  {author} {\bibinfo {author} {\bibfnamefont {A.}~\bibnamefont
  {de~Roubin}}, \bibinfo {author} {\bibfnamefont {J.}~\bibnamefont
  {Kostensalo}}, \bibinfo {author} {\bibfnamefont {T.}~\bibnamefont {Eronen}},
  \bibinfo {author} {\bibfnamefont {L.}~\bibnamefont {Canete}}, \bibinfo
  {author} {\bibfnamefont {R.~P.}\ \bibnamefont {de~Groote}}, \bibinfo {author}
  {\bibfnamefont {A.}~\bibnamefont {Jokinen}}, \bibinfo {author} {\bibfnamefont
  {A.}~\bibnamefont {Kankainen}}, \bibinfo {author} {\bibfnamefont {D.~A.}\
  \bibnamefont {Nesterenko}}, \bibinfo {author} {\bibfnamefont {I.~D.}\
  \bibnamefont {Moore}}, \bibinfo {author} {\bibfnamefont {S.}~\bibnamefont
  {Rinta-Antila}}, \bibinfo {author} {\bibfnamefont {J.}~\bibnamefont
  {Suhonen}},\ and\ \bibinfo {author} {\bibfnamefont {M.}~\bibnamefont
  {Vil\'en}},\ }\href {https://doi.org/10.1103/PhysRevLett.124.222503}
  {\bibfield  {journal} {\bibinfo  {journal} {Phys. Rev. Lett.}\ }\textbf
  {\bibinfo {volume} {124}},\ \bibinfo {pages} {222503} (\bibinfo {year}
  {2020})}\BibitemShut {NoStop}%
\bibitem [{\citenamefont {Carney}\ \emph {et~al.}(2022)\citenamefont {Carney},
  \citenamefont {Leach},\ and\ \citenamefont {Moore}}]{Carney:2022pku}%
  \BibitemOpen
  \bibfield  {author} {\bibinfo {author} {\bibfnamefont {D.}~\bibnamefont
  {Carney}}, \bibinfo {author} {\bibfnamefont {K.~G.}\ \bibnamefont {Leach}},\
  and\ \bibinfo {author} {\bibfnamefont {D.~C.}\ \bibnamefont {Moore}},\
  }\href@noop {} {\  (\bibinfo {year} {2022})},\ \Eprint
  {https://arxiv.org/abs/2207.05883} {arXiv:2207.05883 [hep-ex]} \BibitemShut
  {NoStop}%
\bibitem [{\citenamefont {Acero}\ \emph {et~al.}(2022)\citenamefont {Acero}
  \emph {et~al.}}]{Acero:2022wqg}%
  \BibitemOpen
  \bibfield  {author} {\bibinfo {author} {\bibfnamefont {M.~A.}\ \bibnamefont
  {Acero}} \emph {et~al.},\ }\href@noop {} {\  (\bibinfo {year} {2022})},\
  \Eprint {https://arxiv.org/abs/2203.07323} {arXiv:2203.07323 [hep-ex]}
  \BibitemShut {NoStop}%
\bibitem [{\citenamefont {Dodelson}\ and\ \citenamefont
  {Widrow}(1994)}]{Dod94}%
  \BibitemOpen
  \bibfield  {author} {\bibinfo {author} {\bibfnamefont {S.}~\bibnamefont
  {Dodelson}}\ and\ \bibinfo {author} {\bibfnamefont {L.~M.}\ \bibnamefont
  {Widrow}},\ }\href {https://doi.org/10.1103/PhysRevLett.72.17} {\bibfield
  {journal} {\bibinfo  {journal} {Phys. Rev. Lett.}\ }\textbf {\bibinfo
  {volume} {72}},\ \bibinfo {pages} {17} (\bibinfo {year} {1994})}\BibitemShut
  {NoStop}%
\bibitem [{\citenamefont {Adhikari}\ \emph {et~al.}(2017)\citenamefont
  {Adhikari}, \citenamefont {Agostini}, \citenamefont {Ky}, \citenamefont
  {Araki}, \citenamefont {Archidiacono}, \citenamefont {Bahr}, \citenamefont
  {Baur}, \citenamefont {Behrens}, \citenamefont {Bezrukov}, \citenamefont
  {Dev}, \citenamefont {Borah}, \citenamefont {Boyarsky}, \citenamefont
  {de~Gouvea}, \citenamefont {de~S.~Pires}, \citenamefont {de~Vega},
  \citenamefont {Dias}, \citenamefont {Bari}, \citenamefont {Djurcic},
  \citenamefont {Dolde}, \citenamefont {Dorrer}, \citenamefont {Durero},
  \citenamefont {Dragoun}, \citenamefont {Drewes}, \citenamefont {Drexlin},
  \citenamefont {Düllmann}, \citenamefont {Eberhardt}, \citenamefont
  {Eliseev}, \citenamefont {Enss}, \citenamefont {Evans}, \citenamefont
  {Faessler}, \citenamefont {Filianin}, \citenamefont {Fischer}, \citenamefont
  {Fleischmann}, \citenamefont {Formaggio}, \citenamefont {Franse},
  \citenamefont {Fraenkle}, \citenamefont {Frenk}, \citenamefont {Fuller},
  \citenamefont {Gastaldo}, \citenamefont {Garzilli}, \citenamefont {Giunti},
  \citenamefont {Gl\"uck}, \citenamefont {Goodman}, \citenamefont
  {Gonzalez-Garcia}, \citenamefont {Gorbunov}, \citenamefont {Hamann},
  \citenamefont {Hannen}, \citenamefont {Hannestad}, \citenamefont {Hansen},
  \citenamefont {Hassel}, \citenamefont {Heeck}, \citenamefont {Hofmann},
  \citenamefont {Houdy}, \citenamefont {Huber}, \citenamefont {Iakubovskyi},
  \citenamefont {Ianni}, \citenamefont {Ibarra}, \citenamefont {Jacobsson},
  \citenamefont {Jeltema}, \citenamefont {Jochum}, \citenamefont {Kempf},
  \citenamefont {Kieck}, \citenamefont {Korzeczek}, \citenamefont {Kornoukhov},
  \citenamefont {Lachenmaier}, \citenamefont {Laine}, \citenamefont
  {Langacker}, \citenamefont {Lasserre}, \citenamefont {Lesgourgues},
  \citenamefont {Lhuillier}, \citenamefont {Li}, \citenamefont {Liao},
  \citenamefont {Long}, \citenamefont {Maltoni}, \citenamefont {Mangano},
  \citenamefont {Mavromatos}, \citenamefont {Menci}, \citenamefont {Merle},
  \citenamefont {Mertens}, \citenamefont {Mirizzi}, \citenamefont {Monreal},
  \citenamefont {Nozik}, \citenamefont {Neronov}, \citenamefont {Niro},
  \citenamefont {Novikov}, \citenamefont {Oberauer}, \citenamefont {Otten},
  \citenamefont {Palanque-Delabrouille}, \citenamefont {Pallavicini},
  \citenamefont {Pantuev}, \citenamefont {Papastergis}, \citenamefont {Parke},
  \citenamefont {Pascoli}, \citenamefont {Pastor}, \citenamefont {Patwardhan},
  \citenamefont {Pilaftsis}, \citenamefont {Radford}, \citenamefont
  {Ranitzsch}, \citenamefont {Rest}, \citenamefont {Robinson}, \citenamefont
  {da~Silva}, \citenamefont {Ruchayskiy}, \citenamefont {Sanchez},
  \citenamefont {Sasaki}, \citenamefont {Saviano}, \citenamefont {Schneider},
  \citenamefont {Schneider}, \citenamefont {Schwetz}, \citenamefont
  {Sch\"onert}, \citenamefont {Scholl}, \citenamefont {Shankar}, \citenamefont
  {Shrock}, \citenamefont {Steinbrink}, \citenamefont {Strigari}, \citenamefont
  {Suekane}, \citenamefont {Suerfu}, \citenamefont {Takahashi}, \citenamefont
  {Van}, \citenamefont {Tkachev}, \citenamefont {Totzauer}, \citenamefont
  {Tsai}, \citenamefont {Tully}, \citenamefont {Valerius}, \citenamefont
  {Valle}, \citenamefont {Venos}, \citenamefont {Viel}, \citenamefont {Vivier},
  \citenamefont {Wang}, \citenamefont {Weinheimer}, \citenamefont {Wendt},
  \citenamefont {Winslow}, \citenamefont {Wolf}, \citenamefont {Wurm},
  \citenamefont {Xing}, \citenamefont {Zhou},\ and\ \citenamefont
  {Zuber}}]{Adh17}%
  \BibitemOpen
  \bibfield  {author} {\bibinfo {author} {\bibfnamefont {R.}~\bibnamefont
  {Adhikari}}, \bibinfo {author} {\bibfnamefont {M.}~\bibnamefont {Agostini}},
  \bibinfo {author} {\bibfnamefont {N.~A.}\ \bibnamefont {Ky}}, \bibinfo
  {author} {\bibfnamefont {T.}~\bibnamefont {Araki}}, \bibinfo {author}
  {\bibfnamefont {M.}~\bibnamefont {Archidiacono}}, \bibinfo {author}
  {\bibfnamefont {M.}~\bibnamefont {Bahr}}, \bibinfo {author} {\bibfnamefont
  {J.}~\bibnamefont {Baur}}, \bibinfo {author} {\bibfnamefont {J.}~\bibnamefont
  {Behrens}}, \bibinfo {author} {\bibfnamefont {F.}~\bibnamefont {Bezrukov}},
  \bibinfo {author} {\bibfnamefont {P.~B.}\ \bibnamefont {Dev}}, \bibinfo
  {author} {\bibfnamefont {D.}~\bibnamefont {Borah}}, \bibinfo {author}
  {\bibfnamefont {A.}~\bibnamefont {Boyarsky}}, \bibinfo {author}
  {\bibfnamefont {A.}~\bibnamefont {de~Gouvea}}, \bibinfo {author}
  {\bibfnamefont {C.}~\bibnamefont {de~S.~Pires}}, \bibinfo {author}
  {\bibfnamefont {H.}~\bibnamefont {de~Vega}}, \bibinfo {author} {\bibfnamefont
  {A.}~\bibnamefont {Dias}}, \bibinfo {author} {\bibfnamefont {P.~D.}\
  \bibnamefont {Bari}}, \bibinfo {author} {\bibfnamefont {Z.}~\bibnamefont
  {Djurcic}}, \bibinfo {author} {\bibfnamefont {K.}~\bibnamefont {Dolde}},
  \bibinfo {author} {\bibfnamefont {H.}~\bibnamefont {Dorrer}}, \bibinfo
  {author} {\bibfnamefont {M.}~\bibnamefont {Durero}}, \bibinfo {author}
  {\bibfnamefont {O.}~\bibnamefont {Dragoun}}, \bibinfo {author} {\bibfnamefont
  {M.}~\bibnamefont {Drewes}}, \bibinfo {author} {\bibfnamefont
  {G.}~\bibnamefont {Drexlin}}, \bibinfo {author} {\bibfnamefont
  {C.}~\bibnamefont {Düllmann}}, \bibinfo {author} {\bibfnamefont
  {K.}~\bibnamefont {Eberhardt}}, \bibinfo {author} {\bibfnamefont
  {S.}~\bibnamefont {Eliseev}}, \bibinfo {author} {\bibfnamefont
  {C.}~\bibnamefont {Enss}}, \bibinfo {author} {\bibfnamefont {N.}~\bibnamefont
  {Evans}}, \bibinfo {author} {\bibfnamefont {A.}~\bibnamefont {Faessler}},
  \bibinfo {author} {\bibfnamefont {P.}~\bibnamefont {Filianin}}, \bibinfo
  {author} {\bibfnamefont {V.}~\bibnamefont {Fischer}}, \bibinfo {author}
  {\bibfnamefont {A.}~\bibnamefont {Fleischmann}}, \bibinfo {author}
  {\bibfnamefont {J.}~\bibnamefont {Formaggio}}, \bibinfo {author}
  {\bibfnamefont {J.}~\bibnamefont {Franse}}, \bibinfo {author} {\bibfnamefont
  {F.}~\bibnamefont {Fraenkle}}, \bibinfo {author} {\bibfnamefont
  {C.}~\bibnamefont {Frenk}}, \bibinfo {author} {\bibfnamefont
  {G.}~\bibnamefont {Fuller}}, \bibinfo {author} {\bibfnamefont
  {L.}~\bibnamefont {Gastaldo}}, \bibinfo {author} {\bibfnamefont
  {A.}~\bibnamefont {Garzilli}}, \bibinfo {author} {\bibfnamefont
  {C.}~\bibnamefont {Giunti}}, \bibinfo {author} {\bibfnamefont
  {F.}~\bibnamefont {Gl\"uck}}, \bibinfo {author} {\bibfnamefont
  {M.}~\bibnamefont {Goodman}}, \bibinfo {author} {\bibfnamefont
  {M.}~\bibnamefont {Gonzalez-Garcia}}, \bibinfo {author} {\bibfnamefont
  {D.}~\bibnamefont {Gorbunov}}, \bibinfo {author} {\bibfnamefont
  {J.}~\bibnamefont {Hamann}}, \bibinfo {author} {\bibfnamefont
  {V.}~\bibnamefont {Hannen}}, \bibinfo {author} {\bibfnamefont
  {S.}~\bibnamefont {Hannestad}}, \bibinfo {author} {\bibfnamefont
  {S.}~\bibnamefont {Hansen}}, \bibinfo {author} {\bibfnamefont
  {C.}~\bibnamefont {Hassel}}, \bibinfo {author} {\bibfnamefont
  {J.}~\bibnamefont {Heeck}}, \bibinfo {author} {\bibfnamefont
  {F.}~\bibnamefont {Hofmann}}, \bibinfo {author} {\bibfnamefont
  {T.}~\bibnamefont {Houdy}}, \bibinfo {author} {\bibfnamefont
  {A.}~\bibnamefont {Huber}}, \bibinfo {author} {\bibfnamefont
  {D.}~\bibnamefont {Iakubovskyi}}, \bibinfo {author} {\bibfnamefont
  {A.}~\bibnamefont {Ianni}}, \bibinfo {author} {\bibfnamefont
  {A.}~\bibnamefont {Ibarra}}, \bibinfo {author} {\bibfnamefont
  {R.}~\bibnamefont {Jacobsson}}, \bibinfo {author} {\bibfnamefont
  {T.}~\bibnamefont {Jeltema}}, \bibinfo {author} {\bibfnamefont
  {J.}~\bibnamefont {Jochum}}, \bibinfo {author} {\bibfnamefont
  {S.}~\bibnamefont {Kempf}}, \bibinfo {author} {\bibfnamefont
  {T.}~\bibnamefont {Kieck}}, \bibinfo {author} {\bibfnamefont
  {M.}~\bibnamefont {Korzeczek}}, \bibinfo {author} {\bibfnamefont
  {V.}~\bibnamefont {Kornoukhov}}, \bibinfo {author} {\bibfnamefont
  {T.}~\bibnamefont {Lachenmaier}}, \bibinfo {author} {\bibfnamefont
  {M.}~\bibnamefont {Laine}}, \bibinfo {author} {\bibfnamefont
  {P.}~\bibnamefont {Langacker}}, \bibinfo {author} {\bibfnamefont
  {T.}~\bibnamefont {Lasserre}}, \bibinfo {author} {\bibfnamefont
  {J.}~\bibnamefont {Lesgourgues}}, \bibinfo {author} {\bibfnamefont
  {D.}~\bibnamefont {Lhuillier}}, \bibinfo {author} {\bibfnamefont
  {Y.}~\bibnamefont {Li}}, \bibinfo {author} {\bibfnamefont {W.}~\bibnamefont
  {Liao}}, \bibinfo {author} {\bibfnamefont {A.}~\bibnamefont {Long}}, \bibinfo
  {author} {\bibfnamefont {M.}~\bibnamefont {Maltoni}}, \bibinfo {author}
  {\bibfnamefont {G.}~\bibnamefont {Mangano}}, \bibinfo {author} {\bibfnamefont
  {N.}~\bibnamefont {Mavromatos}}, \bibinfo {author} {\bibfnamefont
  {N.}~\bibnamefont {Menci}}, \bibinfo {author} {\bibfnamefont
  {A.}~\bibnamefont {Merle}}, \bibinfo {author} {\bibfnamefont
  {S.}~\bibnamefont {Mertens}}, \bibinfo {author} {\bibfnamefont
  {A.}~\bibnamefont {Mirizzi}}, \bibinfo {author} {\bibfnamefont
  {B.}~\bibnamefont {Monreal}}, \bibinfo {author} {\bibfnamefont
  {A.}~\bibnamefont {Nozik}}, \bibinfo {author} {\bibfnamefont
  {A.}~\bibnamefont {Neronov}}, \bibinfo {author} {\bibfnamefont
  {V.}~\bibnamefont {Niro}}, \bibinfo {author} {\bibfnamefont {Y.}~\bibnamefont
  {Novikov}}, \bibinfo {author} {\bibfnamefont {L.}~\bibnamefont {Oberauer}},
  \bibinfo {author} {\bibfnamefont {E.}~\bibnamefont {Otten}}, \bibinfo
  {author} {\bibfnamefont {N.}~\bibnamefont {Palanque-Delabrouille}}, \bibinfo
  {author} {\bibfnamefont {M.}~\bibnamefont {Pallavicini}}, \bibinfo {author}
  {\bibfnamefont {V.}~\bibnamefont {Pantuev}}, \bibinfo {author} {\bibfnamefont
  {E.}~\bibnamefont {Papastergis}}, \bibinfo {author} {\bibfnamefont
  {S.}~\bibnamefont {Parke}}, \bibinfo {author} {\bibfnamefont
  {S.}~\bibnamefont {Pascoli}}, \bibinfo {author} {\bibfnamefont
  {S.}~\bibnamefont {Pastor}}, \bibinfo {author} {\bibfnamefont
  {A.}~\bibnamefont {Patwardhan}}, \bibinfo {author} {\bibfnamefont
  {A.}~\bibnamefont {Pilaftsis}}, \bibinfo {author} {\bibfnamefont
  {D.}~\bibnamefont {Radford}}, \bibinfo {author} {\bibfnamefont {P.-O.}\
  \bibnamefont {Ranitzsch}}, \bibinfo {author} {\bibfnamefont {O.}~\bibnamefont
  {Rest}}, \bibinfo {author} {\bibfnamefont {D.}~\bibnamefont {Robinson}},
  \bibinfo {author} {\bibfnamefont {P.~R.}\ \bibnamefont {da~Silva}}, \bibinfo
  {author} {\bibfnamefont {O.}~\bibnamefont {Ruchayskiy}}, \bibinfo {author}
  {\bibfnamefont {N.}~\bibnamefont {Sanchez}}, \bibinfo {author} {\bibfnamefont
  {M.}~\bibnamefont {Sasaki}}, \bibinfo {author} {\bibfnamefont
  {N.}~\bibnamefont {Saviano}}, \bibinfo {author} {\bibfnamefont
  {A.}~\bibnamefont {Schneider}}, \bibinfo {author} {\bibfnamefont
  {F.}~\bibnamefont {Schneider}}, \bibinfo {author} {\bibfnamefont
  {T.}~\bibnamefont {Schwetz}}, \bibinfo {author} {\bibfnamefont
  {S.}~\bibnamefont {Sch\"onert}}, \bibinfo {author} {\bibfnamefont
  {S.}~\bibnamefont {Scholl}}, \bibinfo {author} {\bibfnamefont
  {F.}~\bibnamefont {Shankar}}, \bibinfo {author} {\bibfnamefont
  {R.}~\bibnamefont {Shrock}}, \bibinfo {author} {\bibfnamefont
  {N.}~\bibnamefont {Steinbrink}}, \bibinfo {author} {\bibfnamefont
  {L.}~\bibnamefont {Strigari}}, \bibinfo {author} {\bibfnamefont
  {F.}~\bibnamefont {Suekane}}, \bibinfo {author} {\bibfnamefont
  {B.}~\bibnamefont {Suerfu}}, \bibinfo {author} {\bibfnamefont
  {R.}~\bibnamefont {Takahashi}}, \bibinfo {author} {\bibfnamefont {N.~T.~H.}\
  \bibnamefont {Van}}, \bibinfo {author} {\bibfnamefont {I.}~\bibnamefont
  {Tkachev}}, \bibinfo {author} {\bibfnamefont {M.}~\bibnamefont {Totzauer}},
  \bibinfo {author} {\bibfnamefont {Y.}~\bibnamefont {Tsai}}, \bibinfo {author}
  {\bibfnamefont {C.}~\bibnamefont {Tully}}, \bibinfo {author} {\bibfnamefont
  {K.}~\bibnamefont {Valerius}}, \bibinfo {author} {\bibfnamefont
  {J.}~\bibnamefont {Valle}}, \bibinfo {author} {\bibfnamefont
  {D.}~\bibnamefont {Venos}}, \bibinfo {author} {\bibfnamefont
  {M.}~\bibnamefont {Viel}}, \bibinfo {author} {\bibfnamefont {M.}~\bibnamefont
  {Vivier}}, \bibinfo {author} {\bibfnamefont {M.}~\bibnamefont {Wang}},
  \bibinfo {author} {\bibfnamefont {C.}~\bibnamefont {Weinheimer}}, \bibinfo
  {author} {\bibfnamefont {K.}~\bibnamefont {Wendt}}, \bibinfo {author}
  {\bibfnamefont {L.}~\bibnamefont {Winslow}}, \bibinfo {author} {\bibfnamefont
  {J.}~\bibnamefont {Wolf}}, \bibinfo {author} {\bibfnamefont {M.}~\bibnamefont
  {Wurm}}, \bibinfo {author} {\bibfnamefont {Z.}~\bibnamefont {Xing}}, \bibinfo
  {author} {\bibfnamefont {S.}~\bibnamefont {Zhou}},\ and\ \bibinfo {author}
  {\bibfnamefont {K.}~\bibnamefont {Zuber}},\ }\href
  {https://doi.org/10.1088/1475-7516/2017/01/025} {\bibfield  {journal}
  {\bibinfo  {journal} {Journal of Cosmology and Astroparticle Physics}\
  }\textbf {\bibinfo {volume} {2017}}\bibinfo  {number} { (01)},\ \bibinfo
  {pages} {025}}\BibitemShut {NoStop}%
\bibitem [{\citenamefont {Boyarsky}\ \emph {et~al.}(2019)\citenamefont
  {Boyarsky}, \citenamefont {Drewes}, \citenamefont {Lasserre}, \citenamefont
  {Mertens},\ and\ \citenamefont {Ruchayskiy}}]{Boy19}%
  \BibitemOpen
\bibfield  {number} {  }\bibfield  {author} {\bibinfo {author} {\bibfnamefont
  {A.}~\bibnamefont {Boyarsky}}, \bibinfo {author} {\bibfnamefont
  {M.}~\bibnamefont {Drewes}}, \bibinfo {author} {\bibfnamefont
  {T.}~\bibnamefont {Lasserre}}, \bibinfo {author} {\bibfnamefont
  {S.}~\bibnamefont {Mertens}},\ and\ \bibinfo {author} {\bibfnamefont
  {O.}~\bibnamefont {Ruchayskiy}},\ }\href
  {https://doi.org/https://doi.org/10.1016/j.ppnp.2018.07.004} {\bibfield
  {journal} {\bibinfo  {journal} {Progress in Particle and Nuclear Physics}\
  }\textbf {\bibinfo {volume} {104}},\ \bibinfo {pages} {1 } (\bibinfo {year}
  {2019})}\BibitemShut {NoStop}%
\bibitem [{\citenamefont {Leach}\ and\ \citenamefont
  {Friedrich}(2021)}]{Leach:2021bvh}%
  \BibitemOpen
  \bibfield  {author} {\bibinfo {author} {\bibfnamefont {K.~G.}\ \bibnamefont
  {Leach}}\ and\ \bibinfo {author} {\bibfnamefont {S.}~\bibnamefont
  {Friedrich}} (\bibinfo {collaboration} {BeEST}),\ }in\ \href
  {https://doi.org/10.1007/s10909-022-02759-z} {\emph {\bibinfo {booktitle}
  {{19th International Workshop on Low Temperature Detectors}}}}\ (\bibinfo
  {year} {2021})\ \Eprint {https://arxiv.org/abs/2112.02029} {arXiv:2112.02029
  [nucl-ex]} \BibitemShut {NoStop}%
\bibitem [{\citenamefont {Martoff}\ \emph {et~al.}(2021)\citenamefont {Martoff}
  \emph {et~al.}}]{Martoff:2021vxp}%
  \BibitemOpen
  \bibfield  {author} {\bibinfo {author} {\bibfnamefont {C.~J.}\ \bibnamefont
  {Martoff}} \emph {et~al.},\ }\href {https://doi.org/10.1088/2058-9565/abdb9b}
  {\bibfield  {journal} {\bibinfo  {journal} {Quantum Sci. Technol.}\ }\textbf
  {\bibinfo {volume} {6}},\ \bibinfo {pages} {024008} (\bibinfo {year}
  {2021})}\BibitemShut {NoStop}%
\bibitem [{\citenamefont {Fallot}\ \emph {et~al.}(2012)\citenamefont {Fallot},
  \citenamefont {Cormon}, \citenamefont {Estienne}, \citenamefont {Algora},
  \citenamefont {Bui}, \citenamefont {Cucoanes}, \citenamefont {Elnimr},
  \citenamefont {Giot}, \citenamefont {Jordan}, \citenamefont {Martino},
  \citenamefont {Onillon}, \citenamefont {Porta}, \citenamefont {Pronost},
  \citenamefont {Remoto}, \citenamefont {Ta\'{i}n}, \citenamefont {Yermia},\
  and\ \citenamefont {Zakari-Issoufou}}]{fallot2012}%
  \BibitemOpen
  \bibfield  {author} {\bibinfo {author} {\bibfnamefont {M.}~\bibnamefont
  {Fallot}}, \bibinfo {author} {\bibfnamefont {S.}~\bibnamefont {Cormon}},
  \bibinfo {author} {\bibfnamefont {M.}~\bibnamefont {Estienne}}, \bibinfo
  {author} {\bibfnamefont {A.}~\bibnamefont {Algora}}, \bibinfo {author}
  {\bibfnamefont {V.~M.}\ \bibnamefont {Bui}}, \bibinfo {author} {\bibfnamefont
  {A.}~\bibnamefont {Cucoanes}}, \bibinfo {author} {\bibfnamefont
  {M.}~\bibnamefont {Elnimr}}, \bibinfo {author} {\bibfnamefont
  {L.}~\bibnamefont {Giot}}, \bibinfo {author} {\bibfnamefont {D.}~\bibnamefont
  {Jordan}}, \bibinfo {author} {\bibfnamefont {J.}~\bibnamefont {Martino}},
  \bibinfo {author} {\bibfnamefont {A.}~\bibnamefont {Onillon}}, \bibinfo
  {author} {\bibfnamefont {A.}~\bibnamefont {Porta}}, \bibinfo {author}
  {\bibfnamefont {G.}~\bibnamefont {Pronost}}, \bibinfo {author} {\bibfnamefont
  {A.}~\bibnamefont {Remoto}}, \bibinfo {author} {\bibfnamefont {J.~L.}\
  \bibnamefont {Ta\'{i}n}}, \bibinfo {author} {\bibfnamefont {F.}~\bibnamefont
  {Yermia}},\ and\ \bibinfo {author} {\bibfnamefont {A.-A.}\ \bibnamefont
  {Zakari-Issoufou}},\ }\href {https://doi.org/10.1103/PhysRevLett.109.202504}
  {\bibfield  {journal} {\bibinfo  {journal} {Phys. Rev. Lett.}\ }\textbf
  {\bibinfo {volume} {109}},\ \bibinfo {pages} {202504} (\bibinfo {year}
  {2012})}\BibitemShut {NoStop}%
\bibitem [{\citenamefont {Hayes}\ \emph {et~al.}(2014)\citenamefont {Hayes},
  \citenamefont {Friar}, \citenamefont {Garvey}, \citenamefont {Jungman},\ and\
  \citenamefont {Jonkmans}}]{hayes2014}%
  \BibitemOpen
  \bibfield  {author} {\bibinfo {author} {\bibfnamefont {A.~C.}\ \bibnamefont
  {Hayes}}, \bibinfo {author} {\bibfnamefont {J.~L.}\ \bibnamefont {Friar}},
  \bibinfo {author} {\bibfnamefont {G.~T.}\ \bibnamefont {Garvey}}, \bibinfo
  {author} {\bibfnamefont {G.}~\bibnamefont {Jungman}},\ and\ \bibinfo {author}
  {\bibfnamefont {G.}~\bibnamefont {Jonkmans}},\ }\href
  {https://doi.org/10.1103/PhysRevLett.112.202501} {\bibfield  {journal}
  {\bibinfo  {journal} {Phys. Rev. Lett.}\ }\textbf {\bibinfo {volume} {112}},\
  \bibinfo {pages} {202501} (\bibinfo {year} {2014})}\BibitemShut {NoStop}%
\bibitem [{\citenamefont {Sonzogni}\ \emph {et~al.}(2015)\citenamefont
  {Sonzogni}, \citenamefont {Johnson},\ and\ \citenamefont
  {McCutchan}}]{sonzogni2015}%
  \BibitemOpen
  \bibfield  {author} {\bibinfo {author} {\bibfnamefont {A.~A.}\ \bibnamefont
  {Sonzogni}}, \bibinfo {author} {\bibfnamefont {T.~D.}\ \bibnamefont
  {Johnson}},\ and\ \bibinfo {author} {\bibfnamefont {E.~A.}\ \bibnamefont
  {McCutchan}},\ }\href {https://doi.org/10.1103/PhysRevC.91.011301} {\bibfield
   {journal} {\bibinfo  {journal} {Phys. Rev. C}\ }\textbf {\bibinfo {volume}
  {91}},\ \bibinfo {pages} {011301} (\bibinfo {year} {2015})}\BibitemShut
  {NoStop}%
\bibitem [{\citenamefont {Rasco}\ \emph {et~al.}(2016)\citenamefont {Rasco},
  \citenamefont {Woli\'{n}ska-Cichocka}, \citenamefont {Fija\l{}kowska},
  \citenamefont {Rykaczewski}, \citenamefont {Karny}, \citenamefont {Grzywacz},
  \citenamefont {Goetz}, \citenamefont {Gross}, \citenamefont {Stracener},
  \citenamefont {Zganjar}, \citenamefont {Batchelder}, \citenamefont
  {Blackmon}, \citenamefont {Brewer}, \citenamefont {Go}, \citenamefont
  {Heffron}, \citenamefont {King}, \citenamefont {Matta}, \citenamefont
  {Miernik}, \citenamefont {Nesaraja}, \citenamefont {Paulauskas},
  \citenamefont {Rajabali}, \citenamefont {Wang}, \citenamefont {Winger},
  \citenamefont {Xiao},\ and\ \citenamefont {Zachary}}]{rasco2016}%
  \BibitemOpen
  \bibfield  {author} {\bibinfo {author} {\bibfnamefont {B.~C.}\ \bibnamefont
  {Rasco}}, \bibinfo {author} {\bibfnamefont {M.}~\bibnamefont
  {Woli\'{n}ska-Cichocka}}, \bibinfo {author} {\bibfnamefont {A.}~\bibnamefont
  {Fija\l{}kowska}}, \bibinfo {author} {\bibfnamefont {K.~P.}\ \bibnamefont
  {Rykaczewski}}, \bibinfo {author} {\bibfnamefont {M.}~\bibnamefont {Karny}},
  \bibinfo {author} {\bibfnamefont {R.~K.}\ \bibnamefont {Grzywacz}}, \bibinfo
  {author} {\bibfnamefont {K.~C.}\ \bibnamefont {Goetz}}, \bibinfo {author}
  {\bibfnamefont {C.~J.}\ \bibnamefont {Gross}}, \bibinfo {author}
  {\bibfnamefont {D.~W.}\ \bibnamefont {Stracener}}, \bibinfo {author}
  {\bibfnamefont {E.~F.}\ \bibnamefont {Zganjar}}, \bibinfo {author}
  {\bibfnamefont {J.~C.}\ \bibnamefont {Batchelder}}, \bibinfo {author}
  {\bibfnamefont {J.~C.}\ \bibnamefont {Blackmon}}, \bibinfo {author}
  {\bibfnamefont {N.~T.}\ \bibnamefont {Brewer}}, \bibinfo {author}
  {\bibfnamefont {S.}~\bibnamefont {Go}}, \bibinfo {author} {\bibfnamefont
  {B.}~\bibnamefont {Heffron}}, \bibinfo {author} {\bibfnamefont
  {T.}~\bibnamefont {King}}, \bibinfo {author} {\bibfnamefont {J.~T.}\
  \bibnamefont {Matta}}, \bibinfo {author} {\bibfnamefont {K.}~\bibnamefont
  {Miernik}}, \bibinfo {author} {\bibfnamefont {C.~D.}\ \bibnamefont
  {Nesaraja}}, \bibinfo {author} {\bibfnamefont {S.~V.}\ \bibnamefont
  {Paulauskas}}, \bibinfo {author} {\bibfnamefont {M.~M.}\ \bibnamefont
  {Rajabali}}, \bibinfo {author} {\bibfnamefont {E.~H.}\ \bibnamefont {Wang}},
  \bibinfo {author} {\bibfnamefont {J.~A.}\ \bibnamefont {Winger}}, \bibinfo
  {author} {\bibfnamefont {Y.}~\bibnamefont {Xiao}},\ and\ \bibinfo {author}
  {\bibfnamefont {C.~J.}\ \bibnamefont {Zachary}},\ }\href
  {https://doi.org/10.1103/PhysRevLett.117.092501} {\bibfield  {journal}
  {\bibinfo  {journal} {Phys. Rev. Lett.}\ }\textbf {\bibinfo {volume} {117}},\
  \bibinfo {pages} {092501} (\bibinfo {year} {2016})}\BibitemShut {NoStop}%
\bibitem [{\citenamefont {Fija\l{}kowska}\ \emph {et~al.}(2017)\citenamefont
  {Fija\l{}kowska}, \citenamefont {Karny}, \citenamefont {Rykaczewski},
  \citenamefont {Rasco}, \citenamefont {Grzywacz}, \citenamefont {Gross},
  \citenamefont {Woli\ifmmode \acute{n}\else~\'{n}\fi{}ska Cichocka},
  \citenamefont {Goetz}, \citenamefont {Stracener}, \citenamefont {Bielewski},
  \citenamefont {Goans}, \citenamefont {Hamilton}, \citenamefont {Johnson},
  \citenamefont {Jost}, \citenamefont {Madurga}, \citenamefont {Miernik},
  \citenamefont {Miller}, \citenamefont {Padgett}, \citenamefont {Paulauskas},
  \citenamefont {Ramayya},\ and\ \citenamefont {Zganjar}}]{fijalkowska2017}%
  \BibitemOpen
  \bibfield  {author} {\bibinfo {author} {\bibfnamefont {A.}~\bibnamefont
  {Fija\l{}kowska}}, \bibinfo {author} {\bibfnamefont {M.}~\bibnamefont
  {Karny}}, \bibinfo {author} {\bibfnamefont {K.~P.}\ \bibnamefont
  {Rykaczewski}}, \bibinfo {author} {\bibfnamefont {B.}~\bibnamefont {Rasco}},
  \bibinfo {author} {\bibfnamefont {R.}~\bibnamefont {Grzywacz}}, \bibinfo
  {author} {\bibfnamefont {C.~J.}\ \bibnamefont {Gross}}, \bibinfo {author}
  {\bibfnamefont {M.}~\bibnamefont {Woli\ifmmode \acute{n}\else~\'{n}\fi{}ska
  Cichocka}}, \bibinfo {author} {\bibfnamefont {K.~C.}\ \bibnamefont {Goetz}},
  \bibinfo {author} {\bibfnamefont {D.~W.}\ \bibnamefont {Stracener}}, \bibinfo
  {author} {\bibfnamefont {W.}~\bibnamefont {Bielewski}}, \bibinfo {author}
  {\bibfnamefont {R.}~\bibnamefont {Goans}}, \bibinfo {author} {\bibfnamefont
  {J.~H.}\ \bibnamefont {Hamilton}}, \bibinfo {author} {\bibfnamefont {J.~W.}\
  \bibnamefont {Johnson}}, \bibinfo {author} {\bibfnamefont {C.}~\bibnamefont
  {Jost}}, \bibinfo {author} {\bibfnamefont {M.}~\bibnamefont {Madurga}},
  \bibinfo {author} {\bibfnamefont {K.}~\bibnamefont {Miernik}}, \bibinfo
  {author} {\bibfnamefont {D.}~\bibnamefont {Miller}}, \bibinfo {author}
  {\bibfnamefont {S.~W.}\ \bibnamefont {Padgett}}, \bibinfo {author}
  {\bibfnamefont {S.~V.}\ \bibnamefont {Paulauskas}}, \bibinfo {author}
  {\bibfnamefont {A.~V.}\ \bibnamefont {Ramayya}},\ and\ \bibinfo {author}
  {\bibfnamefont {E.~F.}\ \bibnamefont {Zganjar}},\ }\href
  {https://doi.org/10.1103/PhysRevLett.119.052503} {\bibfield  {journal}
  {\bibinfo  {journal} {Phys. Rev. Lett.}\ }\textbf {\bibinfo {volume} {119}},\
  \bibinfo {pages} {052503} (\bibinfo {year} {2017})}\BibitemShut {NoStop}%
\bibitem [{\citenamefont {Estienne}\ \emph {et~al.}(2019)\citenamefont
  {Estienne}, \citenamefont {Fallot}, \citenamefont {Algora}, \citenamefont
  {Briz-Monago}, \citenamefont {Bui}, \citenamefont {Cormon}, \citenamefont
  {Gelletly}, \citenamefont {Giot}, \citenamefont {Guadilla}, \citenamefont
  {Jordan}, \citenamefont {Le~Meur}, \citenamefont {Porta}, \citenamefont
  {Rice}, \citenamefont {Rubio}, \citenamefont {Ta\'{\i}n}, \citenamefont
  {Valencia},\ and\ \citenamefont {Zakari-Issoufou}}]{estienne2019}%
  \BibitemOpen
  \bibfield  {author} {\bibinfo {author} {\bibfnamefont {M.}~\bibnamefont
  {Estienne}}, \bibinfo {author} {\bibfnamefont {M.}~\bibnamefont {Fallot}},
  \bibinfo {author} {\bibfnamefont {A.}~\bibnamefont {Algora}}, \bibinfo
  {author} {\bibfnamefont {J.}~\bibnamefont {Briz-Monago}}, \bibinfo {author}
  {\bibfnamefont {V.~M.}\ \bibnamefont {Bui}}, \bibinfo {author} {\bibfnamefont
  {S.}~\bibnamefont {Cormon}}, \bibinfo {author} {\bibfnamefont
  {W.}~\bibnamefont {Gelletly}}, \bibinfo {author} {\bibfnamefont
  {L.}~\bibnamefont {Giot}}, \bibinfo {author} {\bibfnamefont {V.}~\bibnamefont
  {Guadilla}}, \bibinfo {author} {\bibfnamefont {D.}~\bibnamefont {Jordan}},
  \bibinfo {author} {\bibfnamefont {L.}~\bibnamefont {Le~Meur}}, \bibinfo
  {author} {\bibfnamefont {A.}~\bibnamefont {Porta}}, \bibinfo {author}
  {\bibfnamefont {S.}~\bibnamefont {Rice}}, \bibinfo {author} {\bibfnamefont
  {B.}~\bibnamefont {Rubio}}, \bibinfo {author} {\bibfnamefont {J.~L.}\
  \bibnamefont {Ta\'{\i}n}}, \bibinfo {author} {\bibfnamefont {E.}~\bibnamefont
  {Valencia}},\ and\ \bibinfo {author} {\bibfnamefont {A.-A.}\ \bibnamefont
  {Zakari-Issoufou}},\ }\href {https://doi.org/10.1103/PhysRevLett.123.022502}
  {\bibfield  {journal} {\bibinfo  {journal} {Phys. Rev. Lett.}\ }\textbf
  {\bibinfo {volume} {123}},\ \bibinfo {pages} {022502} (\bibinfo {year}
  {2019})}\BibitemShut {NoStop}%
\bibitem [{\citenamefont {Hayen}\ \emph {et~al.}(2019)\citenamefont {Hayen},
  \citenamefont {Kostensalo}, \citenamefont {Severijns},\ and\ \citenamefont
  {Suhonen}}]{hayen2019}%
  \BibitemOpen
  \bibfield  {author} {\bibinfo {author} {\bibfnamefont {L.}~\bibnamefont
  {Hayen}}, \bibinfo {author} {\bibfnamefont {J.}~\bibnamefont {Kostensalo}},
  \bibinfo {author} {\bibfnamefont {N.}~\bibnamefont {Severijns}},\ and\
  \bibinfo {author} {\bibfnamefont {J.}~\bibnamefont {Suhonen}},\ }\href
  {https://doi.org/10.1103/PhysRevC.100.054323} {\bibfield  {journal} {\bibinfo
   {journal} {Phys. Rev. C}\ }\textbf {\bibinfo {volume} {100}},\ \bibinfo
  {pages} {054323} (\bibinfo {year} {2019})}\BibitemShut {NoStop}%
\bibitem [{\citenamefont {Stukel}\ \emph {et~al.}(2022)\citenamefont {Stukel},
  \citenamefont {Hariasz}, \citenamefont {Di~Stefano}, \citenamefont {Rasco},
  \citenamefont {Rykaczewski}, \citenamefont {Brewer}, \citenamefont
  {Stracener}, \citenamefont {Liu}, \citenamefont {Gai}, \citenamefont
  {Rouleau}, \citenamefont {Carter}, \citenamefont {Kostensalo}, \citenamefont
  {Suhonen}, \citenamefont {Davis}, \citenamefont {Lukosi}, \citenamefont
  {Goetz}, \citenamefont {Grzywacz}, \citenamefont {Mancuso}, \citenamefont
  {Petricca}, \citenamefont {Fija\l{}kowska}, \citenamefont
  {Woli\'{n}ska-Cichocka}, \citenamefont {Ninkovic}, \citenamefont {Lechner},
  \citenamefont {Ickert}, \citenamefont {Morgan}, \citenamefont {Renne},\ and\
  \citenamefont {Yavin}}]{stuckel2022}%
  \BibitemOpen
  \bibfield  {author} {\bibinfo {author} {\bibfnamefont {M.}~\bibnamefont
  {Stukel}}, \bibinfo {author} {\bibfnamefont {L.}~\bibnamefont {Hariasz}},
  \bibinfo {author} {\bibfnamefont {P.~C.~F.}\ \bibnamefont {Di~Stefano}},
  \bibinfo {author} {\bibfnamefont {B.~C.}\ \bibnamefont {Rasco}}, \bibinfo
  {author} {\bibfnamefont {K.~P.}\ \bibnamefont {Rykaczewski}}, \bibinfo
  {author} {\bibfnamefont {N.~T.}\ \bibnamefont {Brewer}}, \bibinfo {author}
  {\bibfnamefont {D.~W.}\ \bibnamefont {Stracener}}, \bibinfo {author}
  {\bibfnamefont {Y.}~\bibnamefont {Liu}}, \bibinfo {author} {\bibfnamefont
  {Z.}~\bibnamefont {Gai}}, \bibinfo {author} {\bibfnamefont {C.}~\bibnamefont
  {Rouleau}}, \bibinfo {author} {\bibfnamefont {J.}~\bibnamefont {Carter}},
  \bibinfo {author} {\bibfnamefont {J.}~\bibnamefont {Kostensalo}}, \bibinfo
  {author} {\bibfnamefont {J.}~\bibnamefont {Suhonen}}, \bibinfo {author}
  {\bibfnamefont {H.}~\bibnamefont {Davis}}, \bibinfo {author} {\bibfnamefont
  {E.~D.}\ \bibnamefont {Lukosi}}, \bibinfo {author} {\bibfnamefont {K.~C.}\
  \bibnamefont {Goetz}}, \bibinfo {author} {\bibfnamefont {R.~K.}\ \bibnamefont
  {Grzywacz}}, \bibinfo {author} {\bibfnamefont {M.}~\bibnamefont {Mancuso}},
  \bibinfo {author} {\bibfnamefont {F.}~\bibnamefont {Petricca}}, \bibinfo
  {author} {\bibfnamefont {A.}~\bibnamefont {Fija\l{}kowska}}, \bibinfo
  {author} {\bibfnamefont {M.}~\bibnamefont {Woli\'{n}ska-Cichocka}}, \bibinfo
  {author} {\bibfnamefont {J.}~\bibnamefont {Ninkovic}}, \bibinfo {author}
  {\bibfnamefont {P.}~\bibnamefont {Lechner}}, \bibinfo {author} {\bibfnamefont
  {R.~B.}\ \bibnamefont {Ickert}}, \bibinfo {author} {\bibfnamefont {L.~E.}\
  \bibnamefont {Morgan}}, \bibinfo {author} {\bibfnamefont {P.~R.}\
  \bibnamefont {Renne}},\ and\ \bibinfo {author} {\bibfnamefont
  {I.}~\bibnamefont {Yavin}},\ }\href
  {https://doi.org/10.48550/ARXIV.2211.10319} {\bibinfo {title} {Rare $^{40}$k
  decay with implications for fundamental physics and geochronology}} (\bibinfo
  {year} {2022})\BibitemShut {NoStop}%
\bibitem [{\citenamefont {Horoi}\ and\ \citenamefont
  {Neacsu}(2018)}]{HoroiNeacsu-prc18}%
  \BibitemOpen
  \bibfield  {author} {\bibinfo {author} {\bibfnamefont {M.}~\bibnamefont
  {Horoi}}\ and\ \bibinfo {author} {\bibfnamefont {A.}~\bibnamefont {Neacsu}},\
  }\href {https://doi.org/10.1103/PhysRevC.98.035502} {\bibfield  {journal}
  {\bibinfo  {journal} {Phys. Rev. C}\ }\textbf {\bibinfo {volume} {98}},\
  \bibinfo {pages} {035502} (\bibinfo {year} {2018})}\BibitemShut {NoStop}%
\bibitem [{\citenamefont {Horoi}\ and\ \citenamefont
  {Neacsu}(2016)}]{Horoi-prd16}%
  \BibitemOpen
  \bibfield  {author} {\bibinfo {author} {\bibfnamefont {M.}~\bibnamefont
  {Horoi}}\ and\ \bibinfo {author} {\bibfnamefont {A.}~\bibnamefont {Neacsu}},\
  }\href {https://doi.org/10.1103/PhysRevD.93.113014} {\bibfield  {journal}
  {\bibinfo  {journal} {Phys. Rev. D}\ }\textbf {\bibinfo {volume} {93}},\
  \bibinfo {pages} {113014} (\bibinfo {year} {2016})}\BibitemShut {NoStop}%
\bibitem [{\citenamefont {Sen'kov}\ and\ \citenamefont
  {Horoi}(2016)}]{SenkovHoroi-prc16}%
  \BibitemOpen
  \bibfield  {author} {\bibinfo {author} {\bibfnamefont {R.~A.}\ \bibnamefont
  {Sen'kov}}\ and\ \bibinfo {author} {\bibfnamefont {M.}~\bibnamefont
  {Horoi}},\ }\href {https://doi.org/10.1103/PhysRevC.93.044334} {\bibfield
  {journal} {\bibinfo  {journal} {Phys. Rev. C}\ }\textbf {\bibinfo {volume}
  {93}},\ \bibinfo {pages} {044334} (\bibinfo {year} {2016})}\BibitemShut
  {NoStop}%
\bibitem [{\citenamefont {Brown}\ \emph {et~al.}(2015)\citenamefont {Brown},
  \citenamefont {Fang},\ and\ \citenamefont {Horoi}}]{BrownFangHoroi2015}%
  \BibitemOpen
  \bibfield  {author} {\bibinfo {author} {\bibfnamefont {B.~A.}\ \bibnamefont
  {Brown}}, \bibinfo {author} {\bibfnamefont {D.~L.}\ \bibnamefont {Fang}},\
  and\ \bibinfo {author} {\bibfnamefont {M.}~\bibnamefont {Horoi}},\ }\href
  {https://doi.org/10.1103/PhysRevC.92.041301} {\bibfield  {journal} {\bibinfo
  {journal} {Phys. Rev. C}\ }\textbf {\bibinfo {volume} {92}},\ \bibinfo
  {pages} {041301} (\bibinfo {year} {2015})}\BibitemShut {NoStop}%
\bibitem [{\citenamefont {Ramalho}\ \emph
  {et~al.}(2022{\natexlab{b}})\citenamefont {Ramalho}, \citenamefont {Suhonen},
  \citenamefont {Kostensalo}, \citenamefont {Alcal\'a}, \citenamefont {Algora},
  \citenamefont {Fallot}, \citenamefont {Porta},\ and\ \citenamefont
  {Zakari-Issoufou}}]{ramalho2022}%
  \BibitemOpen
  \bibfield  {author} {\bibinfo {author} {\bibfnamefont {M.}~\bibnamefont
  {Ramalho}}, \bibinfo {author} {\bibfnamefont {J.}~\bibnamefont {Suhonen}},
  \bibinfo {author} {\bibfnamefont {J.}~\bibnamefont {Kostensalo}}, \bibinfo
  {author} {\bibfnamefont {G.~A.}\ \bibnamefont {Alcal\'a}}, \bibinfo {author}
  {\bibfnamefont {A.}~\bibnamefont {Algora}}, \bibinfo {author} {\bibfnamefont
  {M.}~\bibnamefont {Fallot}}, \bibinfo {author} {\bibfnamefont
  {A.}~\bibnamefont {Porta}},\ and\ \bibinfo {author} {\bibfnamefont {A.-A.}\
  \bibnamefont {Zakari-Issoufou}},\ }\href
  {https://doi.org/10.1103/PhysRevC.106.024315} {\bibfield  {journal} {\bibinfo
   {journal} {Phys. Rev. C}\ }\textbf {\bibinfo {volume} {106}},\ \bibinfo
  {pages} {024315} (\bibinfo {year} {2022}{\natexlab{b}})}\BibitemShut
  {NoStop}%
\bibitem [{\citenamefont {Fallot}\ \emph {et~al.}(2019)\citenamefont {Fallot},
  \citenamefont {Littlejohn},\ and\ \citenamefont {Dimitriou}}]{Fallot2019}%
  \BibitemOpen
  \bibfield  {author} {\bibinfo {author} {\bibfnamefont {M.}~\bibnamefont
  {Fallot}}, \bibinfo {author} {\bibfnamefont {B.}~\bibnamefont {Littlejohn}},\
  and\ \bibinfo {author} {\bibfnamefont {P.}~\bibnamefont {Dimitriou}},\ }\href
  {https://www-nds.iaea.org/publications/indc/indc-nds-0786.pdf} {\emph
  {\bibinfo {title} {{Antineutrino spectra and their applications}}}},\
  \bibinfo {type} {Tech. Rep.}\ \bibinfo {number} {INDC(NDS)-0786}\ (\bibinfo
  {institution} {International Atomic Energy Agency},\ \bibinfo {year}
  {2019})\BibitemShut {NoStop}%
\bibitem [{\citenamefont {Fornal}\ and\ \citenamefont
  {Grinstein}(2018)}]{Fornal2018}%
  \BibitemOpen
  \bibfield  {author} {\bibinfo {author} {\bibfnamefont {B.}~\bibnamefont
  {Fornal}}\ and\ \bibinfo {author} {\bibfnamefont {B.}~\bibnamefont
  {Grinstein}},\ }\href {https://doi.org/10.1103/PhysRevLett.120.191801}
  {\bibfield  {journal} {\bibinfo  {journal} {Phys. Rev. Lett.}\ }\textbf
  {\bibinfo {volume} {120}},\ \bibinfo {pages} {191801} (\bibinfo {year}
  {2018})}\BibitemShut {NoStop}%
\bibitem [{\citenamefont {Berryman}\ \emph {et~al.}(2022)\citenamefont
  {Berryman}, \citenamefont {Gardner},\ and\ \citenamefont
  {Zakeri}}]{Berryman2022}%
  \BibitemOpen
  \bibfield  {author} {\bibinfo {author} {\bibfnamefont {J.~M.}\ \bibnamefont
  {Berryman}}, \bibinfo {author} {\bibfnamefont {S.}~\bibnamefont {Gardner}},\
  and\ \bibinfo {author} {\bibfnamefont {M.}~\bibnamefont {Zakeri}},\
  }\bibfield  {journal} {\bibinfo  {journal} {Symmetry}\ }\textbf {\bibinfo
  {volume} {14}},\ \href {https://doi.org/10.3390/sym14030518}
  {10.3390/sym14030518} (\bibinfo {year} {2022})\BibitemShut {NoStop}%
\bibitem [{\citenamefont {Pf\"utzner}\ and\ \citenamefont
  {Riisager}(2018)}]{Pfutzner2018}%
  \BibitemOpen
  \bibfield  {author} {\bibinfo {author} {\bibfnamefont {M.}~\bibnamefont
  {Pf\"utzner}}\ and\ \bibinfo {author} {\bibfnamefont {K.}~\bibnamefont
  {Riisager}},\ }\href {https://doi.org/10.1103/PhysRevC.97.042501} {\bibfield
  {journal} {\bibinfo  {journal} {Phys. Rev. C}\ }\textbf {\bibinfo {volume}
  {97}},\ \bibinfo {pages} {042501} (\bibinfo {year} {2018})}\BibitemShut
  {NoStop}%
\bibitem [{\citenamefont {Ayyad}\ \emph {et~al.}(2019)\citenamefont {Ayyad},
  \citenamefont {Olaizola}, \citenamefont {Mittig}, \citenamefont {Potel},
  \citenamefont {Zelevinsky}, \citenamefont {Horoi}, \citenamefont
  {Beceiro-Novo}, \citenamefont {Alcorta}, \citenamefont {Andreoiu},
  \citenamefont {Ahn}, \citenamefont {Anholm}, \citenamefont {Atar},
  \citenamefont {Babu}, \citenamefont {Bazin}, \citenamefont {Bernier},
  \citenamefont {Bhattacharjee}, \citenamefont {Bowry}, \citenamefont
  {Caballero-Folch}, \citenamefont {Cortesi}, \citenamefont {Dalitz},
  \citenamefont {Dunling}, \citenamefont {Garnsworthy}, \citenamefont {Holl},
  \citenamefont {Kootte}, \citenamefont {Leach}, \citenamefont {Randhawa},
  \citenamefont {Saito}, \citenamefont {Santamaria}, \citenamefont {\ifmmode
  \check{S}\else \v{S}\fi{}iuryt\ifmmode~\dot{e}\else \.{e}\fi{}},
  \citenamefont {Svensson}, \citenamefont {Umashankar}, \citenamefont
  {Watwood},\ and\ \citenamefont {Yates}}]{Ayyad2019}%
  \BibitemOpen
  \bibfield  {author} {\bibinfo {author} {\bibfnamefont {Y.}~\bibnamefont
  {Ayyad}}, \bibinfo {author} {\bibfnamefont {B.}~\bibnamefont {Olaizola}},
  \bibinfo {author} {\bibfnamefont {W.}~\bibnamefont {Mittig}}, \bibinfo
  {author} {\bibfnamefont {G.}~\bibnamefont {Potel}}, \bibinfo {author}
  {\bibfnamefont {V.}~\bibnamefont {Zelevinsky}}, \bibinfo {author}
  {\bibfnamefont {M.}~\bibnamefont {Horoi}}, \bibinfo {author} {\bibfnamefont
  {S.}~\bibnamefont {Beceiro-Novo}}, \bibinfo {author} {\bibfnamefont
  {M.}~\bibnamefont {Alcorta}}, \bibinfo {author} {\bibfnamefont
  {C.}~\bibnamefont {Andreoiu}}, \bibinfo {author} {\bibfnamefont
  {T.}~\bibnamefont {Ahn}}, \bibinfo {author} {\bibfnamefont {M.}~\bibnamefont
  {Anholm}}, \bibinfo {author} {\bibfnamefont {L.}~\bibnamefont {Atar}},
  \bibinfo {author} {\bibfnamefont {A.}~\bibnamefont {Babu}}, \bibinfo {author}
  {\bibfnamefont {D.}~\bibnamefont {Bazin}}, \bibinfo {author} {\bibfnamefont
  {N.}~\bibnamefont {Bernier}}, \bibinfo {author} {\bibfnamefont {S.~S.}\
  \bibnamefont {Bhattacharjee}}, \bibinfo {author} {\bibfnamefont
  {M.}~\bibnamefont {Bowry}}, \bibinfo {author} {\bibfnamefont
  {R.}~\bibnamefont {Caballero-Folch}}, \bibinfo {author} {\bibfnamefont
  {M.}~\bibnamefont {Cortesi}}, \bibinfo {author} {\bibfnamefont
  {C.}~\bibnamefont {Dalitz}}, \bibinfo {author} {\bibfnamefont
  {E.}~\bibnamefont {Dunling}}, \bibinfo {author} {\bibfnamefont {A.~B.}\
  \bibnamefont {Garnsworthy}}, \bibinfo {author} {\bibfnamefont
  {M.}~\bibnamefont {Holl}}, \bibinfo {author} {\bibfnamefont {B.}~\bibnamefont
  {Kootte}}, \bibinfo {author} {\bibfnamefont {K.~G.}\ \bibnamefont {Leach}},
  \bibinfo {author} {\bibfnamefont {J.~S.}\ \bibnamefont {Randhawa}}, \bibinfo
  {author} {\bibfnamefont {Y.}~\bibnamefont {Saito}}, \bibinfo {author}
  {\bibfnamefont {C.}~\bibnamefont {Santamaria}}, \bibinfo {author}
  {\bibfnamefont {P.}~\bibnamefont {\ifmmode \check{S}\else
  \v{S}\fi{}iuryt\ifmmode~\dot{e}\else \.{e}\fi{}}}, \bibinfo {author}
  {\bibfnamefont {C.~E.}\ \bibnamefont {Svensson}}, \bibinfo {author}
  {\bibfnamefont {R.}~\bibnamefont {Umashankar}}, \bibinfo {author}
  {\bibfnamefont {N.}~\bibnamefont {Watwood}},\ and\ \bibinfo {author}
  {\bibfnamefont {D.}~\bibnamefont {Yates}},\ }\href
  {https://doi.org/10.1103/PhysRevLett.123.082501} {\bibfield  {journal}
  {\bibinfo  {journal} {Phys. Rev. Lett.}\ }\textbf {\bibinfo {volume} {123}},\
  \bibinfo {pages} {082501} (\bibinfo {year} {2019})}\BibitemShut {NoStop}%
\bibitem [{\citenamefont {Ayyad}\ \emph {et~al.}(2022)\citenamefont {Ayyad},
  \citenamefont {Mittig}, \citenamefont {Tang}, \citenamefont {Olaizola},
  \citenamefont {Potel}, \citenamefont {Rijal}, \citenamefont {Watwood},
  \citenamefont {Alvarez-Pol}, \citenamefont {Bazin}, \citenamefont
  {Caama\~no}, \citenamefont {Chen}, \citenamefont {Cortesi}, \citenamefont
  {Fern\'andez-Dom\'{\i}nguez}, \citenamefont {Giraud}, \citenamefont {Gueye},
  \citenamefont {Heinitz}, \citenamefont {Jain}, \citenamefont {Kay},
  \citenamefont {Maugeri}, \citenamefont {Monteagudo}, \citenamefont
  {Ndayisabye}, \citenamefont {Paneru}, \citenamefont {Pereira}, \citenamefont
  {Rubino}, \citenamefont {Santamaria}, \citenamefont {Schumann}, \citenamefont
  {Surbrook}, \citenamefont {Wagner}, \citenamefont {Zamora},\ and\
  \citenamefont {Zelevinsky}}]{Ayyad2022}%
  \BibitemOpen
  \bibfield  {author} {\bibinfo {author} {\bibfnamefont {Y.}~\bibnamefont
  {Ayyad}}, \bibinfo {author} {\bibfnamefont {W.}~\bibnamefont {Mittig}},
  \bibinfo {author} {\bibfnamefont {T.}~\bibnamefont {Tang}}, \bibinfo {author}
  {\bibfnamefont {B.}~\bibnamefont {Olaizola}}, \bibinfo {author}
  {\bibfnamefont {G.}~\bibnamefont {Potel}}, \bibinfo {author} {\bibfnamefont
  {N.}~\bibnamefont {Rijal}}, \bibinfo {author} {\bibfnamefont
  {N.}~\bibnamefont {Watwood}}, \bibinfo {author} {\bibfnamefont
  {H.}~\bibnamefont {Alvarez-Pol}}, \bibinfo {author} {\bibfnamefont
  {D.}~\bibnamefont {Bazin}}, \bibinfo {author} {\bibfnamefont
  {M.}~\bibnamefont {Caama\~no}}, \bibinfo {author} {\bibfnamefont
  {J.}~\bibnamefont {Chen}}, \bibinfo {author} {\bibfnamefont {M.}~\bibnamefont
  {Cortesi}}, \bibinfo {author} {\bibfnamefont {B.}~\bibnamefont
  {Fern\'andez-Dom\'{\i}nguez}}, \bibinfo {author} {\bibfnamefont
  {S.}~\bibnamefont {Giraud}}, \bibinfo {author} {\bibfnamefont
  {P.}~\bibnamefont {Gueye}}, \bibinfo {author} {\bibfnamefont
  {S.}~\bibnamefont {Heinitz}}, \bibinfo {author} {\bibfnamefont
  {R.}~\bibnamefont {Jain}}, \bibinfo {author} {\bibfnamefont {B.~P.}\
  \bibnamefont {Kay}}, \bibinfo {author} {\bibfnamefont {E.~A.}\ \bibnamefont
  {Maugeri}}, \bibinfo {author} {\bibfnamefont {B.}~\bibnamefont {Monteagudo}},
  \bibinfo {author} {\bibfnamefont {F.}~\bibnamefont {Ndayisabye}}, \bibinfo
  {author} {\bibfnamefont {S.~N.}\ \bibnamefont {Paneru}}, \bibinfo {author}
  {\bibfnamefont {J.}~\bibnamefont {Pereira}}, \bibinfo {author} {\bibfnamefont
  {E.}~\bibnamefont {Rubino}}, \bibinfo {author} {\bibfnamefont
  {C.}~\bibnamefont {Santamaria}}, \bibinfo {author} {\bibfnamefont
  {D.}~\bibnamefont {Schumann}}, \bibinfo {author} {\bibfnamefont
  {J.}~\bibnamefont {Surbrook}}, \bibinfo {author} {\bibfnamefont
  {L.}~\bibnamefont {Wagner}}, \bibinfo {author} {\bibfnamefont {J.~C.}\
  \bibnamefont {Zamora}},\ and\ \bibinfo {author} {\bibfnamefont
  {V.}~\bibnamefont {Zelevinsky}},\ }\href
  {https://doi.org/10.1103/PhysRevLett.129.012501} {\bibfield  {journal}
  {\bibinfo  {journal} {Phys. Rev. Lett.}\ }\textbf {\bibinfo {volume} {129}},\
  \bibinfo {pages} {012501} (\bibinfo {year} {2022})}\BibitemShut {NoStop}%
\bibitem [{\citenamefont {Lopez-Saavedra}\ \emph {et~al.}(2022)\citenamefont
  {Lopez-Saavedra}, \citenamefont {Almaraz-Calderon}, \citenamefont {Asher},
  \citenamefont {Baby}, \citenamefont {Gerken}, \citenamefont {Hanselman},
  \citenamefont {Kemper}, \citenamefont {Kuchera}, \citenamefont {Morelock},
  \citenamefont {Perello}, \citenamefont {Temanson}, \citenamefont {Volya},\
  and\ \citenamefont {Wiedenh\"over}}]{Lopez2022}%
  \BibitemOpen
  \bibfield  {author} {\bibinfo {author} {\bibfnamefont {E.}~\bibnamefont
  {Lopez-Saavedra}}, \bibinfo {author} {\bibfnamefont {S.}~\bibnamefont
  {Almaraz-Calderon}}, \bibinfo {author} {\bibfnamefont {B.~W.}\ \bibnamefont
  {Asher}}, \bibinfo {author} {\bibfnamefont {L.~T.}\ \bibnamefont {Baby}},
  \bibinfo {author} {\bibfnamefont {N.}~\bibnamefont {Gerken}}, \bibinfo
  {author} {\bibfnamefont {K.}~\bibnamefont {Hanselman}}, \bibinfo {author}
  {\bibfnamefont {K.~W.}\ \bibnamefont {Kemper}}, \bibinfo {author}
  {\bibfnamefont {A.~N.}\ \bibnamefont {Kuchera}}, \bibinfo {author}
  {\bibfnamefont {A.~B.}\ \bibnamefont {Morelock}}, \bibinfo {author}
  {\bibfnamefont {J.~F.}\ \bibnamefont {Perello}}, \bibinfo {author}
  {\bibfnamefont {E.~S.}\ \bibnamefont {Temanson}}, \bibinfo {author}
  {\bibfnamefont {A.}~\bibnamefont {Volya}},\ and\ \bibinfo {author}
  {\bibfnamefont {I.}~\bibnamefont {Wiedenh\"over}},\ }\href
  {https://doi.org/10.1103/PhysRevLett.129.012502} {\bibfield  {journal}
  {\bibinfo  {journal} {Phys. Rev. Lett.}\ }\textbf {\bibinfo {volume} {129}},\
  \bibinfo {pages} {012502} (\bibinfo {year} {2022})}\BibitemShut {NoStop}%
\bibitem [{\citenamefont {Riisager}\ \emph {et~al.}(2020)\citenamefont
  {Riisager}, \citenamefont {Borge}, \citenamefont {Briz}, \citenamefont
  {Carmona-Gallardo}, \citenamefont {Forstner}, \citenamefont {Fraile},
  \citenamefont {Fynbo}, \citenamefont {Camacho}, \citenamefont {Johansen},
  \citenamefont {Jonson}, \citenamefont {Lund}, \citenamefont {Lachner},
  \citenamefont {Madurga}, \citenamefont {Merchel}, \citenamefont {Nacher},
  \citenamefont {Nilsson}, \citenamefont {Steier}, \citenamefont {Tengblad},\
  and\ \citenamefont {Vedia}}]{Riisager2020}%
  \BibitemOpen
  \bibfield  {author} {\bibinfo {author} {\bibfnamefont {K.}~\bibnamefont
  {Riisager}}, \bibinfo {author} {\bibfnamefont {M.~J.~G.}\ \bibnamefont
  {Borge}}, \bibinfo {author} {\bibfnamefont {J.~A.}\ \bibnamefont {Briz}},
  \bibinfo {author} {\bibfnamefont {M.}~\bibnamefont {Carmona-Gallardo}},
  \bibinfo {author} {\bibfnamefont {O.}~\bibnamefont {Forstner}}, \bibinfo
  {author} {\bibfnamefont {L.~M.}\ \bibnamefont {Fraile}}, \bibinfo {author}
  {\bibfnamefont {H.~O.~U.}\ \bibnamefont {Fynbo}}, \bibinfo {author}
  {\bibfnamefont {A.~G.}\ \bibnamefont {Camacho}}, \bibinfo {author}
  {\bibfnamefont {J.~G.}\ \bibnamefont {Johansen}}, \bibinfo {author}
  {\bibfnamefont {B.}~\bibnamefont {Jonson}}, \bibinfo {author} {\bibfnamefont
  {M.~V.}\ \bibnamefont {Lund}}, \bibinfo {author} {\bibfnamefont
  {J.}~\bibnamefont {Lachner}}, \bibinfo {author} {\bibfnamefont
  {M.}~\bibnamefont {Madurga}}, \bibinfo {author} {\bibfnamefont
  {S.}~\bibnamefont {Merchel}}, \bibinfo {author} {\bibfnamefont
  {E.}~\bibnamefont {Nacher}}, \bibinfo {author} {\bibfnamefont
  {T.}~\bibnamefont {Nilsson}}, \bibinfo {author} {\bibfnamefont
  {P.}~\bibnamefont {Steier}}, \bibinfo {author} {\bibfnamefont
  {O.}~\bibnamefont {Tengblad}},\ and\ \bibinfo {author} {\bibfnamefont
  {V.}~\bibnamefont {Vedia}},\ }\href
  {https://doi.org/10.1140/epja/s10050-020-00110-2} {\bibfield  {journal}
  {\bibinfo  {journal} {The European Physical Journal A}\ }\textbf {\bibinfo
  {volume} {56}},\ \bibinfo {pages} {100} (\bibinfo {year} {2020})}\BibitemShut
  {NoStop}%
\bibitem [{\citenamefont {Savajols}()}]{Savajols2022}%
  \BibitemOpen
  \bibfield  {author} {\bibinfo {author} {\bibfnamefont {H.}~\bibnamefont
  {Savajols}},\ }\href@noop {} {}\bibinfo {howpublished} {private
  communication}\BibitemShut {NoStop}%
\end{thebibliography}%


%

\end{document}